\documentclass{article}

\usepackage{arxiv}

\usepackage{etoolbox}

\usepackage[utf8]{inputenc} 
\usepackage[T1]{fontenc}    
\usepackage{hyperref}       
\usepackage{url}            
\usepackage{booktabs}       
\usepackage{amsfonts}       
\usepackage{nicefrac}       
\usepackage{microtype}      
\usepackage{lipsum}
\usepackage{graphicx}
\graphicspath{ {./images/} }

\usepackage[utf8]{inputenc} 
\usepackage[T1]{fontenc}    
\usepackage{hyperref}       
\usepackage{url}            
\usepackage{booktabs}       
\usepackage{amsfonts}       
\usepackage{nicefrac}       
\usepackage{microtype}      
\usepackage{lipsum}
\usepackage{fancyhdr}       
\usepackage{graphicx}       
\graphicspath{{media/}}     

\usepackage{algorithm}%
\usepackage{algorithmic}

\usepackage{tikz}
\usetikzlibrary{arrows.meta, positioning, shadows}
\usepackage{amsmath}
\usepackage{caption}
\usepackage{subcaption}
\usepackage{comment}
\usepackage{bm}
\newcommand{\bu}{{\bm u}}

\usepackage{xcolor}
\newcommand{\meen}[1]{\textcolor{blue}{#1}}

\newtheorem{remark}{Remark}%

\title{Enhancing neural network extrapolation in thermo-fluid systems using steady-state solutions}

\author{
 Sanjeeb Poudel \\
  Department of Scientific Computing \\
  Florida State University\\
  Tallahassee, FL \\
  \texttt{spoudel@fsu.edu} \\
   \And
   Teeratorn Kadeethum\\
  AI Lab\\
  Siemens Energy\\
  Orlando, FL \\
  \texttt{meen.kadeethum@siemens-energy.com} \\
  \And
  Sanghyun Lee\\
  Department of Mathematics \\
  Florida State University\\
  Tallahassee, FL \\
  \texttt{slee17@fsu.edu} \\
}

\begin{document}
\maketitle
\begin{abstract}
Time-dependent partial differential equations (PDEs) arise in many engineering systems, including thermo-fluid applications. Classical numerical simulations of such systems can become computationally expensive for long-time dynamics because they typically require sequential time integration with time steps constrained by stability, accuracy, or nonlinear solvers. Although scientific machine learning provides an alternative for approximating PDE solutions, standard neural network approximations often degrade when extrapolated beyond the training time interval.

In this work, we propose a steady-state-informed neural network representation for dissipative PDE systems whose solutions relax toward a stationary equilibrium. The proposed ansatz decomposes the solution into a steady-state component and a transient correction modulated by a time-dependent decay profile. When the decay profile vanishes at long time and the transient correction remains bounded, the representation embeds convergence to the prescribed steady state directly into the architecture, rather than enforcing it through an additional penalty term. This allows the network to learn the transient dynamics while preserving the correct asymptotic behavior.

We implement the approach within a physics-informed neural network (PINN) framework and train the resulting model using the SOAP optimizer. The method is evaluated on a sequence of problems of increasing physical and geometric complexity, ranging from the one-dimensional heat equation to incompressible Navier–Stokes flow in a lid-driven cavity, natural convection in a square cavity, and a full three-dimensional conjugate heat transfer problem. The numerical results show that the steady-state-informed architecture substantially improves temporal extrapolation beyond the training interval compared with architectures that do not explicitly enforce the asymptotic condition.
\end{abstract}


\section{Introduction}
\label{sec:intro}
Differential equations are the primary mathematical language used to describe
physical phenomena in engineering. While analytical 
solutions exist for a few canonical cases, the nonlinearity of most 
governing equations and the geometric complexity of real-world 
configurations necessitate numerical approximation. Finite 
difference \cite{leveque2007finite}, finite volume 
\cite{leveque2002finite, versteeg2007introduction}, and finite 
element methods \cite{hughes2012finite, zienkiewicz2005finite} 
provide accurate and reliable solutions across a broad range of 
applications \cite{quarteroni2009numerical, reddy2019introduction}, but time-dependent simulations based on these methods can become 
computationally expensive when long-time dynamics must be resolved. 
The cost arises from the sequential nature of time-stepping: each step 
advances the solution by a small increment dictated by stability or accuracy 
constraints, and when the system's relaxation timescale greatly exceeds the allowable time step, the total number of steps becomes prohibitively large.

Many engineering systems of practical interest are governed by 
time-dependent PDEs that ultimately converge to a stationary equilibrium. 
This class of problems is ubiquitous in thermal and fluid systems encountered 
in energy infrastructure. For instance, the thermal management of power transformers involves transient heat conduction and convection processes that, following a load change or fault event, eventually settle into a steady-state temperature distribution determined by the balance between internal 
losses and external cooling 
\cite{susa2005dynamic, seddik2024thermal, zhang2018numerical, 
dorella2024enhancing}. Similarly, the startup and load-transient behavior of gas-turbine cooling passages involves complex conjugate heat transfer 
that relaxes toward thermal equilibrium once operating conditions stabilize 
\cite{cai2022impacts, tang2016numerical}. In fluid dynamics, internal cavity 
flows, such as those arising in sealed transformer tanks or enclosed cooling 
chambers, can exhibit transient vortex development that gradually relaxes to a 
steady recirculating pattern governed by the imposed boundary conditions and 
the Reynolds number \cite{ghia1982high, shankar2000fluid}.

In these applications, the transient phase is the primary 
object of engineering interest, as it governs thermal stresses, peak 
temperatures, and flow-induced loads during operational transitions 
\cite{cai2022impacts, incropera2007fundamentals}. Yet it is precisely the long-time integration required to capture the full evolution from initial conditions to a steady state that makes simulations expensive. We often need a fine mesh to resolve boundary layers, which imposes restrictive time-step limits to maintain stability, accuracy, and nonlinear solver convergence. Therefore, the required computational step size is orders of magnitude smaller than the system's actual relaxation timescale~\cite{tang2016numerical, dorella2024enhancing}. The computational burden 
is further compounded when repeated simulations are needed for design 
optimization or uncertainty quantification 
\cite{eldred2009recent, xiu2010numerical}. These factors motivate 
the development of methods that can bypass the costly long-time 
integration while still capturing the essential transient dynamics.

The heat equation and the incompressible Navier--Stokes 
equations---along with their coupling through buoyancy under the 
Boussinesq approximation---form the mathematical backbone of these energy-infrastructure problems, including transformer cooling, gas-turbine thermal management, and thermally driven cavity flows. A common structural feature of the dissipative regimes considered here is that their governing physics admits a steady-state equilibrium that encodes the system's long-time behavior. This property is not merely a simplifying assumption but an exploitable piece of physical knowledge: if the asymptotic state is known or can be computed cheaply, it can serve as an anchor for predicting the transient evolution. The present work systematically develops this idea.

In recent years, scientific machine learning has emerged as a compelling 
alternative for approximating solutions to differential equations and 
predicting dynamical systems \cite{karniadakis2021physics, brunton2020machine}. 
Physics-informed Neural Networks (PINNs) \cite{raissi2019physics, 
kadeethum2020pinn_forward} and their variants 
\cite{jagtap2020conservative, kadeethum2020pinn_inverse, 
lu2021deepxde} integrate observational data with the governing 
equations directly in the loss function, enabling solutions to 
both forward and inverse problems with limited or no labeled 
simulation data. Concurrently, neural operators---such as the Deep 
Operator Network (DeepONet) \cite{lu2021learning}, the Fourier 
Neural Operator (FNO) \cite{li2021fourier}, and their variants 
\cite{wang2021learning_pideeponet, kovachki2023neural}---have 
shifted the paradigm from learning pointwise function approximations 
to learning mappings between infinite-dimensional function spaces. 
Despite these advances, capturing accurate long-time dynamics 
remains a persistent challenge. PINN-based surrogates may achieve accurate 
approximations within the training domain, but often degrade rapidly during 
temporal extrapolation, as standard network architectures lack the inductive bias needed to represent asymptotic decay. Neural operators face a 
related but distinct difficulty: autoregressive rollout strategies 
can accumulate errors at each prediction step, while models trained on fixed-length trajectories may fail to generalize beyond the training-time horizon.

In this work, we address the extrapolation limitation by 
exploiting a structural property shared by a large class of 
dissipative physical systems: their solutions ultimately 
converge to a stationary equilibrium. The steady-state solution 
encodes the asymptotic behavior of the system. In many cases, especially when 
the relaxation timescale is long, this steady state can be obtained at 
significantly lower cost than the full transient simulation. We propose a 
solution ansatz that decomposes the network prediction into a steady-state 
component and a transient correction modulated by a time-dependent decay 
function. The steady-state component can be supplied from any available 
source, including a classical numerical solver, experimental measurements, or a 
separately trained neural model. The decay function can be chosen as an 
exponential, algebraic, or damped oscillatory profile. As this decay 
function tends to zero, provided that the transient correction remains bounded,
the network prediction converges to the prescribed steady state while retaining 
the flexibility to learn the transient dynamics within the training window. 
Because the ansatz imposes convergence at the architectural level 
rather than through the loss function, it is modular and can be 
integrated with different neural network frameworks. Thus, the main 
contribution is not an additional penalty term, but a steady-state-informed 
representation that embeds the correct long-time limit directly into the neural 
network architecture.

We instantiate this approach within a PINN framework, embedding the governing differential equations directly into the training loss. Our longer-term objective is a parametric surrogate that maps system parameters to transient solutions; the present non-parametric experiments serve to establish the validity of the steady-state-informed ansatz before its extension to that parametric setting. Training PINNs remains challenging due to the ill-conditioned, multi-objective nature of the composite loss landscape \cite{wang2022when, krishnapriyan2021characterizing}, and several remedies have been proposed, including curriculum learning \cite{krishnapriyan2021characterizing}, sequence-to-sequence training \cite{mattey2022novel}, adaptive loss weighting \cite{wang2021understanding}, and residual-based adaptive sampling \cite{wu2023comprehensive}. Complementary to these training strategies, we adopt the recently proposed SOAP optimizer \cite{vyas2024soap}, a Kronecker-factored preconditioned method that is designed to improve gradient conditioning and reduce directional conflicts within the multi-term loss landscapes characteristic of PINNs \cite{wang2025gradient}.

To validate the proposed method, we conduct a sequence of numerical experiments. We begin with the one-dimensional heat equation, followed by the 
two-dimensional incompressible Navier--Stokes equations in a lid-driven cavity. 
Next, we evaluate the method on natural convection in a square cavity in an 
advection-dominated regime. We conclude our evaluation with a three-dimensional 
conjugate heat transfer problem.

\section{Background}\label{sec:background}
This section provides a brief overview of neural networks and physics-informed 
neural networks \cite{raissi2019physics}. We begin by introducing the 
fundamental architecture of feedforward neural networks, followed by a discussion 
of the PINN framework and its loss function. Finally, we discuss the SOAP optimizer \cite{vyas2024soap} used throughout this work. 

\subsection{Neural Networks}

A feedforward neural network (FNN), also known as a multilayer perceptron, 
is a parametric function that maps an input vector 
$\bm{z} \in \mathbb{R}^{d_{\text{in}}}$ to an output 
$\bm{y} \in \mathbb{R}^{d_{\text{out}}}$ through a composition of affine 
transformations and nonlinear activation functions 
\cite{goodfellow2016deep}. For a network with $L$ hidden layers, the 
computation proceeds as
\begin{equation}
    \bm{h}^{(0)} = \bm{z}, \qquad
    \bm{h}^{(\ell)} = \sigma\!\left( \bm{W}^{(\ell)} \bm{h}^{(\ell-1)} 
    + \bm{b}^{(\ell)} \right), \quad \ell = 1, \ldots, L, \qquad
    \bm{y} = \bm{W}^{(L+1)} \bm{h}^{(L)} + \bm{b}^{(L+1)},
\end{equation}
where $\bm{W}^{(\ell)} \in \mathbb{R}^{n_\ell \times n_{\ell-1}}$ and 
$\bm{b}^{(\ell)} \in \mathbb{R}^{n_\ell}$ are the trainable weight 
matrices and bias vectors, $n_\ell$ denotes the width of the $\ell$-th 
layer, and $\sigma(\cdot)$ is a nonlinear activation function applied 
element-wise. Common choices for $\sigma$ include the hyperbolic tangent 
($\tanh$), the rectified linear unit (ReLU), and the sigmoid linear unit 
(SiLU) \cite{ramachandran2017searching}. The collection of all trainable 
parameters is denoted by 
$\bm{\theta} = \{ \bm{W}^{(\ell)}, \bm{b}^{(\ell)} \}_{\ell=1}^{L+1}$.
 
The universal approximation theorem guarantees that, under mild conditions 
on the activation function, a sufficiently wide single-hidden-layer 
network can approximate any continuous function on a compact domain to 
arbitrary accuracy \cite{hornik1989multilayer, cybenko1989approximation}. 
In practice, deeper architectures with moderate width are often preferred, as 
they can represent complex functions more efficiently 
\cite{lu2017expressive}. The parameters $\bm{\theta}$ are determined by 
minimizing a loss function $\mathcal{L}(\bm{\theta})$ via gradient-based 
optimization, where gradients are computed efficiently using automatic 
differentiation \cite{baydin2018automatic}.

\subsection{Physics-Informed Neural Network (PINN)}

Physics-informed neural networks (PINNs), introduced by Raissi et al.\ 
\cite{raissi2019physics}, embed the governing PDE directly into the 
training loss, enabling the network to approximate PDE solutions without 
requiring labeled simulation data. Let $\Omega \subset \mathbb{R}^d$ be a spatial domain in $d$ dimensions and $I := (0, t_f]$ be a time interval with final time $t_f$. Consider a general partial differential equation (PDE) of the form:
\begin{equation}
    \frac{\partial u}{\partial t} + \mathcal{N}[u; \xi] = \gamma(\bm{x}, t), \quad (\bm{x}, t) \in \Omega \times I,
    \label{eq:general_pde}
\end{equation}
subject to the boundary condition $\mathcal{B}[u] = g$ on $\partial\Omega \times I$ and the initial condition $u(\bm{x}, 0) = u_0(\bm{x})$ for $\bm{x} \in \Omega$. Here, $\mathcal{N}[\cdot; \cdot]$ is a (possibly nonlinear) spatial differential operator with physical parameters $\xi$, and $\gamma(\cdot, \cdot)$ is the source term. In the PINN framework, the solution $u(\bm{x}, t)$ is approximated by a neural network $u_\theta(\bm{x}, t)$, and the required partial derivatives appearing in~\eqref{eq:general_pde} are computed by automatic differentiation through the network's computational graph \cite{baydin2018automatic}.
 
The network parameters $\bm{\theta}$ are obtained by minimizing a 
composite loss function of the form
\begin{equation}
    \mathcal{L}(\bm{\theta}) = 
    \lambda_r \mathcal{L}_r(\bm{\theta}) + 
    \lambda_b \mathcal{L}_b(\bm{\theta}) + 
    \lambda_0 \mathcal{L}_0(\bm{\theta}),
    \label{eq:pinn_loss}
\end{equation}
where $\mathcal{L}_r$, $\mathcal{L}_b$, and $\mathcal{L}_0$ denote the 
PDE residual, boundary, and initial condition losses, respectively, and 
$\lambda_r$, $\lambda_b$, $\lambda_0 > 0$ are weighting coefficients. 
These loss terms are evaluated at a set of collocation points sampled 
within the domain. Specifically, the PDE residual loss is given by
\begin{equation}
    \mathcal{L}_r(\bm{\theta}) = \frac{1}{N_r} \sum_{i=1}^{N_r} 
    \left| \frac{\partial u_\theta}{\partial t}(\bm{x}_i^r, t_i^r) + 
    \mathcal{N}[u_\theta; \xi](\bm{x}_i^r, t_i^r) - \gamma(\bm{x}_i^r, t_i^r) \right|^2,
\end{equation}
where $\{(\bm{x}_i^r, t_i^r)\}_{i=1}^{N_r}$ are collocation points in 
the interior of the space--time domain. The boundary and initial condition 
losses $\mathcal{L}_b$ and $\mathcal{L}_0$ are constructed analogously by 
penalizing deviations from the prescribed boundary data and initial state 
at their respective collocation points.
 
A key advantage of PINNs is that they are mesh-free in the sense that they 
do not require a fixed computational mesh: the collocation points can be 
sampled randomly or quasi-randomly, and the method does not require an 
explicit discretization of the differential operators. This makes PINNs 
particularly attractive for problems in complex geometries and high-dimensional 
spaces \cite{karniadakis2021physics}. The framework has been successfully 
applied to a broad range of forward and inverse problems, including nonlinear 
diffusivity and poromechanics governed by Biot's equations 
\cite{kadeethum2020pinn_forward, kadeethum2020pinn_inverse}. However, 
training PINNs remains challenging, especially for stiff systems and 
problems requiring long-time integration. The multi-objective nature of the 
composite loss \eqref{eq:pinn_loss} introduces competing gradients, and the  
PDE residual loss landscape is often ill-conditioned 
\cite{wang2022when, krishnapriyan2021characterizing}. Various strategies 
have been proposed to address these difficulties, including adaptive loss 
weighting \cite{wang2021understanding}, curriculum training 
\cite{krishnapriyan2021characterizing}, causal training 
\cite{wang2024respecting}, and sequence-to-sequence learning 
\cite{mattey2022novel}. To optimize our PINN framework, we employ the recently 
developed SOAP algorithm \cite{vyas2024soap,wang2025gradient}, which is 
designed to improve gradient conditioning and reduce directional conflicts 
between loss terms.

\subsection{Shampoo with Adam in the Preconditioner's Eigenbasis (SOAP)}
\label{sec:SOAP}

PINNs are typically trained with first-order methods. Adam 
\cite{kingma2014adam} is the most common choice, sometimes combined 
with L-BFGS for fine-tuning \cite{raissi2019physics}. Adam is efficient, 
but it uses only a diagonal approximation of the second-moment matrix. 
This approximation ignores parameter correlations. As a result, Adam 
can converge slowly on ill-conditioned loss landscapes.

Shampoo \cite{gupta2018shampoo} addresses this limitation through 
Kronecker-factored preconditioning. For a weight matrix 
$\bm{W}_k \in \mathbb{R}^{m \times n}$ at iteration $k$, it maintains 
two preconditioners, $\bm{L}_k \in \mathbb{R}^{m \times m}$ and 
$\bm{R}_k \in \mathbb{R}^{n \times n}$. These are updated as
\begin{equation}
\begin{split}
    \bm{L}_k &\leftarrow \bm{L}_{k-1} + \bm{G}_k \bm{G}_k^\top, \\
    \bm{R}_k &\leftarrow \bm{R}_{k-1} + \bm{G}_k^\top \bm{G}_k,
\end{split}
\end{equation}
where $\bm{G}_k$ is the gradient. With learning rate $\eta > 0$, the 
preconditioned update is
\begin{equation}
    \bm{W}_{k+1} \leftarrow \bm{W}_k - \eta \, \bm{L}_k^{-1/4}\,\bm{G}_k\,\bm{R}_k^{-1/4}.
    \label{eq:shampoo_update}
\end{equation}
To avoid a singular inverse at the first step, the preconditioners are 
regularized at initialization, e.g., $\bm{L}_0 = \bm{R}_0 = \epsilon \bm{I}$ 
with $0 < \epsilon \ll 1$. This non-diagonal preconditioning captures 
inter-parameter correlations that Adam misses. Shampoo has been shown to 
improve over Adam on several optimization benchmarks \cite{gupta2018shampoo}.

Shampoo, however, incurs much higher overhead than Adam. The bottleneck 
is the repeated eigendecomposition of $\bm{L}_k$ and $\bm{R}_k$ during 
training. Computing these eigendecompositions less frequently reduces 
cost, but degrades performance \cite{vyas2024soap}. SOAP 
\cite{vyas2024soap} resolves this trade-off. It performs the 
eigendecomposition only periodically. Between eigendecomposition steps, 
it maintains Adam-type moment estimates in the current rotated coordinate 
system. In effect, SOAP runs Adam in a slowly varying eigenspace defined 
by Shampoo's preconditioner.

SOAP rests on a formal equivalence between a variant of Shampoo and 
Adafactor \cite{shazeer2018adafactor}, a memory-efficient variant of Adam. 
The variant applies the preconditioner with power $-1/2$, i.e., 
$\bm{L}_k^{-1/2} \bm{G}_k \bm{R}_k^{-1/2}$, rather than the $-1/4$ form 
in~\eqref{eq:shampoo_update}. The key insight is that this form of Shampoo 
is equivalent to an adaptive optimizer running in the eigenbasis of its 
own preconditioner. SOAP exploits this equivalence in three steps. First, 
it projects the gradient into the preconditioner's eigenbasis. Second, it 
updates Adam-style first- and second-moment estimates in this rotated 
space. Third, it projects the update direction back to the original 
parameter space. The eigendecomposition is computed periodically rather 
than at every step. Between updates, the moment estimates continue to 
accumulate in the current rotated basis.

The preconditioner matrices are updated using moving averages:
\begin{equation}
    \begin{split}
        \bm{L}_k &\leftarrow \beta_2 \bm{L}_{k-1}
        + (1 - \beta_2) \bm{G}_k \bm{G}_k^\top, \\
        \bm{R}_k &\leftarrow \beta_2 \bm{R}_{k-1}
        + (1 - \beta_2)\bm{G}_k^\top \bm{G}_k,
    \end{split}
\end{equation}
where $\beta_2 \in [0, 1)$ controls the moving average. We initialize 
$\bm{L}_0 = \bm{R}_0 = \bm{0}$. The exponential moving average and 
subsequent bias correction, analogous to Adam's moment estimates, 
remove the need for identity regularization. The matrices are then 
eigendecomposed as
\begin{equation}
    \begin{split}
        \bm{L}_k &= \bm{Q}_L \Lambda_L \bm{Q}_L^\top, \\
        \bm{R}_k &= \bm{Q}_R \Lambda_R \bm{Q}_R^\top,
    \end{split}
\end{equation}
where $\Lambda_L$ and $\Lambda_R$ are diagonal matrices of eigenvalues. 
The gradient is then projected into the eigenspace:
\begin{equation}
    \widetilde{\bm{G}}_k = \bm{Q}_L^\top \bm{G}_k \bm{Q}_R .
\end{equation}

SOAP then applies an Adam-style update in the rotated space, with an 
asymmetric treatment of the two moments. The first moment $\bm{M}_k$ 
is accumulated from the original gradient $\bm{G}_k$. It is then 
projected as $\widetilde{\bm{M}}_k = \bm{Q}_L^\top \bm{M}_k \bm{Q}_R$. 
The second moment $\bm{V}_k$ is accumulated directly from the projected 
gradient $\widetilde{\bm{G}}_k$. After bias correction, the normalized 
update direction $\widehat{\bm{N}}$ in the rotated space is mapped back 
to the original parameter space:
\begin{equation}
    \bm{N}_k = \bm{Q}_L \widehat{\bm{N}} \, \bm{Q}_R^\top,
    \qquad
    \bm{W}_{k+1} = \bm{W}_k - \eta \, \bm{N}_k .
\end{equation}

SOAP introduces only one extra hyperparameter relative to Adam: the 
preconditioning frequency. Empirical results in \cite{vyas2024soap} 
show that SOAP reduces both the iteration count and the wall-clock time 
compared to AdamW and Shampoo. We use SOAP as the primary optimizer in 
this work. Algorithm~\ref{alg:soap} summarizes the procedure.

\begin{algorithm}[h!]
    \caption{Single step of SOAP for an $m \times n$ layer \cite{vyas2024soap}}
    \label{alg:soap}
    For each layer, we have $\bm{L}_k \in \mathbb{R}^{m \times m}$, 
    $\bm{R}_k \in \mathbb{R}^{n \times n}$, learning rate 
    $\eta \in \mathbb{R}^+$, $\beta_1, \beta_2 \in [0, 1)$, 
    $0 < \epsilon \ll 1$, preconditioning frequency 
    $n_p \in \mathbb{Z}^+$, and $\bm{V}_k, \bm{M}_k \in \mathbb{R}^{m\times n}$. 
    We initialize $\bm{L}_0 = \bm{R}_0 = \bm{0}$, 
    $\bm{M}_0 = \bm{V}_0 = \bm{0}$, and $\bm{Q}_L = \bm{Q}_R = \bm{I}$.
    \begin{algorithmic}[1]
        \STATE Compute gradient $\bm{G}_k \in \mathbb{R}^{m \times n}$
        \STATE $\bm{L}_k \leftarrow \beta_2 \bm{L}_{k-1} + (1 - \beta_2) \bm{G}_k \bm{G}_k^\top$
        \STATE $\bm{R}_k \leftarrow \beta_2 \bm{R}_{k-1} + (1 - \beta_2)\bm{G}_k^\top \bm{G}_k$
        \IF {$k \mod n_p = 0$}
            \STATE Compute $\bm{Q}_L$ and $\bm{Q}_R$ such that \\
                $\quad\quad \bm{L}_k = \bm{Q}_L \Lambda_L \bm{Q}_L^\top$ \\
                $\quad\quad \bm{R}_k = \bm{Q}_R \Lambda_R \bm{Q}_R^\top$
        \ENDIF
        \STATE $\widetilde{\bm{G}}_k \leftarrow \bm{Q}_L^\top \bm{G}_k \bm{Q}_R$
        \STATE $\bm{M}_k \leftarrow \beta_1 \bm{M}_{k-1} + (1 - \beta_1) \bm{G}_k$
        \STATE $\widetilde{\bm{M}}_k \leftarrow \bm{Q}_L^\top \bm{M}_k \bm{Q}_R$
        \STATE $\bm{V}_k \leftarrow \beta_2 \bm{V}_{k-1} + (1 - \beta_2)(\widetilde{\bm{G}}_k \odot \widetilde{\bm{G}}_k)$ \quad (element-wise)
        \STATE $\widehat{\bm{M}}_k \leftarrow \widetilde{\bm{M}}_k / (1 - \beta_1^k)$
        \STATE $\widehat{\bm{V}}_k \leftarrow \bm{V}_k / (1 - \beta_2^k)$
        \STATE $\widehat{\bm{N}} \leftarrow \widehat{\bm{M}}_k / (\sqrt{\widehat{\bm{V}}_k} + \epsilon)$ \quad (element-wise)
        \STATE $\bm{N} \leftarrow \bm{Q}_L \widehat{\bm{N}} \bm{Q}_R^\top$
        \STATE $\bm{W}_{k+1} \leftarrow \bm{W}_k - \eta \bm{N}$
    \end{algorithmic}
\end{algorithm}

\begin{remark}
SOAP's preconditioners $\bm{L} \in \mathbb{R}^{m \times m}$ and 
$\bm{R} \in \mathbb{R}^{n \times n}$ exploit the two-axis structure 
of matrix-shaped parameters. This Kronecker factorization does not 
apply to one-dimensional parameters such as bias vectors $\bm{b}$ 
or normalization parameters. We therefore apply SOAP only to 
two-dimensional weight matrices $\bm{W}$. All one-dimensional 
parameters are updated with standard Adam, following the convention 
of the reference SOAP implementation~\cite{vyas2024soap}.
\end{remark}

\section{Methodology}
\label{sec:method}

This section details our proposed methodology. We begin by 
representing the solution using a steady-state component and a transient term 
factored into a temporal profile and a space--time correction. 
We then describe the network architecture designed to integrate the steady-state solution into the training process, forming a unified framework for 
time-dependent prediction. Finally, we outline the optimization and training 
procedures for the neural networks. 

\subsection{Solution Ansatz}
Consider a general form of PDE from~\eqref{eq:general_pde}. Suppose the forcing term asymptotically approaches a time-independent function $\gamma_s(\bm{x}) = \lim_{t \rightarrow \infty} \gamma(\bm{x}, t)$, and all boundary conditions on $\partial \Omega$ similarly converge to steady limits. Assuming the resulting steady state is asymptotically stable, the PDE's solution settles into a steady-state limit, 
\[
\lim_{t \rightarrow \infty} u(\bm{x}, t) = u_s(\bm{x}),
\]
where $u_s(\bm{x})$ satisfies the steady-state equation
\begin{equation}
    \mathcal{N}[u_s(\bm{x}); \xi] = \gamma_s(\bm{x}), \quad \text{in } \Omega. 
\end{equation}
Motivated by this asymptotic structure, we approximate the solution using the ansatz
\begin{equation}
    u(\bm{x}, t) = u_s(\bm{x}) + f(t)g(\bm{x}, t), 
    \label{eqn:solution_ansatz}
\end{equation}
where $f(t)$ is a time-dependent function satisfying the asymptotic condition 
\[
\lim_{t\rightarrow\infty} f(t) = 0,
\]
and $g(\bm{x}, t)$ is a function of both space and time. If $g(\bm{x},t)$ 
remains bounded as $t\to\infty$, then the ansatz implies that 
$u(\bm{x}, t) \to u_s(\bm{x})$, so the approximation converges to the 
steady-state solution.

\subsection{Choices of the temporal profile $f(t)$}

As a first step, the temporal profile $f(t)$ in the ansatz 
\eqref{eqn:solution_ansatz} must be specified or learned. 
This choice is inherently problem-dependent and should be guided by the 
expected asymptotic behavior of the solution. For systems that relax toward 
a steady state, a natural strategy is to let $f(t)$ capture the leading-order 
temporal decay, so that the function $g(\bm x,t)$ describes the remaining 
spatial and transient structure.

An attractive feature of the proposed decomposition is that the parameters 
defining the temporal profile $f(t)$ can be learned jointly with the neural 
network parameters. For example, when the underlying dynamics are expected 
to decay monotonically toward equilibrium, one may adopt the exponential form
\begin{equation}
f(t)=e^{-\lambda t},    
\end{equation}
with decay rate $\lambda>0$ treated as a trainable parameter. On the other 
hand, if the transient response exhibits oscillatory behavior before relaxing, 
one may instead use a damped oscillatory profile of the form
\begin{equation}
f(t)=e^{-\lambda t}\cos(\omega t),
\end{equation}
or, more generally,
\begin{equation}
f(t)=e^{-\lambda t}\bigl(a+\cos(\omega t)\bigr),
\end{equation}
where $\lambda$, $a$, and $\omega$ are optimized together with the network 
weights.

To evaluate the sensitivity of the proposed framework to the choice of 
temporal profile, we consider the following four representative cases:
\begin{itemize}
    \item \textbf{Constant profile (baseline):}
    \[
    f(t):=1.
    \]
    This choice does not enforce decay toward the steady state and is therefore included only as a reference case.

    \item \textbf{Exponential decay:}
    \[
    f(t):=e^{-\lambda t}.
    \]

    \item \textbf{Algebraic decay:}
    \[
    f(t):=(1+\lambda t)^{-p}.
    \]

    \item \textbf{Damped oscillatory decay:}
    \[
    f(t):=e^{-\lambda t}\bigl(a+\cos(\omega t)\bigr).
    \]
\end{itemize}

Here, $\lambda>0$ denotes the decay rate, $p>0$ is the algebraic decay 
exponent, $a$ is a shift parameter controlling the magnitude, and 
$\omega>0$ is the angular frequency. 
When a nonnegative oscillatory profile is desired, we enforce 
$a \ge 1$, which guarantees $a+\cos(\omega t)\ge 0$ for all $t$. In our 
implementation, the positivity constraints on $\lambda$, $p$, and $\omega$, 
as well as the lower bound on $a$, are enforced by post-update projection.
With the exception of the constant baseline, all of these profiles satisfy 
the asymptotic condition
\[
f(t)\to 0 \qquad \text{as } t\to\infty,
\]
and are therefore consistent with the steady-state ansatz. Collectively, 
they allow us to represent several qualitatively distinct transient behaviors, 
including no prescribed decay, monotone exponential relaxation, slower 
algebraic decay, and damped oscillatory convergence.

In our implementation, all free parameters appearing in $f(t)$ are treated 
as trainable variables and are optimized jointly with the neural network 
weights. A practical issue arises when the temporal profile $f(t)$ decays 
rapidly. In that case, the network may compensate by producing increasingly 
large values of the correction term $g(\bm{x}, t)$, especially at later times, 
which can adversely affect optimization and reduce training stability. To 
alleviate this difficulty, we constrain the output of the transient network 
by applying a bounded activation function at the final layer. In particular, 
when a hyperbolic tangent activation is used, the network output satisfies
\[
g_\theta(\bm x,t)\in(-1,1).
\]
This boundedness also provides the condition needed for the 
ansatz to converge to the prescribed steady state as $t\to\infty$. At the 
same time, the bounded output limits the maximum representable transient 
amplitude. Therefore, the variables should be nondimensionalized or scaled 
appropriately; for problems with larger transient amplitudes, one may include 
an additional trainable amplitude factor or use a linear output layer with 
suitable regularization. In practice, this improves robustness and stabilizes 
training, particularly for profiles with strong decay.

\subsection{Network Architecture}

Based on the solution ansatz in \eqref{eqn:solution_ansatz}, we construct 
a neural network architecture to approximate the unknown space--time 
correction $g(\bm{x}, t)$ while explicitly incorporating the steady-state 
component $u_s(\bm{x})$ to predict the solution $u(\bm{x}, t)$. The stationary part may be handled in different ways, depending on the available information. It can be precomputed via interpolation from classical numerical solvers or experimental measurements. Alternatively, the steady-state component may be learned with a neural network. In this work, we choose to parameterize $u_s(\bm{x})$ directly using a dedicated neural network sub-architecture.

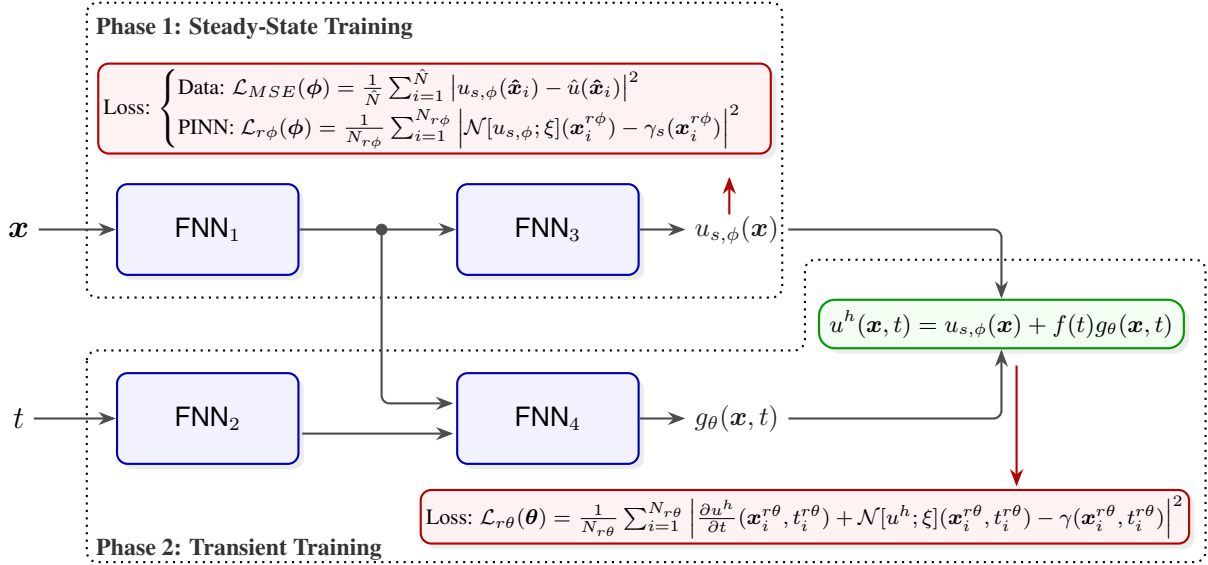
\begin{figure}[!h]
    \centering
    
    \begin{tikzpicture}[
        >=Stealth,
        node distance=2.5cm,
        io/.style={font=\large\bfseries, text=black},
        nnbox/.style={
            draw=blue!70!black,
            fill=blue!5,
            thick,
            rounded corners=4pt,
            minimum width=2.4cm,
            minimum height=1.2cm,
            align=center,
            font=\normalsize\sffamily,
            drop shadow={opacity=0.15, shadow xshift=2pt, shadow yshift=-2pt}
        },
        opnode/.style={
            circle,
            draw=orange!80!black,
            fill=orange!10,
            thick,
            minimum size=0.7cm,
            inner sep=0pt,
            font=\large,
            drop shadow={opacity=0.15, shadow xshift=1.5pt, shadow yshift=-1.5pt}
        },
        variablenode/.style={
            font=\normalsize,
            text=black!90
        },
        arrow/.style={->, thick, draw=black!70, rounded corners=3pt},
        highlightbox/.style={
            draw=green!60!black,
            fill=green!5,
            thick,
            rounded corners=6pt,
            inner sep=4pt,
            font=\footnotesize,
            drop shadow={opacity=0.15, shadow xshift=2pt, shadow yshift=-2pt}
        },
        lossnode/.style={
            draw=red!70!black,
            fill=red!5,
            thick,
            rounded corners=4pt,
            inner sep=2pt,
            align=center,
            font=\fontsize{8.2pt}{9pt}\selectfont,,
            drop shadow={opacity=0.15, shadow xshift=2pt, shadow yshift=-2pt}
        }
    ]
    
    \draw[thick, dotted, rounded corners=4pt] (0.9, 3.0) rectangle (10.1, -0.9);
    \node[above right, font=\small\bfseries, text=black!80] at (0.9, 2.4) {Phase 1: Steady-State Training};

    \draw[thick, dotted, rounded corners=4pt] 
        (0.9, -1.6) --  
        (0.9, -4.4) --  
        (15.7, -4.4) -- 
        (15.7, -0.4) -- 
        (10.4, -0.4) -- 
        (10.4, -1.6) -- 
        cycle;          
        
    \node[above right, font=\small\bfseries, text=black!80] at (0.9, -4.5) {Phase 2: Transient Training};
        
    \node[io] (x) at (0, 0) {$\bm{x}$};
    \node[io] (t) at (0, -2.5) {$t$};
    
    \node[nnbox] (nn1) at (2.5, 0) {FNN$_1$};
    \node[nnbox] (nn2) at (2.5, -2.5) {FNN$_2$};
    
    \node[nnbox] (nn3) at (7.0, 0) {FNN$_3$};
    \node[nnbox] (nn4) at (7.0, -2.5) {FNN$_4$};
    
    \node[variablenode] (uss) at (9.5, 0) {$u_{s,\phi}(\bm{x})$};
    \node[variablenode] (g) at (9.5, -2.5) {$g_\theta(\bm{x}, t)$};
    
    \node[highlightbox] (u) at (13.0, -1.25) {$u^h(\bm{x}, t) = u_{s,\phi}(\bm{x}) + f(t)g_\theta(\bm{x}, t)$};
    
    \node[lossnode] (loss1) at (5.5, 1.6) {Loss: $\begin{cases}\text{Data: }\mathcal{L}_{MSE}(\bm{\phi}) = \frac{1}{\hat{N}} \sum_{i=1}^{\hat{N}} 
    \left| u_{s,\phi}(\bm{\hat{x}}_i) - \hat{u}(\bm{\hat{x}}_i) \right|^2 \\     \text{PINN: }\mathcal{L}_{r\phi}(\bm{\phi}) = \frac{1}{N_{r\phi}} \sum_{i=1}^{N_{r\phi}} 
    \left| \mathcal{N}[u_{s,\phi}; \xi](\bm{x}_i^{r\phi}) - \gamma_s(\bm{x}_i^{r\phi})\right|^2 \end{cases}$};
    
    \node[lossnode] (loss2) at (10.4, -3.8) {Loss: $\mathcal{L}_{r\theta}(\bm{\theta}) = \frac{1}{N_{r\theta}} \sum_{i=1}^{N_{r\theta}} 
    \left| \frac{\partial u^h}{\partial t}(\bm{x}_i^{r\theta}, t_i^{r\theta}) + 
    \mathcal{N}[u^h; \xi](\bm{x}_i^{r\theta}, t_i^{r\theta}) - \gamma(\bm{x}_i^{r\theta}, t_i^{r\theta}) \right|^2$};
    
    
    \draw[arrow] (x) -- (nn1);
    \draw[arrow] (t) -- (nn2);
    
    \draw[arrow] (nn1) -- (nn3);
    \filldraw[black!70] (4.8, 0) circle (2pt); 
    \draw[arrow] (4.8, 0) |- (5.75, -2.3);

    \draw[arrow] (3.75, -2.7) -- (5.75, -2.7);
    
    \draw[arrow] (nn3) -- (uss);
    \draw[arrow] (nn4) -- (g);
    
    \draw[arrow] (uss) -| (u.north);
    \draw[arrow] (g) -| (u.south);

    \draw[arrow, color=red!70!black] (9.4, 0.2) -- (9.4, 0.7);
    \draw[arrow, color=red!70!black] (13.2, -1.8) -- (13.2, -3.4);

    \end{tikzpicture}  
    \caption{Proposed network architecture and multi-phase training strategy. Phase 1 (Steady-State): Fully connected blocks ($\text{FNN}_1$, $\text{FNN}_3$) map spatial inputs $\bm{x}$ to a steady-state baseline $u_{s,\phi}(\bm{x})$, optimized via data or physics-informed residual loss. Phase 2 (Transient): Temporal inputs $t$ and spatial features are merged to capture the space-time correction $g_\theta(\bm{x}, t)$. The combined spatio-temporal solution $u^h(\bm{x}, t)$ is then trained using a time-dependent physics-informed loss. Initial and boundary conditions are enforced as soft constraints (omitted from the diagram for brevity). 
    }
    \label{fig:architecture}
\end{figure}

To achieve this, we design a unified framework comprising four fully connected feedforward neural network blocks, denoted as FNN$_i$ for $i \in \{1, 2, 3, 4\}$. This configuration is designed to output the steady-state baseline, the transient correction, and the final coupled spatio-temporal solution for any given set of coordinates. Figure~\ref{fig:architecture} illustrates the proposed network architecture 
for approximating the PDE solution $u^h(\bm{x}, t)$. The model takes the 
spatial coordinates $\bm{x}$ and time $t$ as inputs. First, $\bm{x}$ and 
$t$ are passed through two parallel subnetworks, FNN$_1$ and FNN$_2$, 
which extract spatial and temporal feature embeddings, respectively. The 
spatial embedding is then passed to FNN$_3$ to approximate the steady-state 
field $u_{s,\phi}(\bm{x})$. The spatial and temporal embeddings are also 
combined and passed to FNN$_4$ to approximate the transient correction 
$g_\theta(\bm{x}, t)$. Finally, the two outputs are coupled through the 
ansatz
\begin{equation}
u^h(\bm{x}, t)=u_{s,\phi}(\bm{x})+f(t)\,g_\theta(\bm{x}, t).
\end{equation}
This formulation explicitly separates the stationary solution from the transient dynamics, enabling a targeted, two-phase optimization strategy.

\subsection{Training}
The model is trained in two phases using the SOAP algorithm detailed in Section \ref{sec:SOAP}. Initially, the two fully connected layers, FNN$_1$ and FNN$_3$, are trained on the steady-state solution. These networks can be trained using either data or a physics-informed loss. When training with data, we use the mean-squared loss
\begin{equation}
    \mathcal{L}_{MSE}(\bm{\phi}) = \frac{1}{\hat{N}} \sum_{i=1}^{\hat{N}} 
    \left| u_{s,\phi}(\bm{\hat{x}}_i) - \hat{u}(\bm{\hat{x}}_i) \right|^2,    
\end{equation}
where $\{\left( \bm{\hat{x}}_i, \hat{u}(\bm{\hat{x}}_i)\right) \}_{i=1}^{\hat{N}}$ are the total available training data.
If we employ the PINN framework, the loss formulation is given by
\begin{equation}
    \mathcal{L}_{r\phi}(\bm{\phi}) = \frac{1}{N_{r\phi}} \sum_{i=1}^{N_{r\phi}} 
    \left| \mathcal{N}[u_{s,\phi}; \xi](\bm{x}_i^{r\phi}) - \gamma_s(\bm{x}_i^{r\phi})\right|^2,
\end{equation}
where $\{ \bm{x}_i^{r\phi} \} \in \Omega$ represents a random collocation point in the domain, and $N_{r\phi}$ denotes the total number of collocation points. Note that the boundary condition is implemented as a soft constraint.  

Once FNN$_1$ and FNN$_3$ are trained, their weights and biases are frozen. Subsequently, FNN$_2$ and FNN$_4$ are optimized based on the transient loss. The output $g_\theta(\bm{x}, t)$ is used to derive the approximate transient solution $u^h(\bm{x}, t)$. Therefore, the physics-informed loss to train FNN$_2$ and FNN$_4$ is given by
\begin{equation}
    \mathcal{L}_{r\theta}(\bm{\theta}) = \frac{1}{N_{r\theta}} \sum_{i=1}^{N_{r\theta}} 
    \left| \frac{\partial u^h}{\partial t}(\bm{x}_i^{r\theta}, t_i^{r\theta}) + 
    \mathcal{N}[u^h; \xi](\bm{x}_i^{r\theta}, t_i^{r\theta}) - \gamma(\bm{x}_i^{r\theta}, t_i^{r\theta}) \right|^2,
\end{equation}
where $\{\left( \bm{x}_i^{r\theta}, t_i^{r\theta} \right)\} \in \Omega \times I$ represent a random spatio-temporal collocation point, and $N_{r\theta}$ denotes the total number of collocation points.

\section{Numerical Experiments}
\label{sec:numerics}

In this section, we present a series of examples to study the method's 
extrapolation capability for different choices of the temporal profile 
$f(t)$ and to apply the method to several heat-transfer and fluid-dynamics problems. We begin with Example~\ref{sec:Eg1}, a one-dimensional diffusion 
equation, and then extend the framework to the lid-driven cavity flow in 
Example~\ref{sec:Eg2} to study the performance of the proposed decomposition for the Navier--Stokes equations. Example~\ref{sec:Eg3} investigates the coupled fluid and heat dynamics in the classical natural convection benchmark problem in a square cavity. Finally, Example~\ref{sec:Eg4} applies the method to a three-dimensional conjugate heat transfer problem.

All examples are implemented using SciPy~\cite{2020SciPy-NMeth}, 
NumPy~\cite{harris2020array}, Keras~\cite{chollet2015keras}, and 
TensorFlow~\cite{tensorflow2015-whitepaper}. Matplotlib~\cite{Hunter:2007} 
is used to generate the figures. For all numerical experiments, the hyperparameters for the SOAP algorithm are fixed as follows: learning rate $\eta = 10^{-3}$, preconditioning frequency $n_p = 10$, $\beta_1 = 0.9$, $\beta_2 = 0.999$, and 
$\epsilon = 10^{-8}$. Unless otherwise stated, we set the loss weights to
$\lambda_r = \lambda_b = \lambda_0 = 1$. The reference solution for fluid and heat dynamics in two or three dimensions is computed using the enriched Galerkin finite element method developed in~\cite{poudel2025pressure} using the deal.II library~\cite{arndt2023deal}.

\subsection{Example 1: Heat Equation}
\label{sec:Eg1}

In this example, we consider the one-dimensional diffusion equation with 
a time-independent forcing term. The solution approaches a
stationary state, and the rate of convergence is determined by the 
diffusion constant and the eigenvalues of the Laplacian operator. 
For the domain $\Omega = (0,1)$, the temperature 
distribution $u: \Omega \times I \rightarrow \mathbb{R}$ satisfies
\begin{equation}
    \frac{\partial u}{\partial t} - \alpha \frac{\partial^2u}{\partial x^2} 
    = \gamma(x), \quad \text{in } \Omega \times I,
\end{equation}
where $\alpha$ is the diffusion constant and $\gamma(x)$ is a prescribed 
time-independent forcing term. For this example, we impose 
homogeneous Dirichlet boundary conditions,
$u(0,t)=u(1,t)=0$ for $t\in I$,
and the initial condition
\begin{equation}
    u(x,0) = \sin(\pi x), \quad x \in \Omega.
\end{equation}
The forcing term is chosen so that the stationary solution is
\begin{equation}
    u_s(x) = x(1-x)\bigl(3 + \sin(3\pi x) + \sin(2\pi x)\bigr), 
\end{equation}
which satisfies the steady-state equation
\begin{equation}
-\alpha \frac{\partial^2u_s}{\partial x^2} 
=\gamma(x).
\end{equation}

First, we define the network architecture. The network architectures are defined below using the notation $[N_{\text{in}}, N_{\text{hidden}} \times D, N_{\text{out}}]$, where $N_{\text{in}}$ and $N_{\text{out}}$ are the input and output widths, $N_{\text{hidden}}$ is the number of neurons per hidden layer, and $D$ is the number of hidden layers. For this experiment, we use the following architectures:
\begin{itemize}
    \item FNN$_1$: $[1, 50 \times 2, 25]$
    \item FNN$_2$: $[1, 50 \times 2, 25]$
    \item FNN$_3$: $[25, 50 \times 3, 5]$
    \item FNN$_4$: $[50, 50 \times 3, 5]$.
\end{itemize}

We employ a two-stage training strategy. Initially, we train $\text{FNN}_1$ and $\text{FNN}_3$ on 1,000 randomly sampled points to approximate the steady-state solution. Next, we train $\text{FNN}_2$ and $\text{FNN}_4$ to capture the transient behavior using 10,000 random collocation points distributed throughout the space-time domain. During this second phase, we enforce the initial and boundary conditions using an additional 1,000 random points for each constraint. To evaluate robustness and account for optimization stochasticity, we perform ten independent training runs with random initializations and compare their performance.

\begin{figure}[h!]
     \centering
     \begin{subfigure}[t]{0.49\textwidth}
         \centering
         \includegraphics[width=\textwidth]{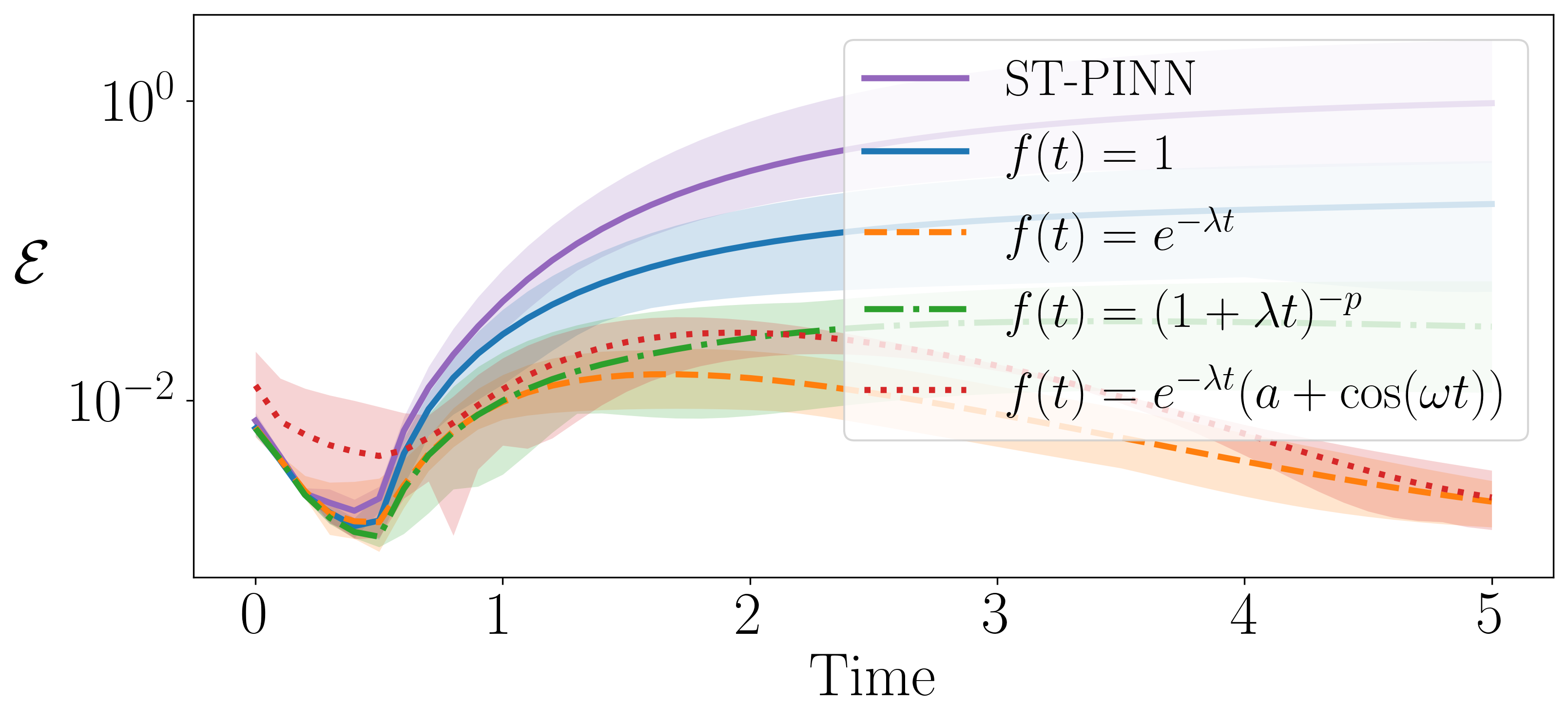}
         \caption{Relative $L_2$ error}         
        \label{fig:heat_compare_error}
     \end{subfigure}
     \centering
     \begin{subfigure}[t]{0.49\textwidth}
         \centering
         \includegraphics[width=\textwidth]{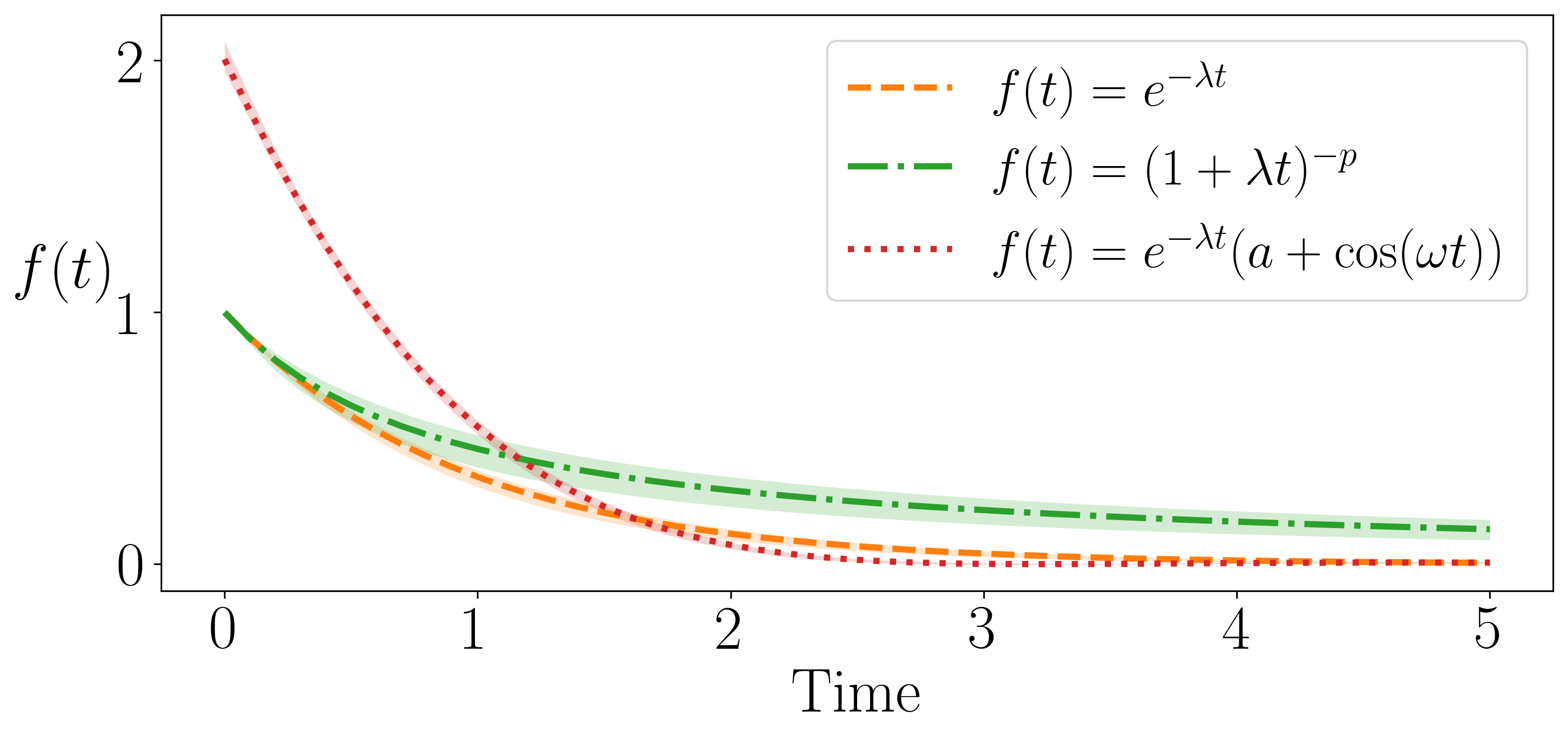}
         \caption{$f(t)$}
         \label{fig:heat_compare_f_t}
     \end{subfigure}
     \caption{One-dimensional heat equation: Extrapolation performance and temporal function $f(t)$. 
     (a) Relative $L_2$ error over time for the temperature field, evaluated 
     across various temporal functions $f(t)$. The network is trained on 
     $t \in [0, 0.5]$ and extrapolated to $t = 5$. 
     (b) Comparison of the learned temporal profiles $f(t)$.}
     \label{fig:heat_compare}
\end{figure}

We train the model over the interval $t \in [0, 0.5]$. Figure~\ref{fig:heat_compare} illustrates the predictive accuracy over time alongside the learned function $f(t)$ and compares the performance against a standard PINN (ST-PINN) featuring 5 hidden layers and 50 neurons per hidden layer. In this figure, the curves
represent the mean trajectory across all 10 realizations, while the 
shaded region indicates the variation across runs. 
Figure~\ref{fig:heat_compare_error} shows the predictive accuracy over time, 
where the $y$-axis represents the relative $L_2$ error between the exact 
solution $u$ and the network prediction $u^h$, defined as
\begin{equation}
    \mathcal{E} := \frac{\|u - u^h\|_{L_2}}{\|u\|_{L_2}}.
\end{equation}

As shown in Figure~\ref{fig:heat_compare_error}, the choice of 
$f(t)$ has a significant impact on extrapolation accuracy. The exponential 
decay function ($f(t)=e^{-\lambda t}$) outperforms all other profiles, consistent with the fact that 
the eigenmodes of the heat equation decay exponentially in time. The algebraic 
decay profile ($f(t)=(1+\lambda t)^{-p}$) also achieves low error, but its convergence to the steady state 
is notably slower. The damped oscillatory function ($f(t)=e^{-\lambda t}\bigl(a+\cos(\omega t)\bigr)$) exhibits higher error at 
intermediate times, but eventually converges as the exponential envelope drives 
$f(t) \to 0$. Finally, the constant profile ($f(t) = 1$) shows a 
monotonically increasing error, since the transient correction 
$g_\theta(\bm{x}, t)$ is never suppressed and the network has no mechanism 
to enforce convergence to $u_s$. The ST-PINN also shows a similar trend: its error increases during extrapolation. This illustrates the well-known limitation of standard PINNs: while they perform well within the training domain, they lack the inductive bias needed for reliable temporal extrapolation. Notably, even in its worst case---the constant profile ($f(t)=1$)---our method is no worse than the ST-PINN, while the decaying profiles improve substantially upon it.

\begin{figure}[h!]
     \centering
     \begin{subfigure}[t]{0.35\textwidth}
         \centering
         \includegraphics[width=\textwidth]{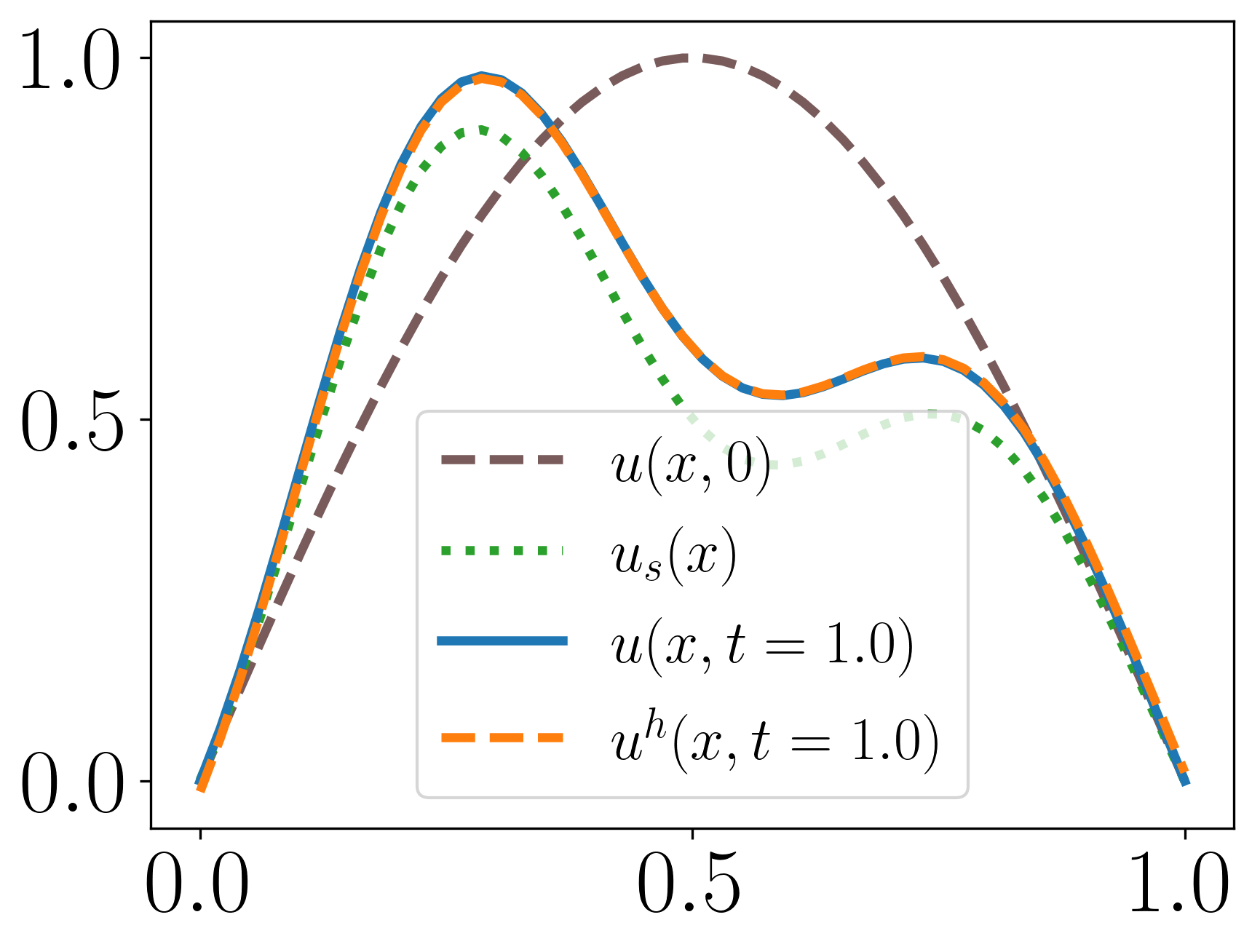}
         \caption{$t = 1$}
     \end{subfigure}
     \centering
     \begin{subfigure}[t]{0.35\textwidth}
         \centering
         \includegraphics[width=\textwidth]{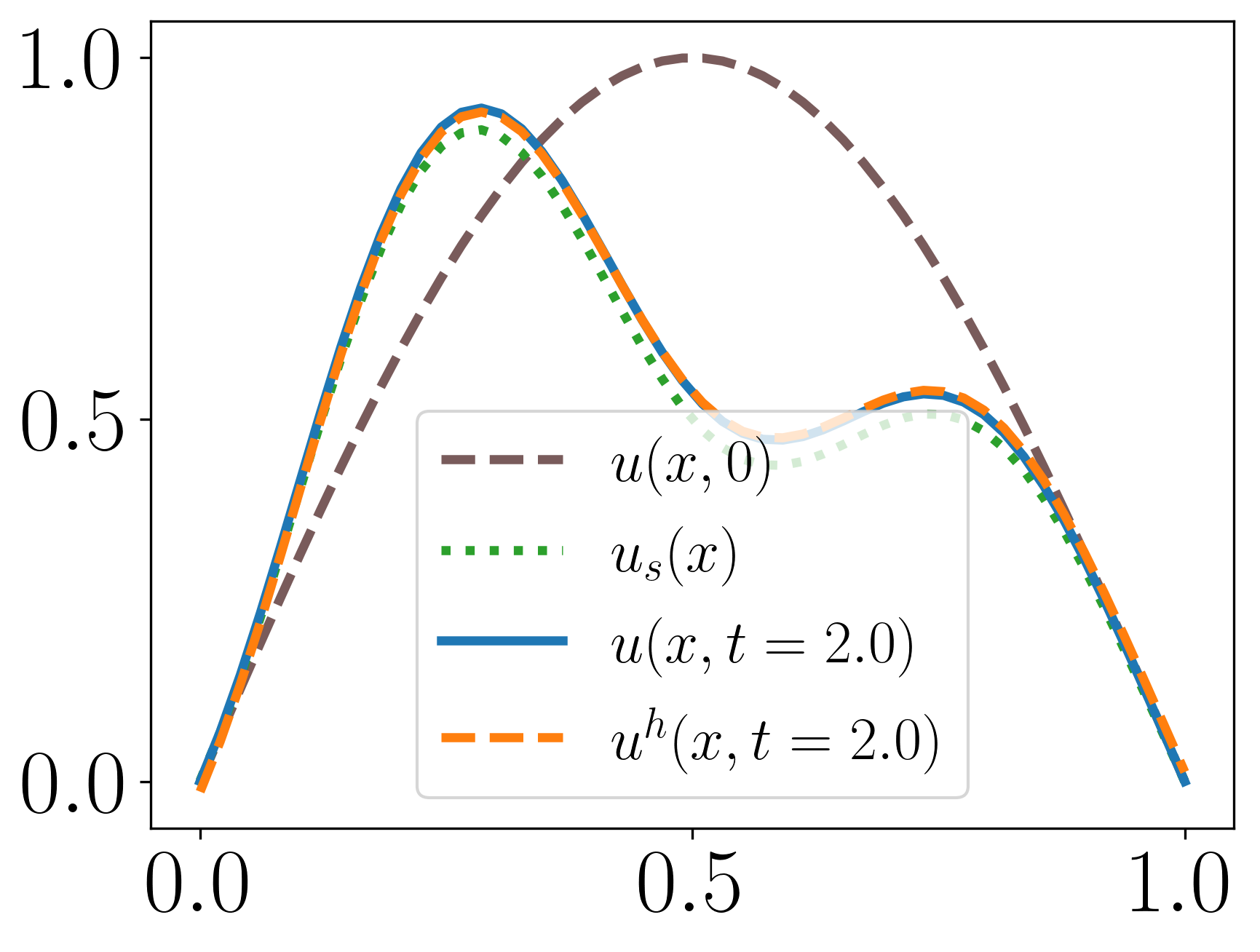}
         \caption{$t = 2$}
     \end{subfigure}
     \caption{One-dimensional heat equation: Extrapolated solution versus the reference solution for 
     $f(t) = e^{-\lambda t}$ at (a)~$t = 1$ and (b)~$t = 2$. The network is trained on $t \in [0, 0.5]$. The initial condition $u(x, 0)$ and the steady-state solution $u_s(x)$ are shown for reference.}
     \label{fig:heat_sol_various_t}
\end{figure}

Figure~\ref{fig:heat_sol_various_t} shows the spatial profiles of the solution 
at $t = 1$ and $t = 2$ for $f(t) = e^{-\lambda t}$. In both cases, the network 
prediction $u^h(x, t)$ is plotted against the reference solution $u(x, t)$, 
with the initial condition $u(x, 0)$ and the steady-state solution $u_s(x)$ 
included to illustrate the dynamical evolution from one to the other. The 
network predictions are in close agreement with the reference, demonstrating 
robust extrapolation performance well beyond the training range.

To understand these results, we examine the analytical structure of the 
transient dynamics. Writing the exact solution as 
$u(x, t) = u_s(x) + v(x, t)$, the transient component $v$ satisfies the 
homogeneous diffusion equation
\begin{equation}
    \frac{\partial v}{\partial t} - \alpha \frac{\partial^2 v}{\partial x^2} 
    = 0, \qquad v(x, 0) = u(x, 0) - u_s(x),
\end{equation}
with homogeneous Dirichlet boundary conditions. By separation of variables, the solution admits the eigenmode expansion
\begin{equation}
    v(x, t) = \sum_{n=1}^{\infty} c_n \sin(n\pi x)\, 
    e^{-\alpha n^2 \pi^2 t},
    \label{eq:eigenmode_expansion}
\end{equation}
where the coefficients $c_n$ are determined by the initial data. The 
transient therefore decays as a superposition of exponentials, with the 
slowest mode corresponding to $n = 1$ and decay rate 
$\alpha \pi^2 \approx 0.987$ for $\alpha = 0.1$. Higher modes decay 
quadratically faster, so for moderate $t$ the transient is well 
approximated by a single exponential. In the proposed ansatz, 
$u^h = u_s + f(t)\,g_\theta(x, t)$, the product $f(t)\,g_\theta$ must 
approximate the expansion~\eqref{eq:eigenmode_expansion}. 

When  $f(t) = e^{-\lambda t}$, the ansatz is structurally compatible with the 
dominant eigenmode: the network needs only to learn 
$g_\theta(x, t) \approx \sum_n c_n \sin(n\pi x)\, 
e^{-\alpha n^2\pi^2 t + \lambda t}$, a function whose time dependence 
reduces to decaying exponentials that the network can represent within the 
training window and that {remains} bounded during extrapolation. The 
learned value of $\lambda$ is expected to approximate $\alpha\pi^2$, aligning 
the ansatz with the slowest-decaying eigenmode.

This structural compatibility argument also explains the behavior of
the remaining profiles.
For algebraic decay, $f(t) = (1 + \lambda t)^{-p}$ decreases as
$\mathcal{O}(t^{-p})$ for large $t$, which is asymptotically slower
than the exponential decline of the eigenmodes
in~\eqref{eq:eigenmode_expansion}. Within the training window, the
network $g_\theta$ can compensate by learning a correction that
accelerates the effective decay of the product $f(t)\,g_\theta$.
However, this compensation is a learned artifact of the training
data, not a structural feature of the ansatz: outside the training
domain, the network has no incentive to maintain it. 

The product
$f(t)\,g_\theta$ therefore converges to zero more slowly than the
true transient, producing the gradual but eventual convergence
observed in Figure~\ref{fig:heat_compare_error}.
For the damped oscillatory profile,
$f(t) = e^{-\lambda t}(a + \cos(\omega t))$, convergence to the
steady state is guaranteed by the exponential envelope regardless of
$g_\theta$. One might expect the cosine factor to introduce spurious
oscillations that the network must cancel; however, the learned
profile in Figure~\ref{fig:heat_compare_f_t} reveals that the
optimizer selects a small angular frequency~$\omega$, producing
only weak oscillation. The elevated extrapolation error at
intermediate times has a different origin: the additional free
parameters ($a$, $\omega$) allow the optimizer to fit a composite
decay shape that matches the training data but whose temporal
structure departs from the pure exponential decay of the
eigenmodes. 

In particular, $f(0) = a + 1$, which generically
differs from unity, and the product
$e^{-\lambda t}(a + \cos(\omega t))$ cannot reduce to a single
exponential for any parameter choice. The network $g_\theta$
compensates for this shape mismatch within the training window,
but because the compensation is learned rather than structurally
imposed, it does not persist during extrapolation. The net result
is a profile that converges to the steady state---as the
exponential envelope guarantees---but less accurately than the
pure exponential ansatz, whose functional form is directly aligned
with the dominant eigenmode.

Finally, for the constant profile $f(t) = 1$, the ansatz reduces
to $u^h = u_s + g_\theta(x, t)$, placing full responsibility for
reproducing the transient decay on the network. Feedforward networks
behave as polynomial-like interpolants with no intrinsic decay
structure, so $g_\theta$ has no mechanism to enforce
$g_\theta \to 0$ as $t \to \infty$. Outside the training domain,
the network simply extrapolates whatever function it has fit,
which generically grows or oscillates, producing the monotonically
increasing error in Figure~\ref{fig:heat_compare_error}. This last
case illustrates the well-known limitation of standard PINNs:
accurate interpolation within the training domain does not imply
reliable temporal extrapolation.

\begin{figure}
    \centering
    \includegraphics[width=0.5\linewidth]{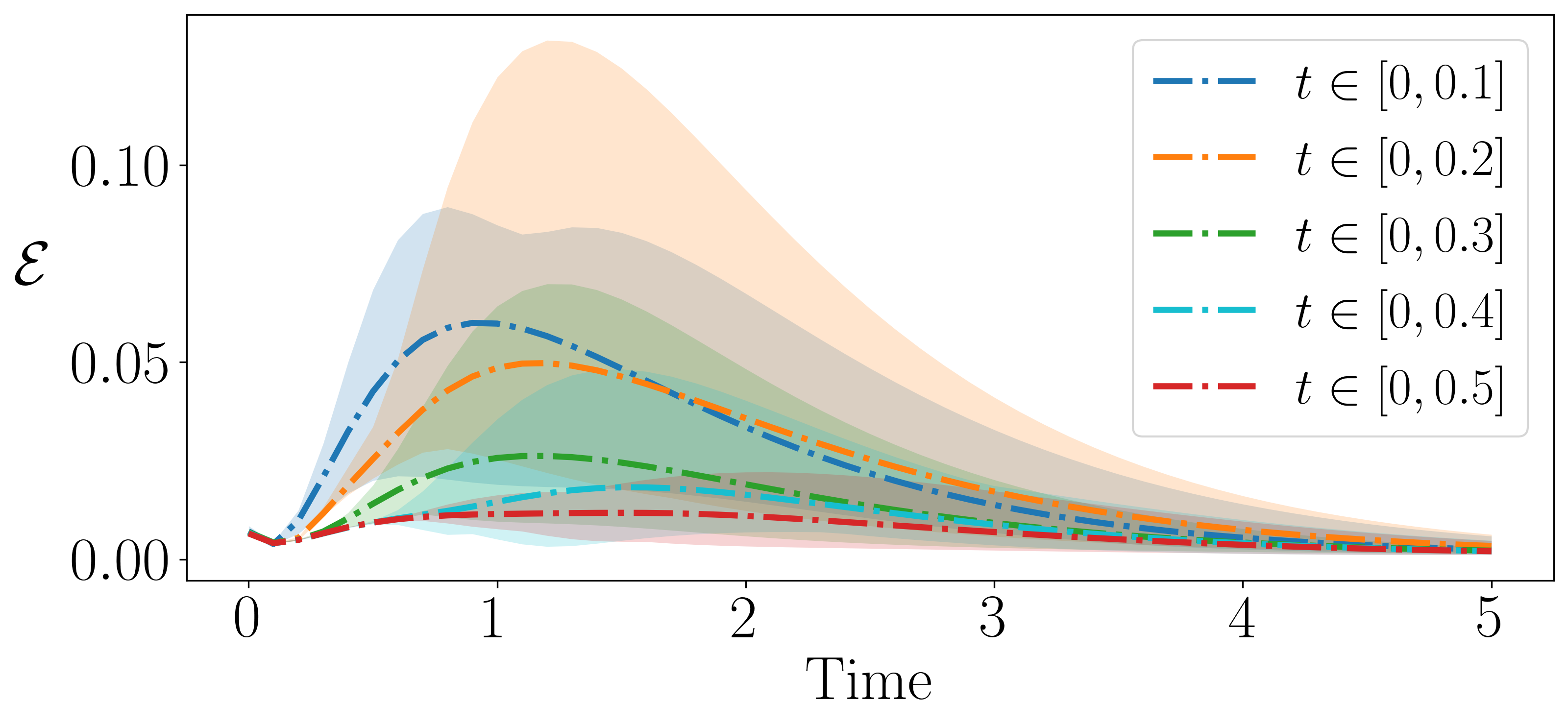}
    \caption{One-dimensional heat equation: Extrapolation error as a function of training duration. The vertical axis shows the relative $L_2$ error $\mathcal{E}$ for the temperature field over time, where $f(t) = e^{-\lambda t}$ with fixed $\lambda  = 0.986$, which is learned from the previous case. }
    \label{fig:heat_compare_fixed_lambda}
\end{figure}
Finally, Figure~\ref{fig:heat_compare_fixed_lambda} illustrates the impact of training duration on extrapolation error. For this evaluation, we use the function $f(t) = e^{-\lambda t}$ and set $\lambda = 0.986$, which is the value learned by the model and used in Figure~\ref{fig:heat_sol_various_t}. As expected, longer training periods have lower errors, with the interval $t \in [0, 0.5]$ achieving the minimum extrapolation error. This error increases as we decrease the training duration. However, due to the underlying exponential decay profile, all solutions ultimately converge to the steady state regardless of the training duration.

In summary, the one-dimensional diffusion example shows that the proposed method captures temporal extrapolation effectively. From the numerical experiment, we observe that the exponential decay function outperforms other choices of function.

\subsection{Example 2: Navier--Stokes equation (Lid-Driven Cavity Flow)}
\label{sec:Eg2}
We now apply the proposed method to the incompressible Navier--Stokes 
equations for lid-driven cavity flow. We consider a square domain 
$\Omega = (0,1)^2$ filled with a viscous fluid, and let $I := (0,t_f]$ 
denote the time interval of interest. The fluid motion is governed by
\begin{subequations}
\begin{alignat}{2}
\frac{\partial \bu}{\partial t} + (\bu\cdot\nabla)\bu
- 2\,\text{Re}^{-1}\nabla\cdot \varepsilon(\bu) + \nabla p
&= 0, \quad &&\text{in } \Omega \times I,\\
\nabla\cdot \bu &= 0, \quad &&\text{in } \Omega \times I,
\end{alignat}
\end{subequations}
where $\bu = [u_1,u_2]^\top : \Omega \times I \rightarrow \mathbb{R}^2$ 
is the fluid velocity, $p : \Omega \times I \rightarrow \mathbb{R}$ is the 
pressure, 
$\varepsilon(\bu) := \tfrac{1}{2}(\nabla \bu + (\nabla \bu)^{\top})$ 
denotes the symmetric part of the velocity gradient, and $\text{Re}$ is the 
Reynolds number. We set $\text{Re}=100$ and impose the boundary conditions
\begin{subequations}
    \begin{alignat}{2}
        \bu &= \bm{0}, \quad && \text{on } \partial \Omega 
        \setminus \{y = 1\}, \\
        \bu &= \begin{bmatrix}
            1 - (2x - 1)^2 \\ 0
        \end{bmatrix}, \quad && \text{on } \partial \Omega \cap \{y = 1\}.
    \end{alignat}
\end{subequations}
Thus, all walls are stationary except the 
top lid, which moves with a prescribed horizontal velocity and generates a 
recirculating flow inside the cavity. 
The parabolic lid profile drives the flow to the right and avoids the 
corner discontinuities associated with a uniform lid velocity. The initial 
condition is
$\bu(\bm{x},0)=\bm{0}, \bm{x}\in\Omega.$
While the theoretical pressure field is mathematically unique only up to an arbitrary additive constant, minimizing the network's loss function generally results in convergence toward one distinct realization of the pressure field.
 
As in Example~\ref{sec:Eg1}, we compare four temporal profiles for $f(t)$: 
constant, exponential decay, algebraic decay, and damped oscillatory. Each 
solution component ($u_1$, $u_2$, $p$) is assigned its own decay rate, amplitude, and 
angular frequency, all of which are learned during training. We report the 
$L_\infty$ error, defined as the maximum absolute pointwise error, rather than the 
relative $L_2$ error used in Example~\ref{sec:Eg1}, as the $L_\infty$ 
norm highlights worst-case deviations near boundaries and in regions of 
steep gradients where extrapolation errors are most likely to concentrate. 
As in the previous example, we use the following architectures:
\begin{itemize}
    \item FNN$_1$: $[2, 50 \times 2, 25]$,
    \item FNN$_2$: $[1, 50 \times 2, 25]$,
    \item FNN$_3$: $[25, 50 \times 3, 3]$,
    \item FNN$_4$: $[50, 50 \times 3, 3]$.
\end{itemize}
We train FNN$_1$ and FNN$_3$ using 10,000 random collocation points to obtain the steady-state solution $\bm{u}_s(\bm{x})$. Once these networks are trained, we freeze the weights of FNN$_1$ and FNN$_3$ and train FNN$_2$ and FNN$_4$ with 30,000 random collocation points. During each optimization step, we employ a mini-batch strategy by subsampling 3,000 collocation points and using 3,000 boundary points. For transient state training, we enforce the initial condition using 1,000 randomly sampled points. 
 
\begin{figure}[h!]
     \centering
     \begin{subfigure}[t]{0.49\textwidth}
         \centering
         \includegraphics[width=\textwidth]{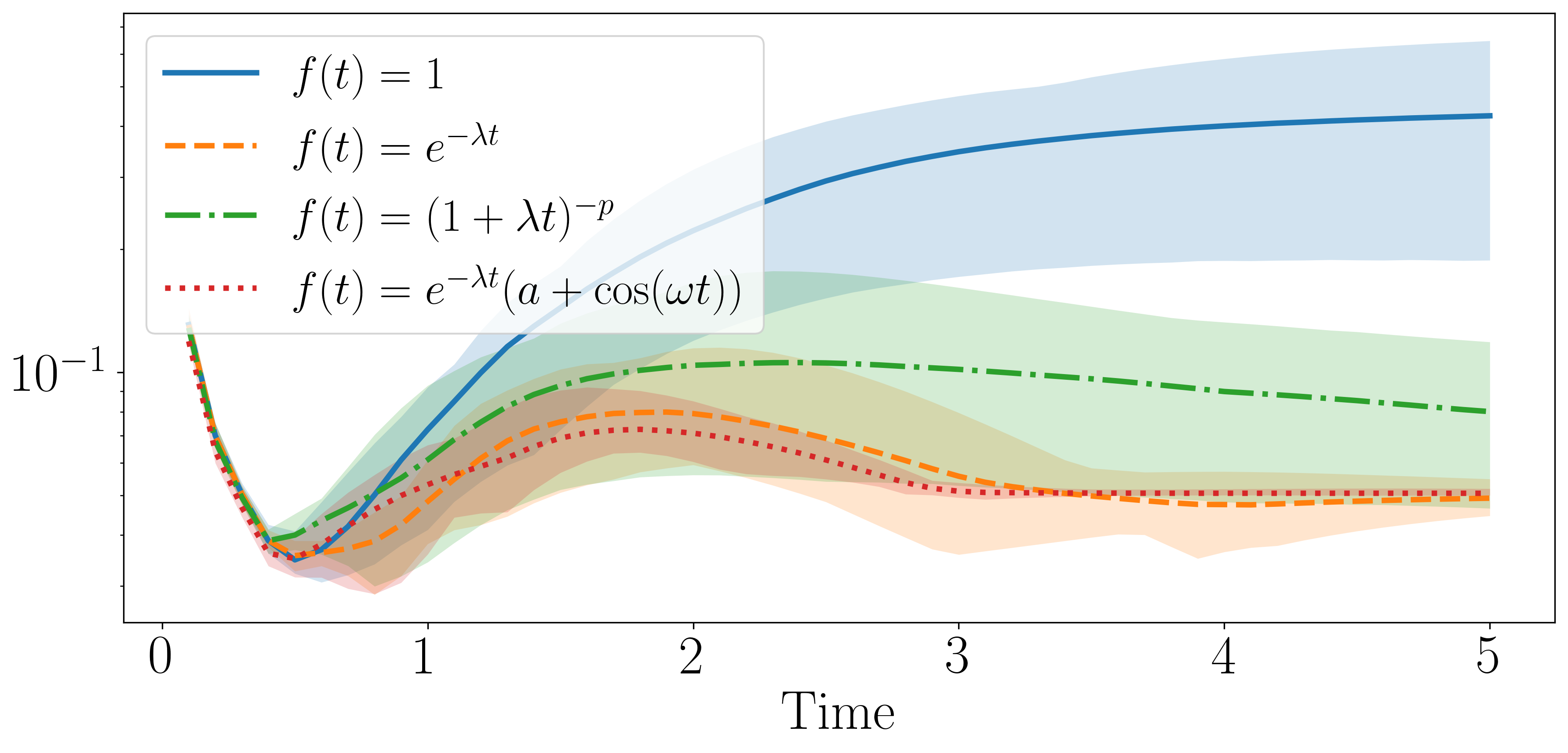}
         \caption{$\|u_1 - u_1^h\|_{L_\infty}$}
     \end{subfigure}
     \begin{subfigure}[t]{0.49\textwidth}
         \centering
         \includegraphics[width=\textwidth]{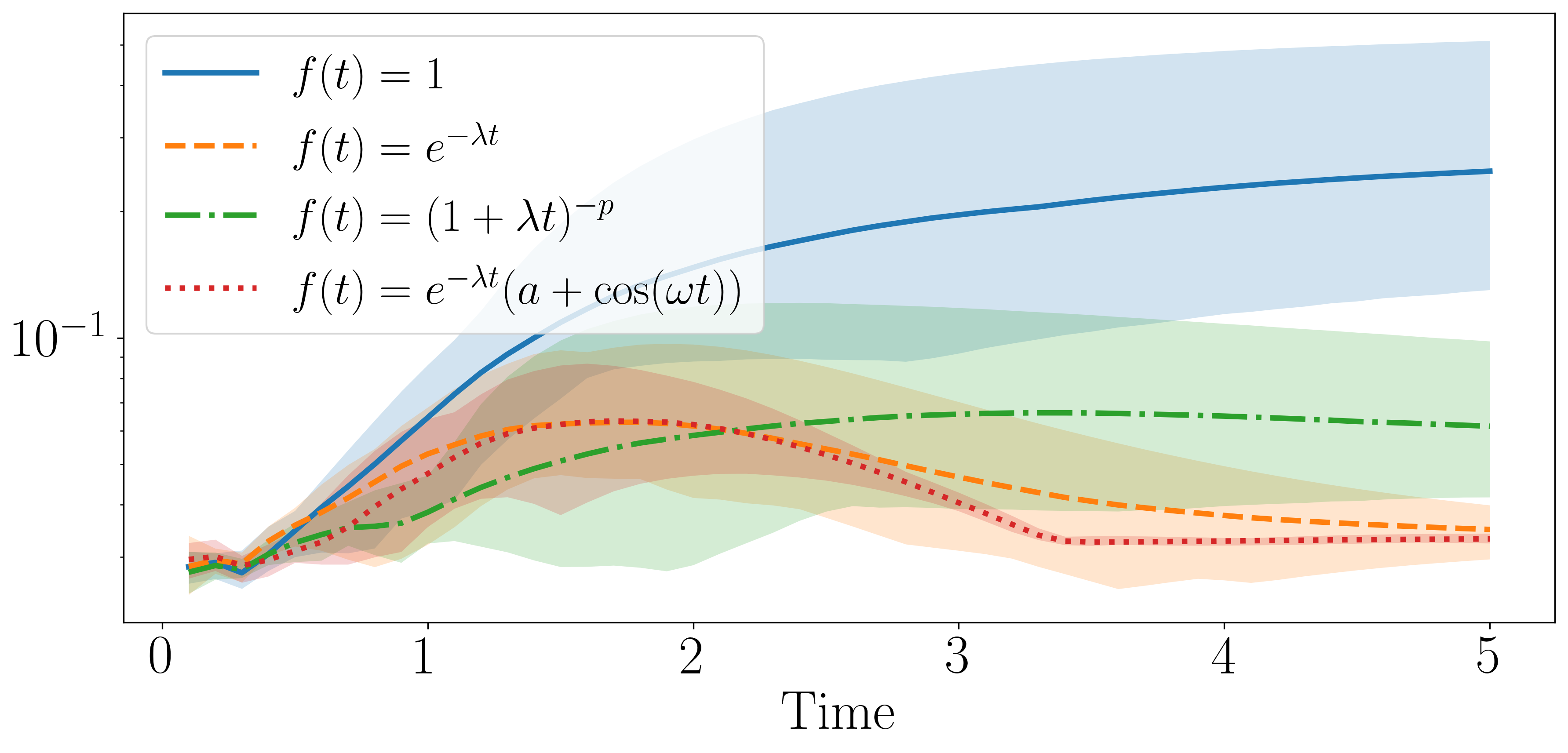}
         \caption{$\|u_2 - u_2^h\|_{L_\infty}$}
     \end{subfigure}
     \caption{Lid-Driven Cavity Flow: $L_\infty$ error over time for the velocity components 
     (a)~$u_1$ and (b)~$u_2$, evaluated across various temporal functions 
     $f(t)$, with the network trained on $t \in [0, 0.5]$.}
     \label{fig:u_compare_L_infty_t_05}
\end{figure}

\subsubsection{Training on $t \in [0,\, 0.5]$.}
Figure~\ref{fig:u_compare_L_infty_t_05} compares the interpolation and 
extrapolation errors for different choices of $f(t)$ when the model is 
trained up to $t = 0.5$. For the horizontal velocity $u_1^h$, the error is 
elevated near $t = 0$ and then decreases within the training window before 
rising again during extrapolation. The large initial error is attributable 
to the incompatibility between the zero initial condition and the non-zero 
lid velocity: at $t = 0$ the solution must develop a thin shear layer near 
the top boundary, which is difficult for the network to resolve. Within the 
training interval, the error decreases regardless of the choice of $f(t)$, 
as expected. During extrapolation, however, the profiles diverge 
significantly. Exponential decay and damped oscillatory profiles 
achieve comparable mean errors. Algebraic decay converges more slowly, and the constant profile shows steadily growing error.
 
The vertical velocity component $u_2^h$ exhibits lower error overall, 
including at early times. This is because $u_2$ satisfies homogeneous 
Dirichlet conditions on all boundaries, so no initial--boundary 
incompatibility arises. The extrapolation trends mirror those of $u_1$: 
exponential and damped oscillatory decay outperform algebraic decay, while 
the constant profile degrades monotonically.

\subsubsection{Training on $t \in [0,\, 1]$.}
\begin{figure}[h!]
     \centering
     \begin{subfigure}[t]{0.49\textwidth}
         \centering
         \includegraphics[width=\textwidth]{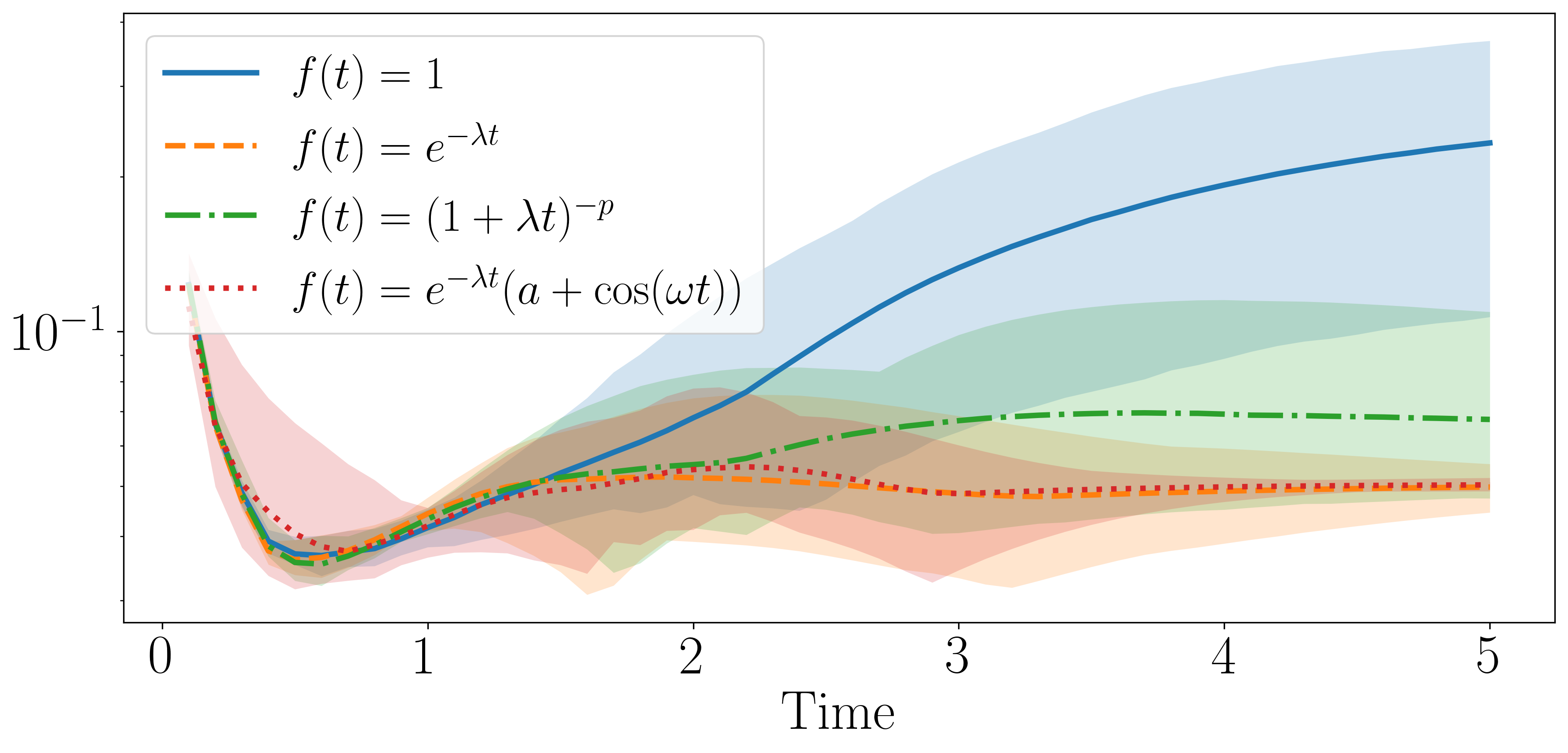}
         \caption{$\|u_1 - u_1^h\|_{L_\infty}$}
     \end{subfigure}
     \begin{subfigure}[t]{0.49\textwidth}
         \centering
         \includegraphics[width=\textwidth]{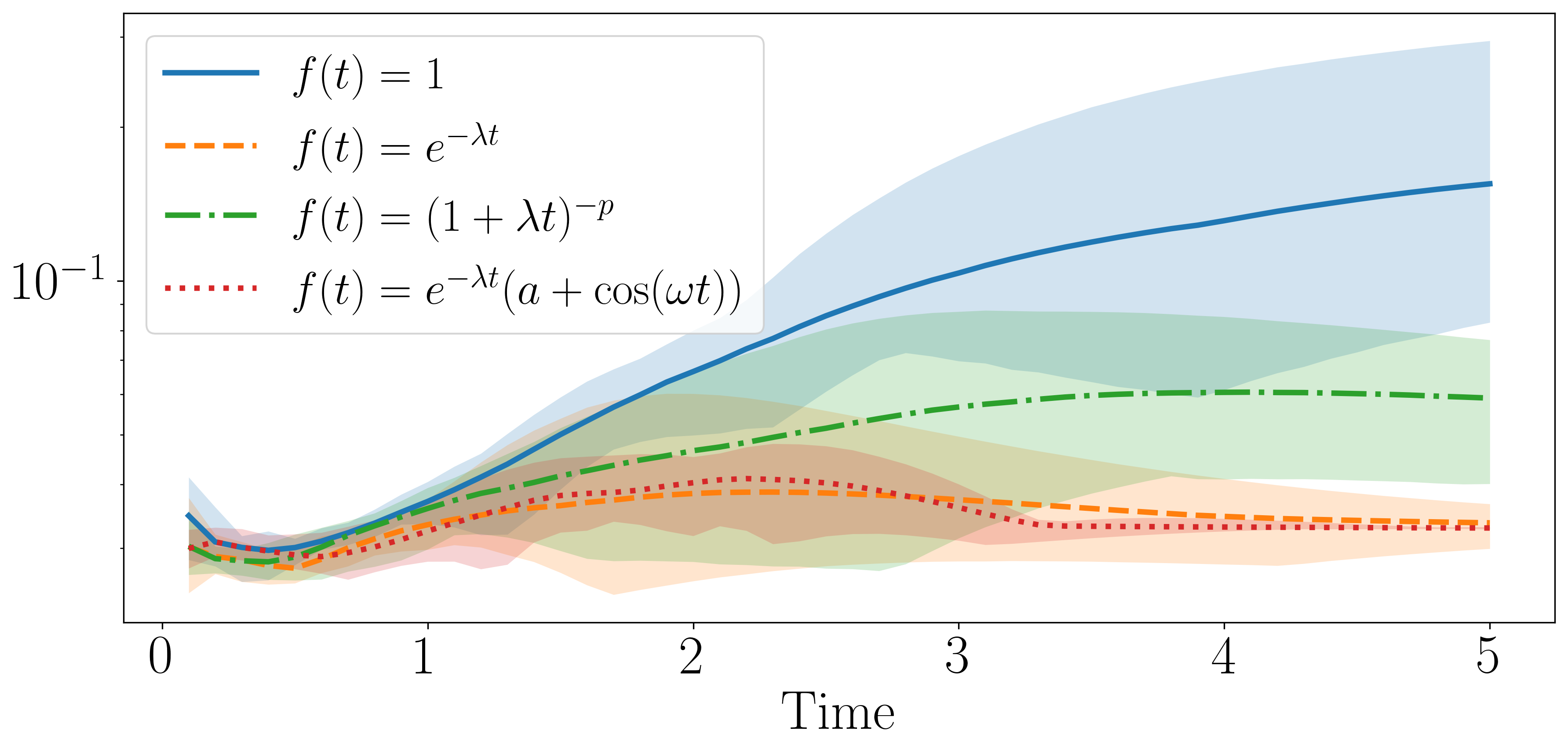}
         \caption{$\|u_2 - u_2^h\|_{L_\infty}$}
     \end{subfigure}
     \caption{Lid-Driven Cavity Flow: $L_\infty$ error over time for the velocity components 
     (a)~$u_1$ and (b)~$u_2$, evaluated across various temporal functions 
     $f(t)$, with the network trained on $t \in [0, 1]$.}
     \label{fig:u_compare_L_infty_t_1}
\end{figure}

Figure~\ref{fig:u_compare_L_infty_t_1} shows the corresponding results 
when the training window is extended to $t = 1$. Within the training 
interval, the interpolation error remains low for all choices of $f(t)$. 
During extrapolation, the relative ranking of the decay profiles is 
preserved, with exponential decay and oscillatory decay achieving the lowest error. 
A comparison of 
Figures~\ref{fig:u_compare_L_infty_t_05} 
and~\ref{fig:u_compare_L_infty_t_1} confirms that extending the training 
horizon improves extrapolation accuracy for exponential decay profiles, as the network has more transient data from which to learn the dynamics before the asymptotic regime takes over.
 
\subsubsection{Spatial error distribution.}
\begin{figure}[h!]
     \centering
     \begin{subfigure}[t]{0.24\textwidth}
         \centering
         \includegraphics[width=\textwidth]{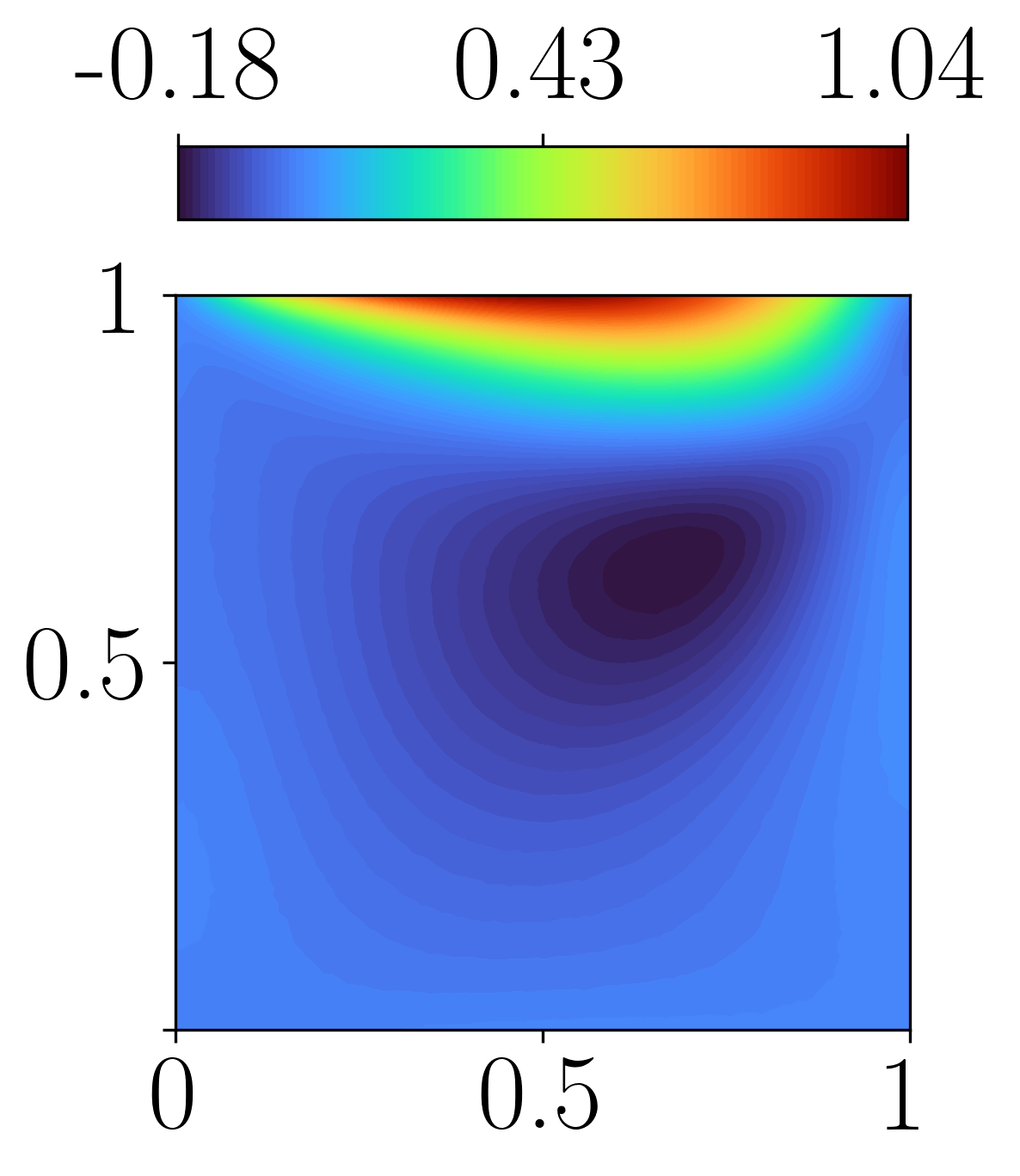}
         \caption{$u_1^h$}
     \end{subfigure}
     \begin{subfigure}[t]{0.24\textwidth}
         \centering
         \includegraphics[width=\textwidth]{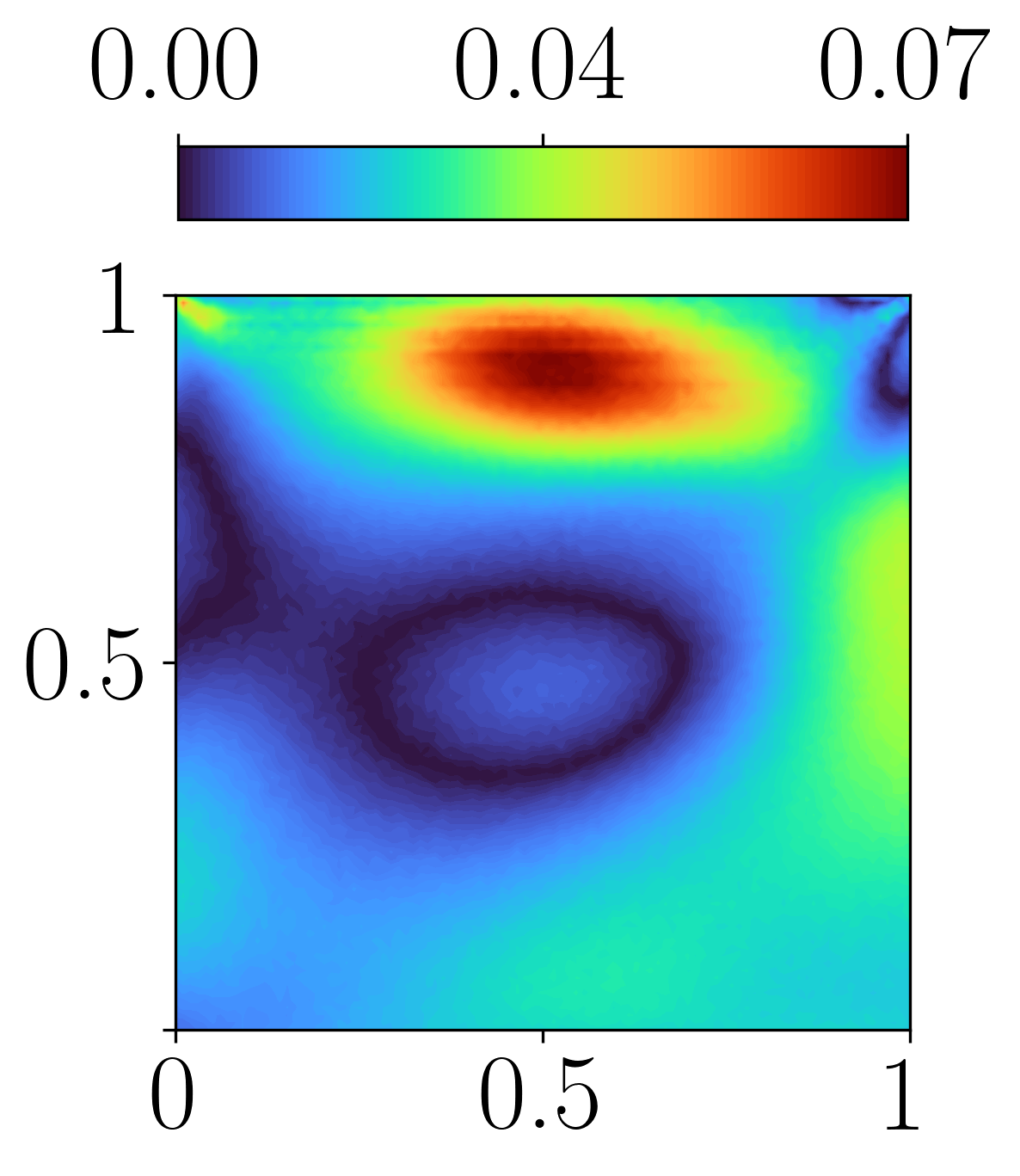}
         \caption{$|u_1 - u_1^h|$}
     \end{subfigure}
     \begin{subfigure}[t]{0.24\textwidth}
         \centering
         \includegraphics[width=\textwidth]{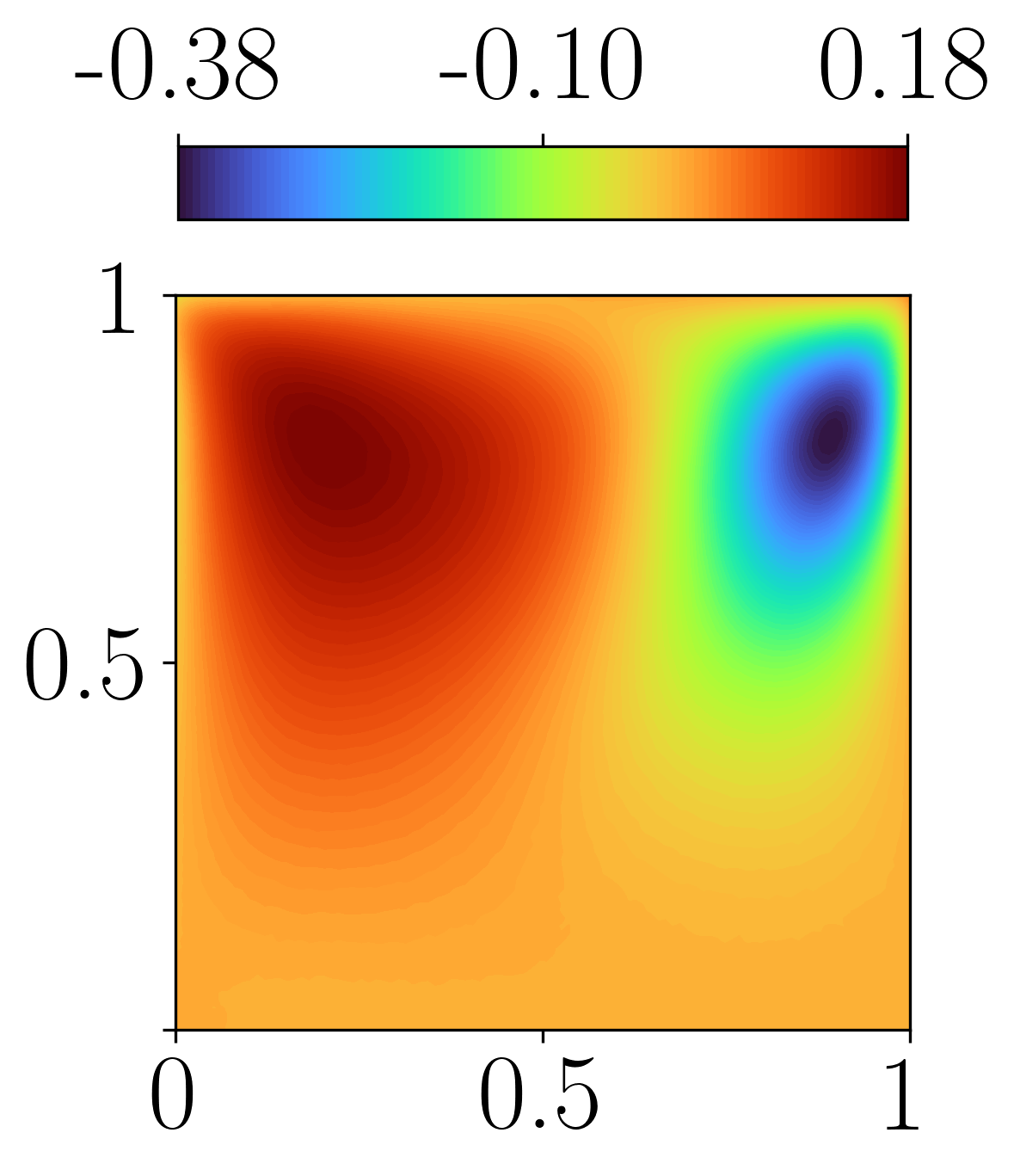}
         \caption{$u_2^h$}
     \end{subfigure}
     \begin{subfigure}[t]{0.24\textwidth}
         \centering
         \includegraphics[width=\textwidth]{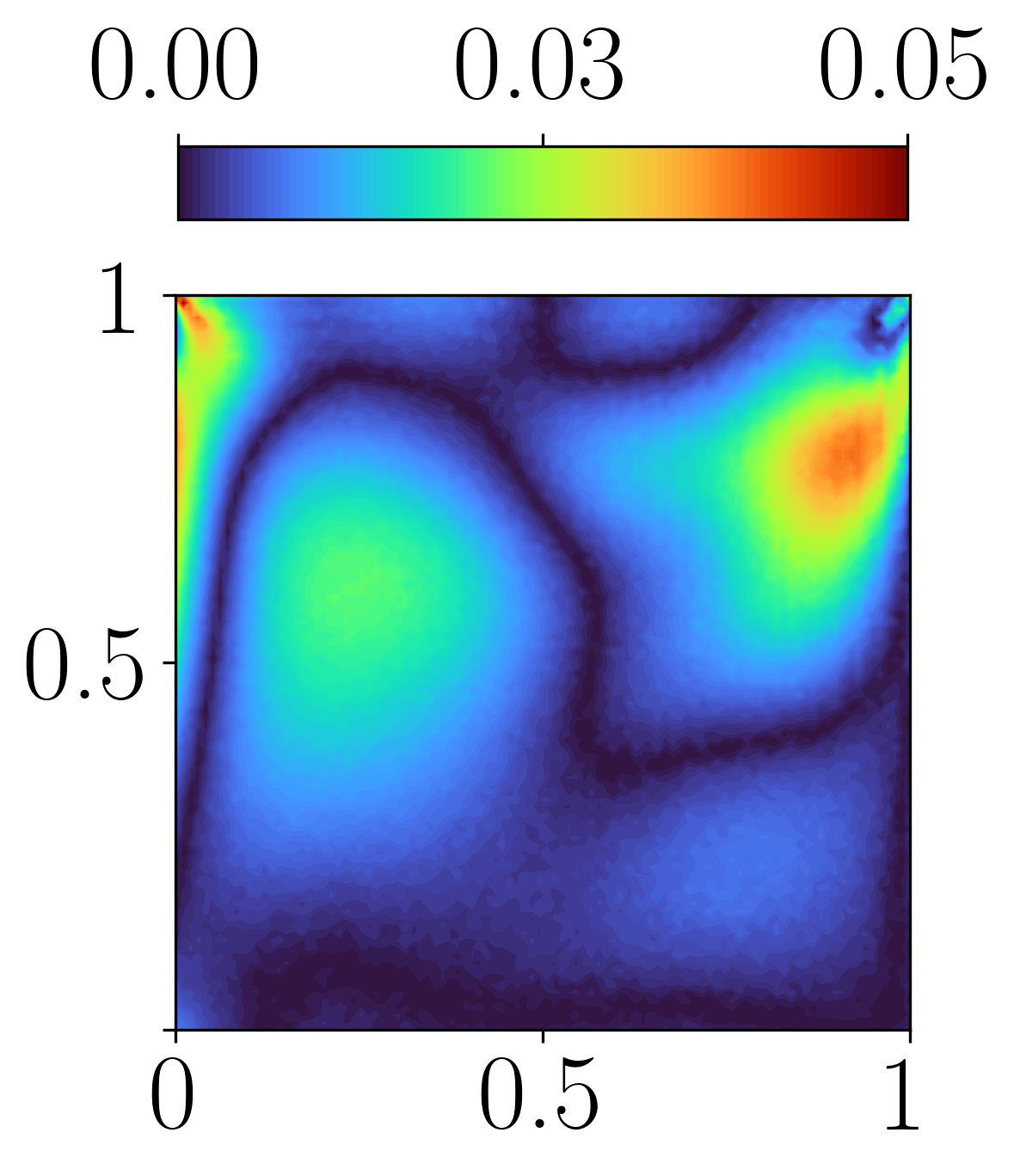}
         \caption{$|u_2 - u_2^h|$}
     \end{subfigure}
     \caption{Lid-Driven Cavity Flow: Velocity predictions and pointwise absolute errors at $t = 1.5$ 
     for $f(t) = e^{-\lambda t}$, with the model trained on 
     $t \in [0, 0.5]$. (a)~Horizontal velocity $u_1^h$. 
     (b)~Absolute error $|u_1 - u_1^h|$. (c)~Vertical velocity $u_2^h$. 
     (d)~Absolute error $|u_2 - u_2^h|$.}
     \label{fig:u_contour_t_1.5}
\end{figure}

Finally, Figure~\ref{fig:u_contour_t_1.5} shows the predicted velocity 
fields and their pointwise absolute errors at $t = 1.5$, using 
$f(t) = e^{-\lambda t}$ with the model trained on $t \in [0, 0.5]$. The 
predicted fields preserve the expected physical structure: a primary 
recirculating vortex driven by the lid, with the horizontal velocity 
concentrated near the top boundary and the vertical velocity marking the 
return flow along the side walls. The largest errors are concentrated 
near the top-left corner and along the boundary of the primary vortex, 
where the velocity gradients are steepest and where the shear layer interacts with the side wall. The absolute errors remain at 
$\mathcal{O}(10^{-2})$ or below, which, given that the model is 
extrapolating three times the training horizon ($t = 1.5$ versus 
training up to $t = 0.5$), demonstrates that the steady-state-informed ansatz provides meaningful predictive capability well outside the training domain.
 
\begin{figure}[h!]
     \centering
     \begin{subfigure}[t]{0.24\textwidth}
         \centering
         \includegraphics[width=\textwidth]{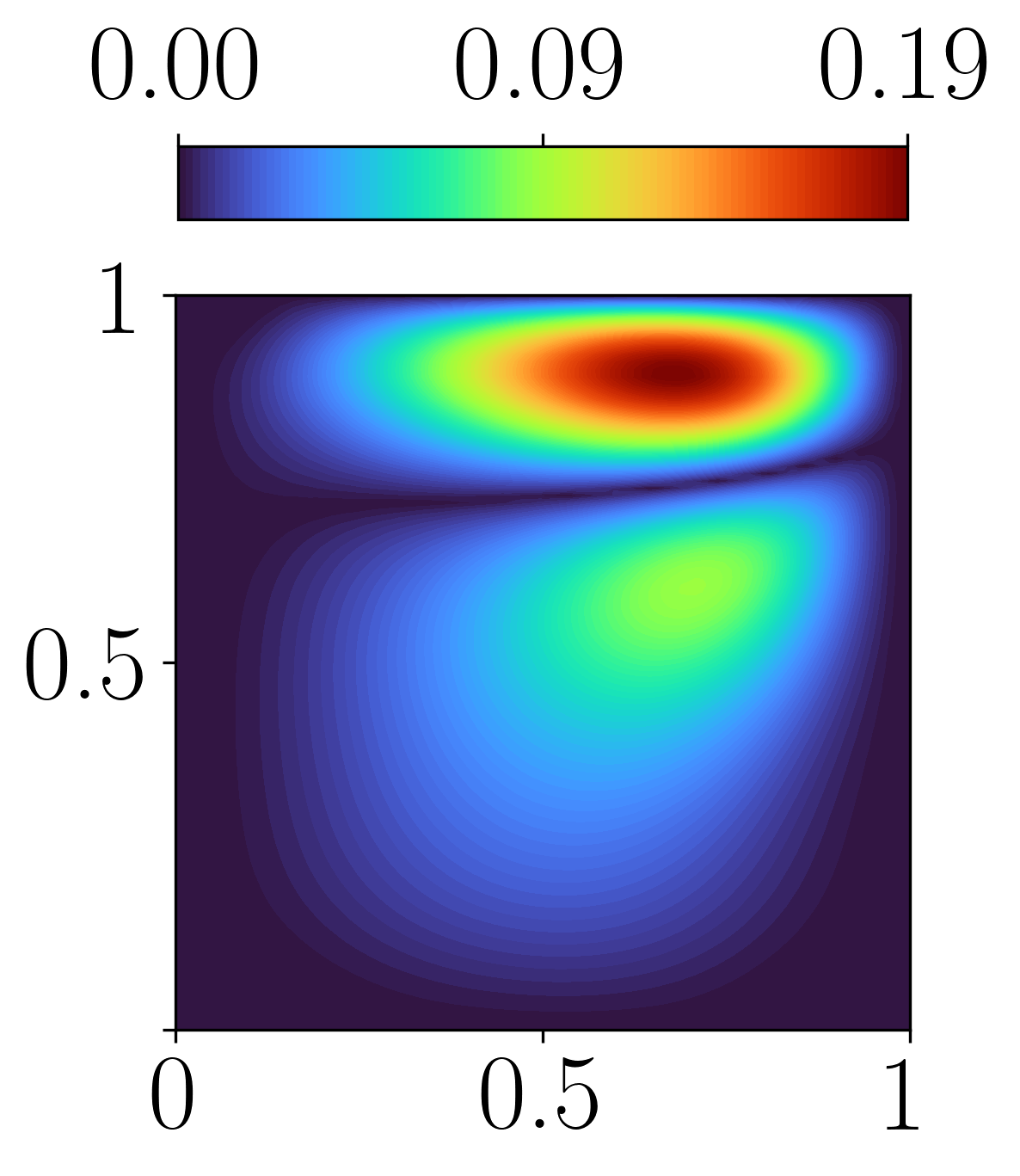}
         \caption{$|u_1(\cdot, t = 0.5) - u_1(\cdot, t=1.5)|$}
     \end{subfigure}
     \begin{subfigure}[t]{0.24\textwidth}
         \centering
         \includegraphics[width=\textwidth]{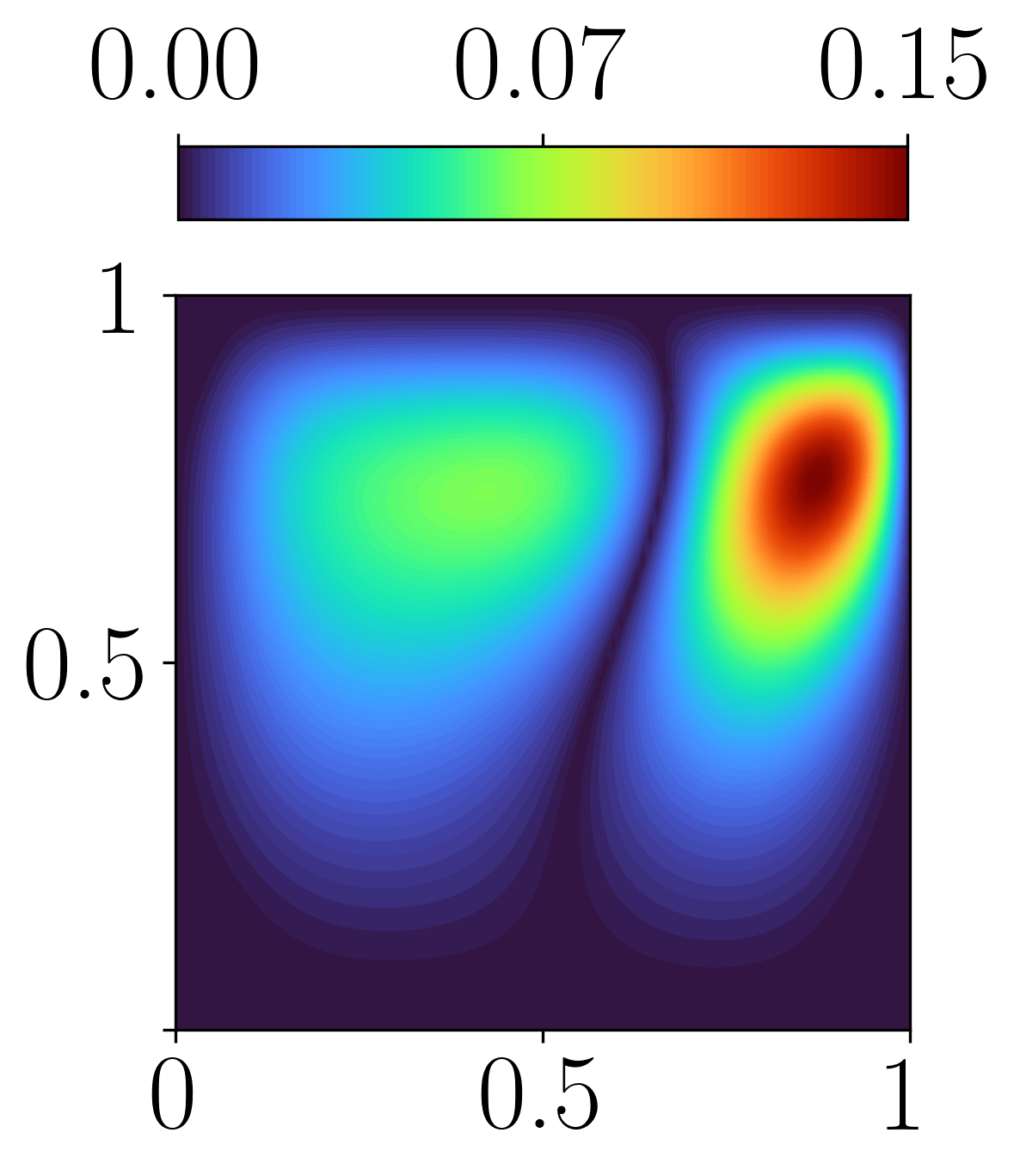}
         \caption{$|u_2(\cdot, t = 0.5) - u_2(\cdot, t=1.5)|$}
     \end{subfigure}
     \begin{subfigure}[t]{0.24\textwidth}
         \centering
         \includegraphics[width=\textwidth]{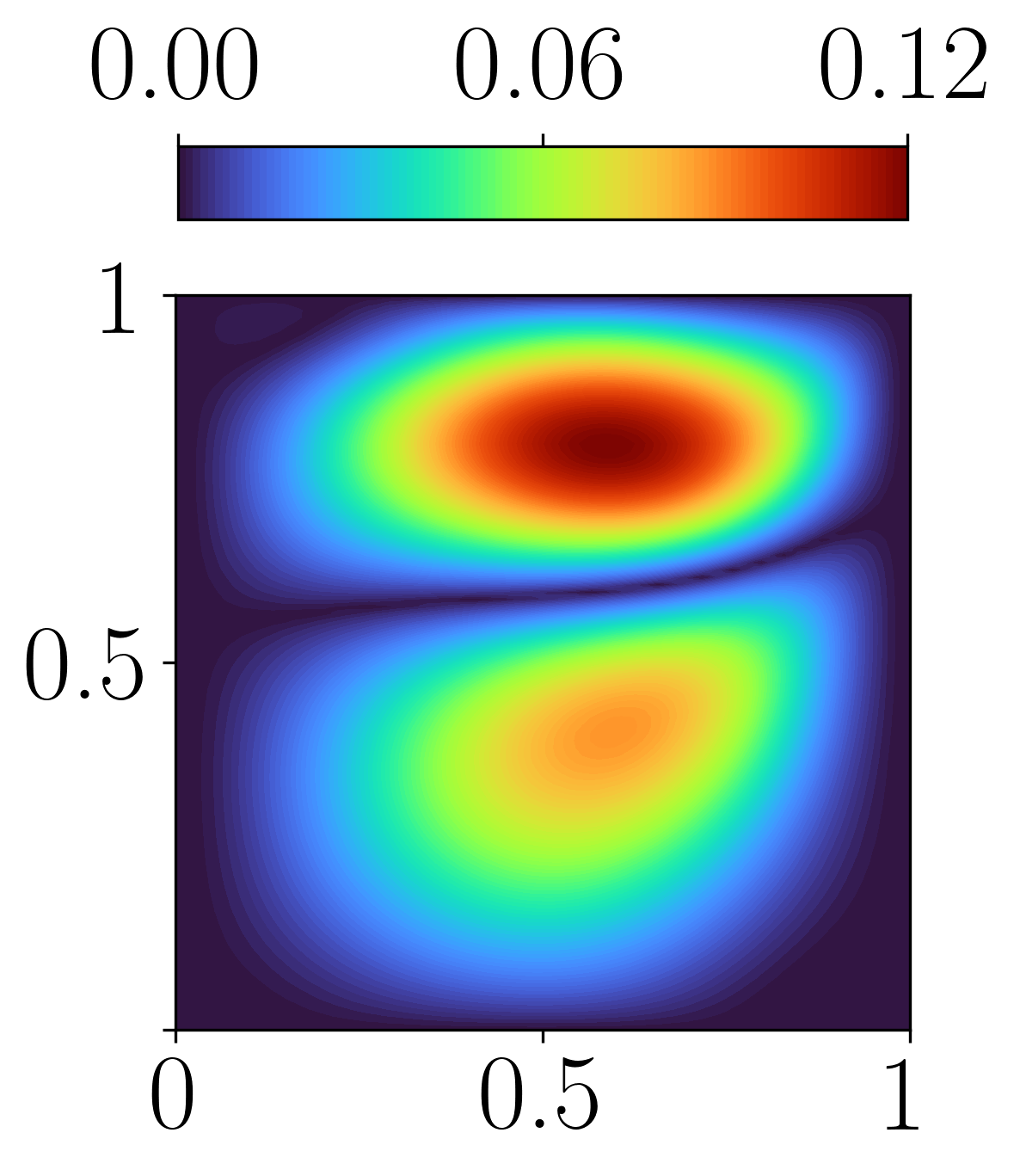}
         \caption{$|u_1(\cdot, t = 1.5) - u_1(\cdot, t=9)|$}
     \end{subfigure}
     \begin{subfigure}[t]{0.24\textwidth}
         \centering
         \includegraphics[width=\textwidth]{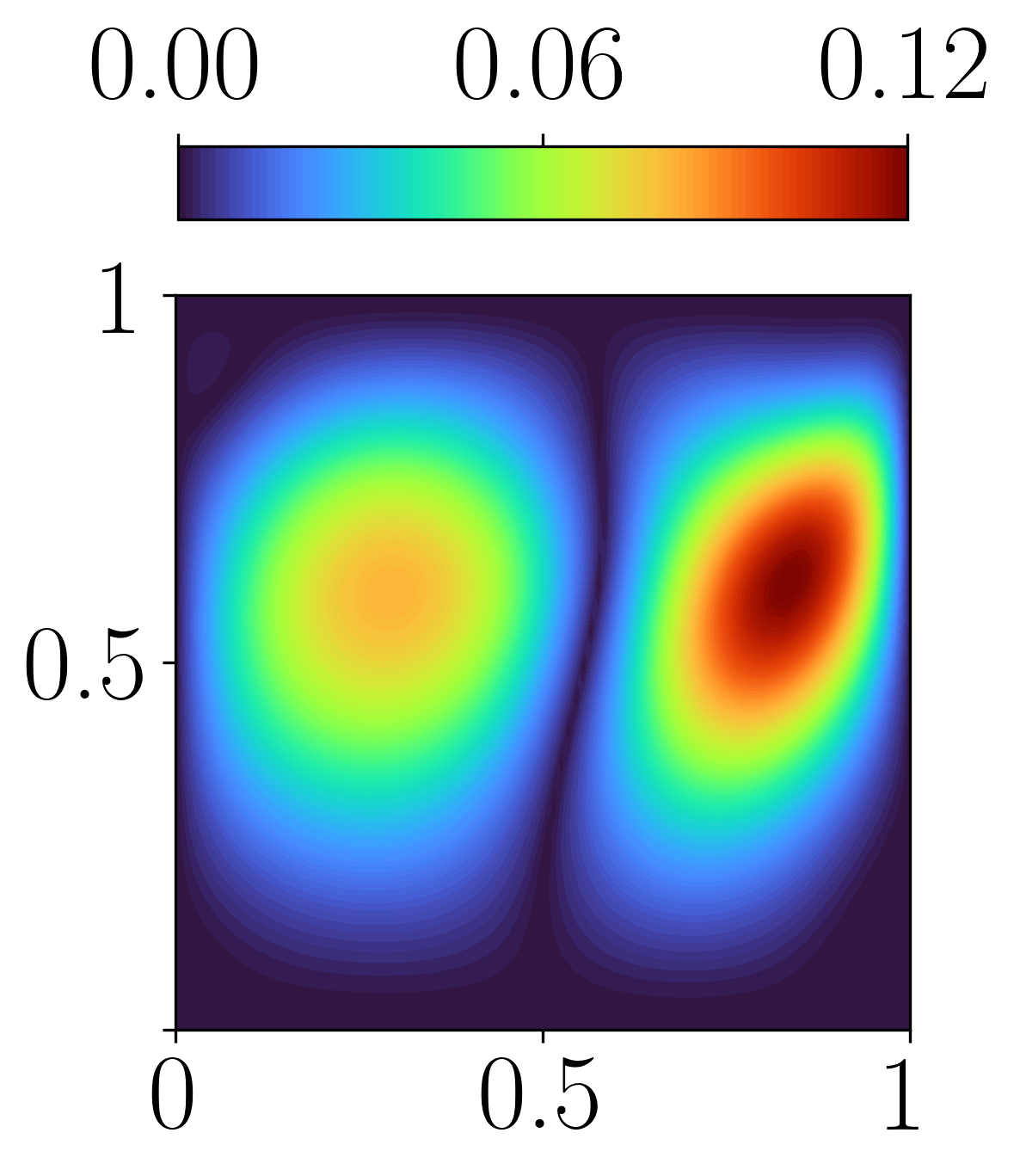}
         \caption{$|u_2(\cdot, t = 1.5) - u_2(\cdot, t=9)|$}
     \end{subfigure}
     \caption{Lid-Driven Cavity Flow: Contour plots showing the absolute spatial differences in the velocity components $u_1$ and $u_2$. Panels (a) and (b) show the evolution of velocity between $t = 0.5$ and $t = 1.5$, while panels (c) and (d) show the evolution of the velocity between $t = 1.5$ and $t = 9$.}
     \label{fig:u_compare_1.5_with_others}
\end{figure}

Figure~\ref{fig:u_compare_1.5_with_others} places the prediction errors 
in context by showing the absolute spatial differences in the velocity 
field between successive time snapshots. Panels~(a) and~(b) display 
$|u_i(\cdot,t=0.5) - u_i(\cdot,t=1.5)|$, which quantify the physical evolution of the flow between the end of the training window and the extrapolation point; these differences reach $\mathcal{O}(0.15$--$0.19)$. Comparing with the prediction errors in Figure~\ref{fig:u_contour_t_1.5}, which are $\mathcal{O}(0.05$--$0.07)$, the model's extrapolation error is roughly three to four times smaller than the actual dynamic changes occurring over the same interval. Panels~(c) and~(d) show 
$|u_i(\cdot,t=1.5) - u_i(\cdot,t=9)|$, confirming that the solution at $t = 1.5$ remains in a genuine transient phase: substantial evolution toward the steady state still lies ahead. Taken together, these comparisons demonstrate that the network predicts an evolving solution that is distinct from both the training-interval data and the final steady-state 
condition, and does so with errors significantly smaller than the 
underlying physical dynamics.
 
\subsection{Example 3: Coupled Navier-Stokes and Heat equations 
(Boussinesq approximation)}
\label{sec:Eg3}

We now study the proposed model for the coupled Navier-Stokes and 
heat equations. This coupling relies on the Boussinesq approximation 
\cite{boussinesq1897theorie}, a standard approach for simulating 
natural convection. Under this approximation, all fluid properties 
are assumed constant except for density variations, which are 
considered solely within the gravitational body force term of the 
momentum equation.

\begin{figure}[h!]
     \centering
     \begin{subfigure}[t]{0.49\textwidth}
         \centering
         \includegraphics[width=\textwidth]{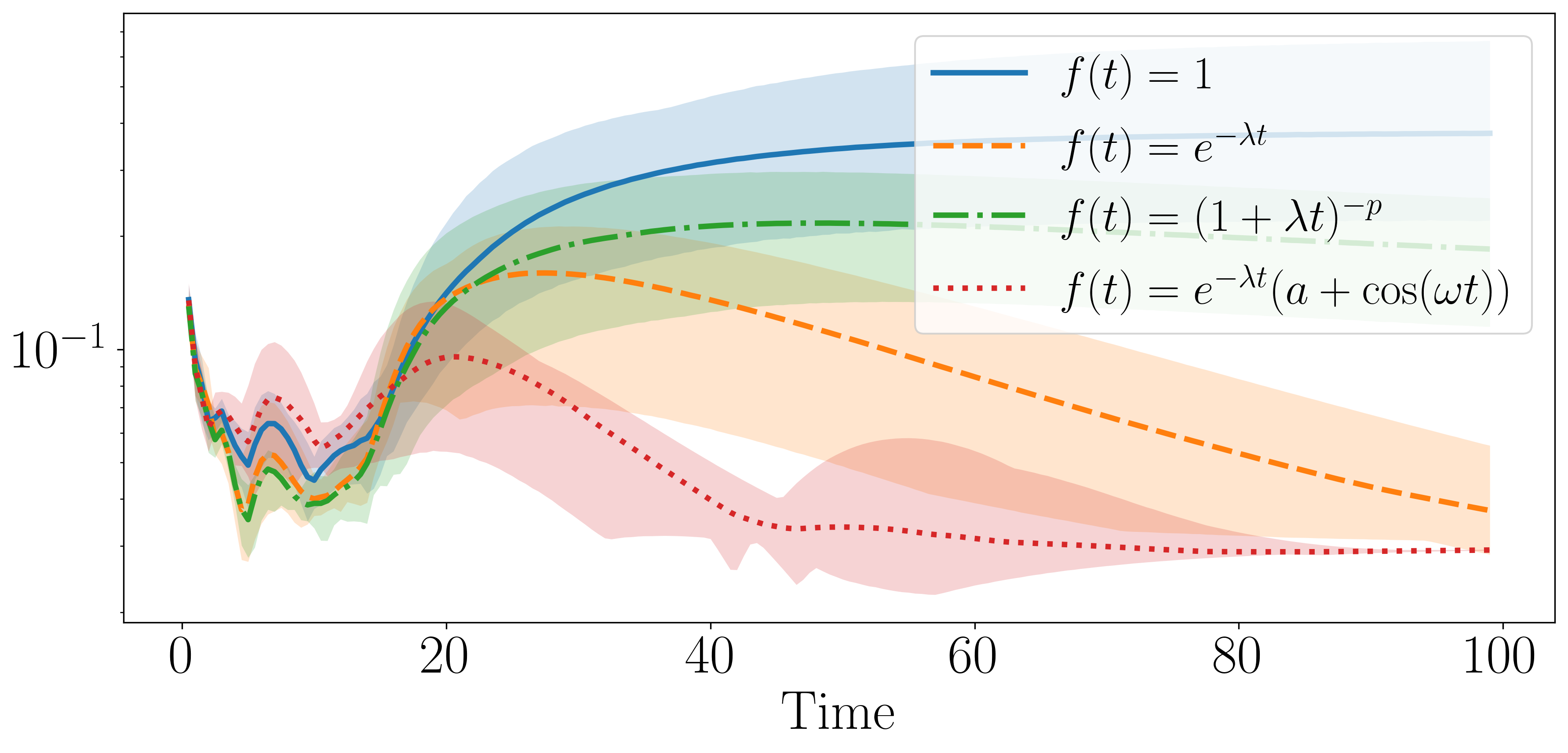}
         \caption{$\|\theta - \theta^h\|_{L_\infty}$}
     \end{subfigure}
     \begin{subfigure}[t]{0.49\textwidth}
         \centering
         \includegraphics[width=\textwidth]{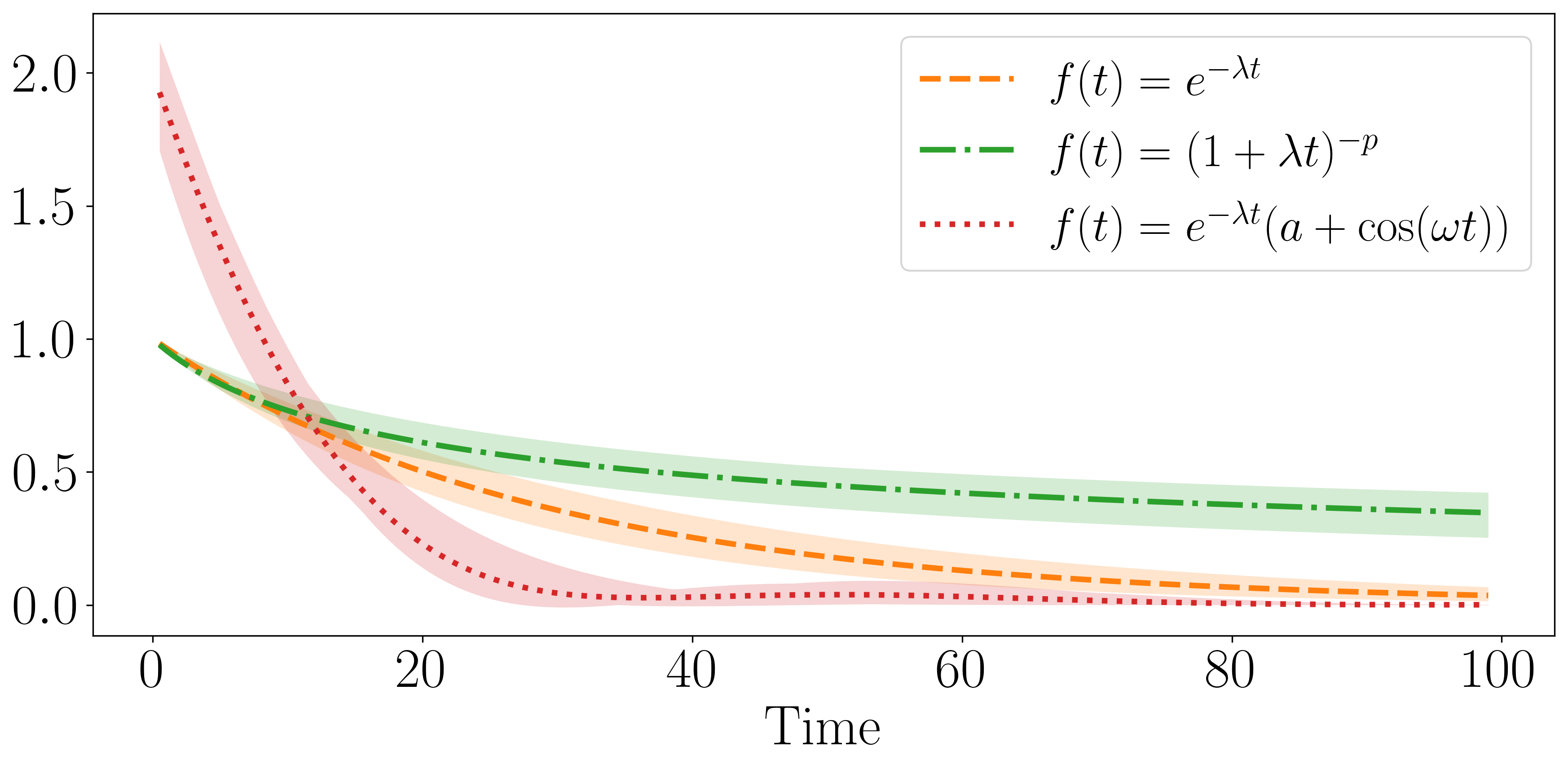}
         \caption{$f(t)$}
         \label{fig:convection_ft}
     \end{subfigure}
     \caption{Natural convection in a square cavity: (a) $L_\infty$ error over time for the temperature field, evaluated across various temporal functions $f(t)$. The network is trained on $t \in [0, 10]$ and extrapolated to $t = 100$. (b) The learned temporal functions for $f(t)$}
     \label{fig:convection_l_infty_n_ft}
\end{figure}

Similar to the previous Example \ref{sec:Eg2}, we consider a square 
domain $\Omega = (0, 1)^2$ filled with viscous fluid. The fluid is initially stationary. Upon imposing a 
high-temperature boundary condition on the left and a 
lower-temperature boundary condition on the right, the fluid begins 
to circulate clockwise due to buoyancy. The thermal gradient causes 
the warmer fluid to rise while the cooler fluid descends. The 
following coupled system governs the fluid motion:
\begin{subequations}\label{eq:coupled}
    \begin{alignat}{2}
        \frac{\partial \bu}{\partial t} + (\bu\cdot\nabla)\bu
        - 2 \text{Re}^{-1}\nabla\cdot\bm{\varepsilon}(\bu) + \nabla p
        &= \text{Ri}\,\theta\,\hat{\bm{e}},
        && \quad \text{in } \Omega \times I, \\
        \nabla\cdot \bu &= 0,
        && \quad \text{in } \Omega \times I, \\
        \frac{\partial \theta}{\partial t} + \bu \cdot \nabla \theta
        - \kappa \Delta \theta &= 0,
        && \quad \text{in } \Omega \times I,
    \end{alignat}
\end{subequations}
where $\bu = [u_1, u_2]^\top : \Omega \times I \rightarrow \mathbb{R}^2$ 
is the fluid velocity, 
$p : \Omega \times I \rightarrow \mathbb{R}$ is the pressure, and 
$\theta : \Omega \times I \rightarrow \mathbb{R}$ is the temperature. 
Here, 
$\varepsilon(\bu) := \tfrac{1}{2}(\nabla \bu + (\nabla \bu)^{\top})$ 
denotes the symmetric part of the velocity gradient. The Reynolds 
number $\text{Re}$ is the ratio of inertial to viscous forces, and 
the Richardson number $\text{Ri}$ measures the ratio of buoyancy to 
shear in the flow. The unit vector $\bm{e}$ is opposite to the 
direction of the gravitational acceleration. The dimensionless 
thermal diffusivity is defined as 
$\kappa := \text{Re}^{-1}\text{Pr}^{-1}$. $\text{Pr}$ denotes the 
Prandtl number, which represents the ratio of kinematic viscosity 
to thermal diffusivity.

In this example, we set $\text{Re} = 500$, $\text{Ri} = 1$, and 
$\text{Pr} = 0.71$. We specify a homogeneous Dirichlet boundary 
condition for the velocity $\bu$ on all boundaries 
$\partial \Omega$. The thermal boundary conditions consist of high 
temperature on the left wall and a low temperature on the right 
wall, while the top and bottom boundaries are treated with 
homogeneous Neumann conditions. These conditions are summarized as 
follows:
\begin{subequations}
    \begin{alignat}{2}
        \bu &= \bm{0}, \quad &&\text{on } \partial\Omega, \\
        \theta &= 1, \quad &&\text{on } \partial \Omega 
            \cap \{x = 0\}, \\
        \theta & = 0, \quad &&\text{on } \partial \Omega 
            \cap \{x = 1 \}, \\
        \frac{\partial \theta}{\partial \bm{n}} & = 0, 
            \quad &&\text{on } \partial \Omega \cap \{y = 0, 1 \}.
    \end{alignat}
\end{subequations}

\begin{figure}[h!]
     \centering
     \begin{subfigure}[t]{0.24\textwidth}
         \centering
         \includegraphics[width=\textwidth]{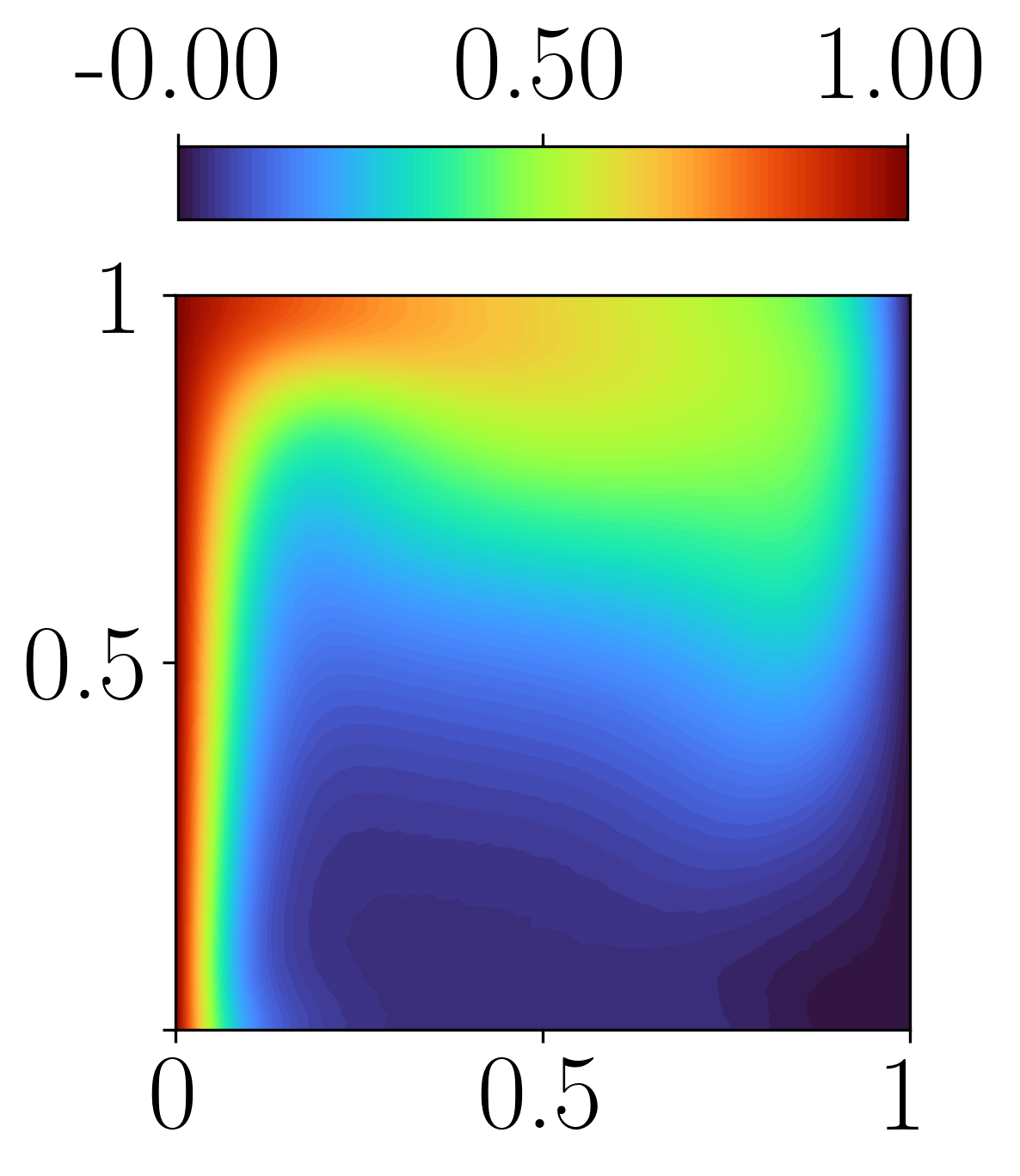}
         \caption{$t = 10$}
     \end{subfigure}
     \begin{subfigure}[t]{0.24\textwidth}
         \centering
         \includegraphics[width=\textwidth]{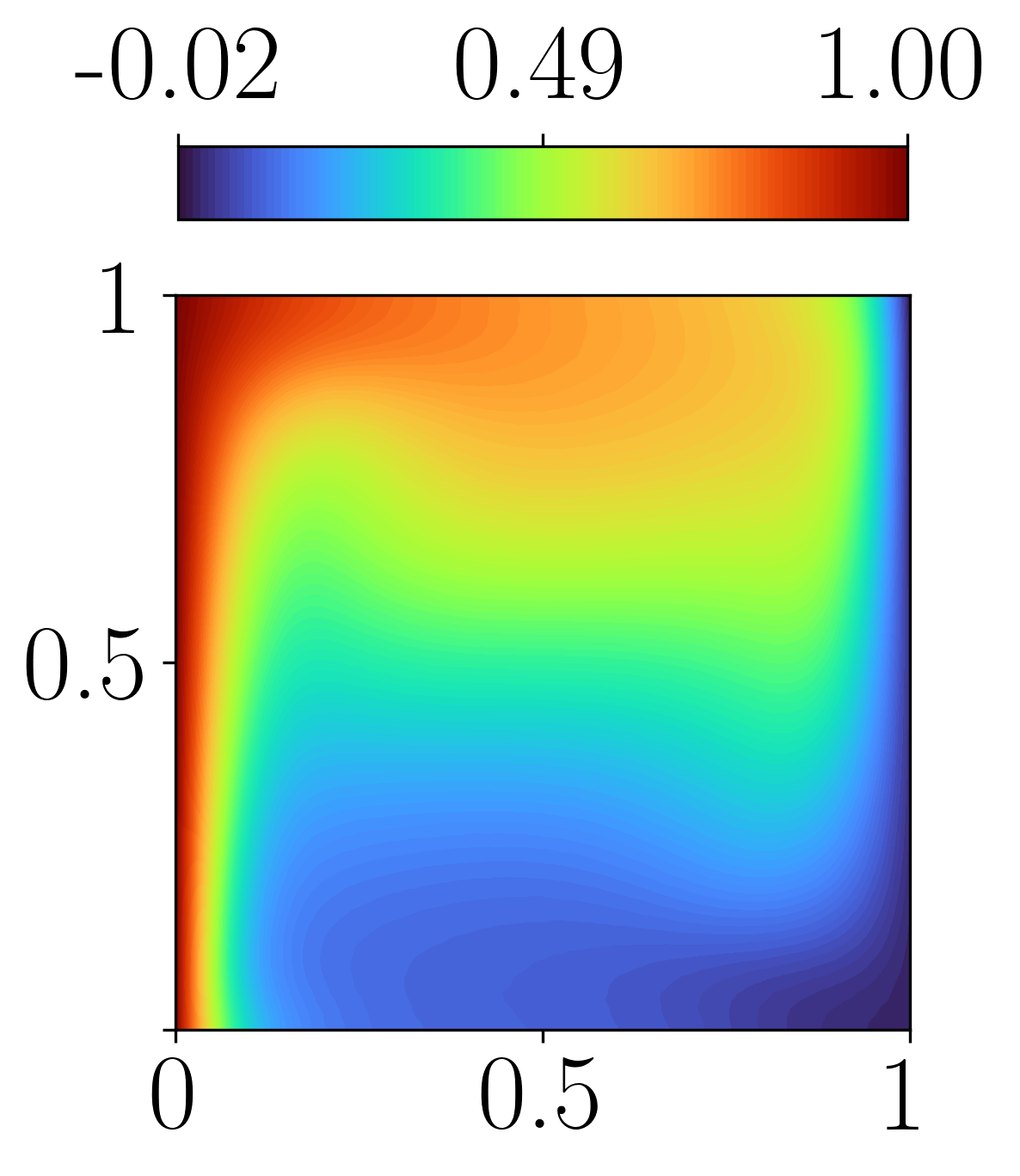}
         \caption{$t = 20$}
     \end{subfigure}
     \begin{subfigure}[t]{0.24\textwidth}
         \centering
         \includegraphics[width=\textwidth]{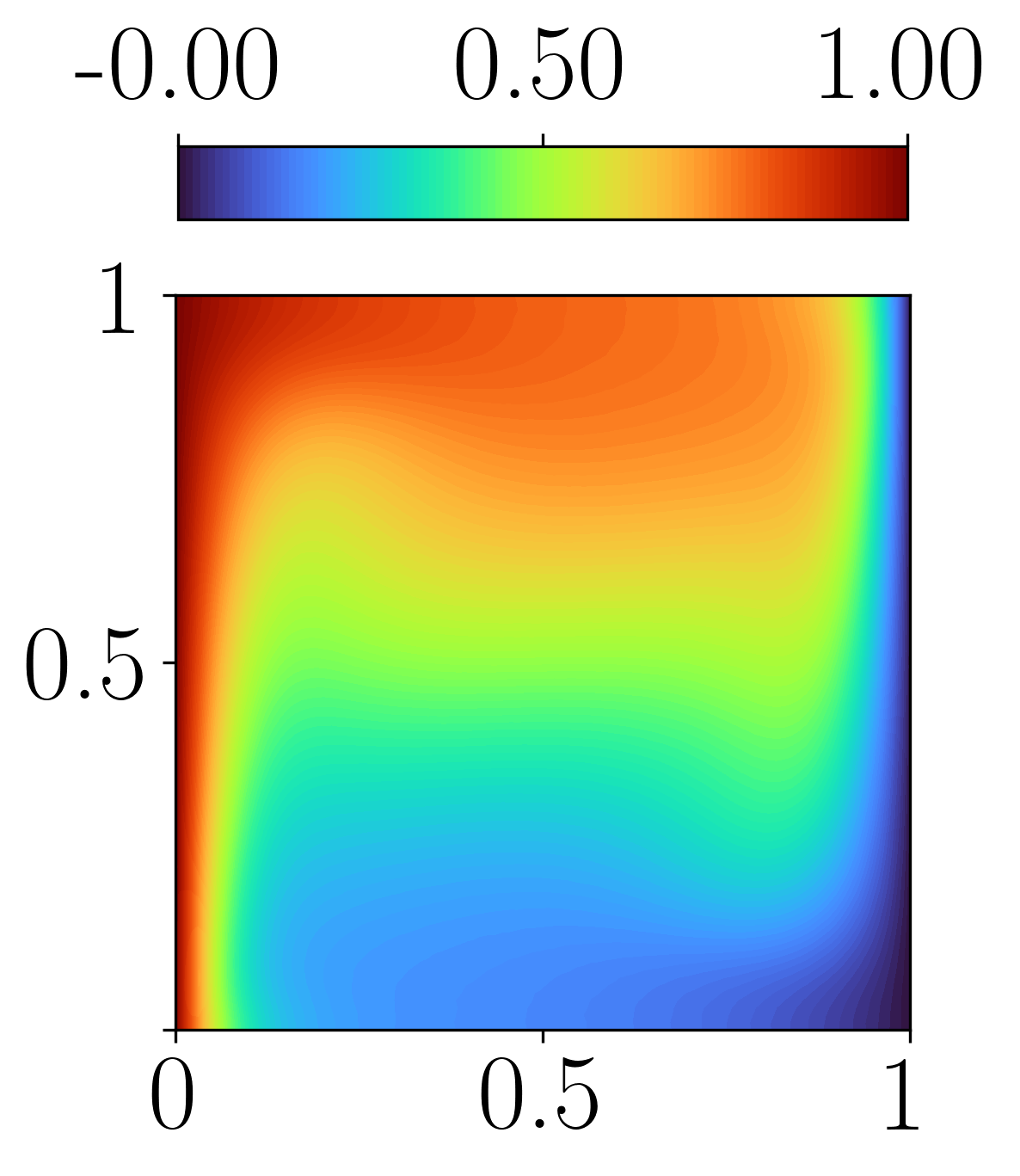}
         \caption{$t = 40$}
     \end{subfigure}
     \begin{subfigure}[t]{0.24\textwidth}
         \centering
         \includegraphics[width=\textwidth]{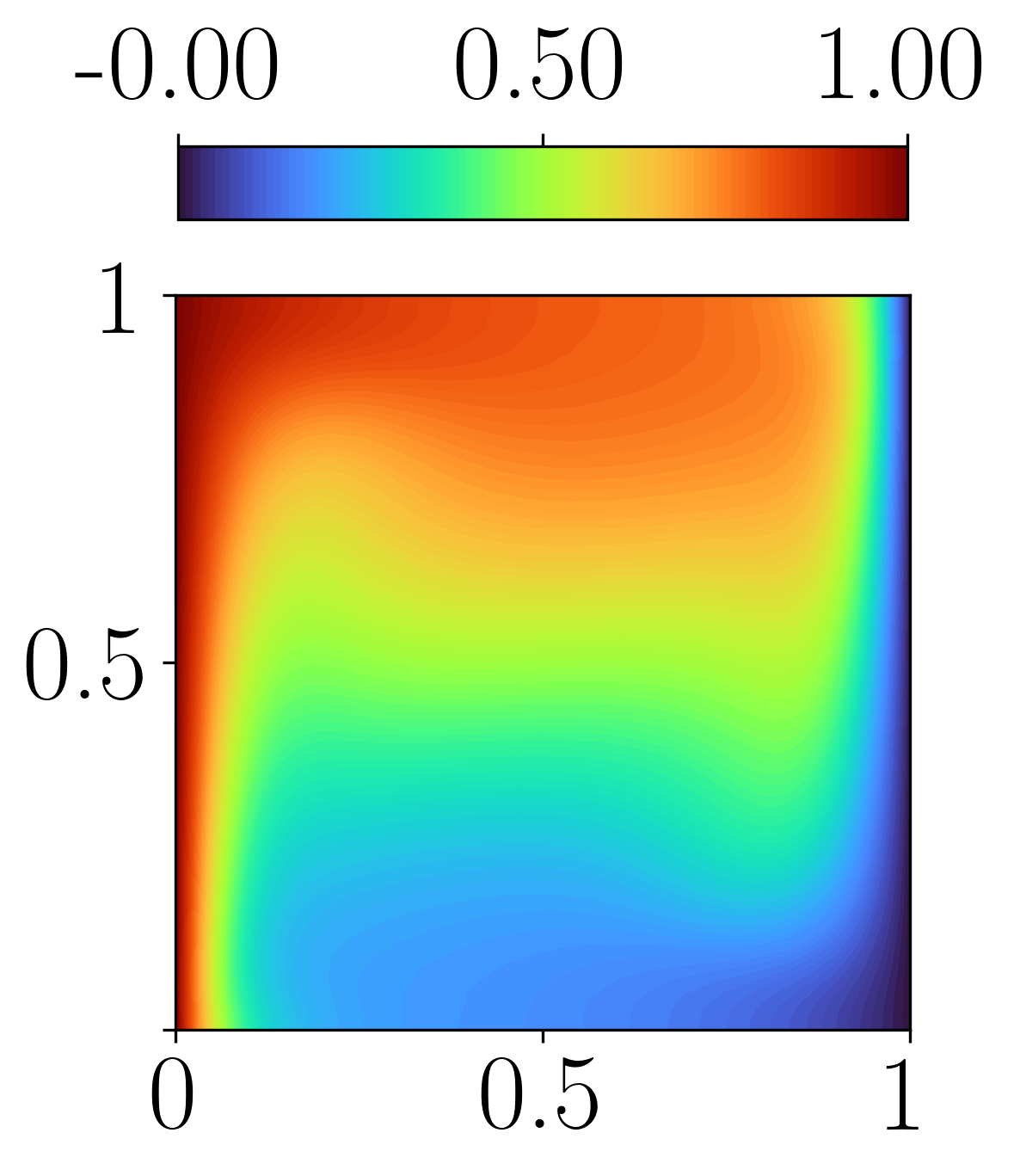}
         \caption{$t = 60$}
     \end{subfigure}     

     \begin{subfigure}[t]{0.24\textwidth}
         \centering
         \includegraphics[width=\textwidth]{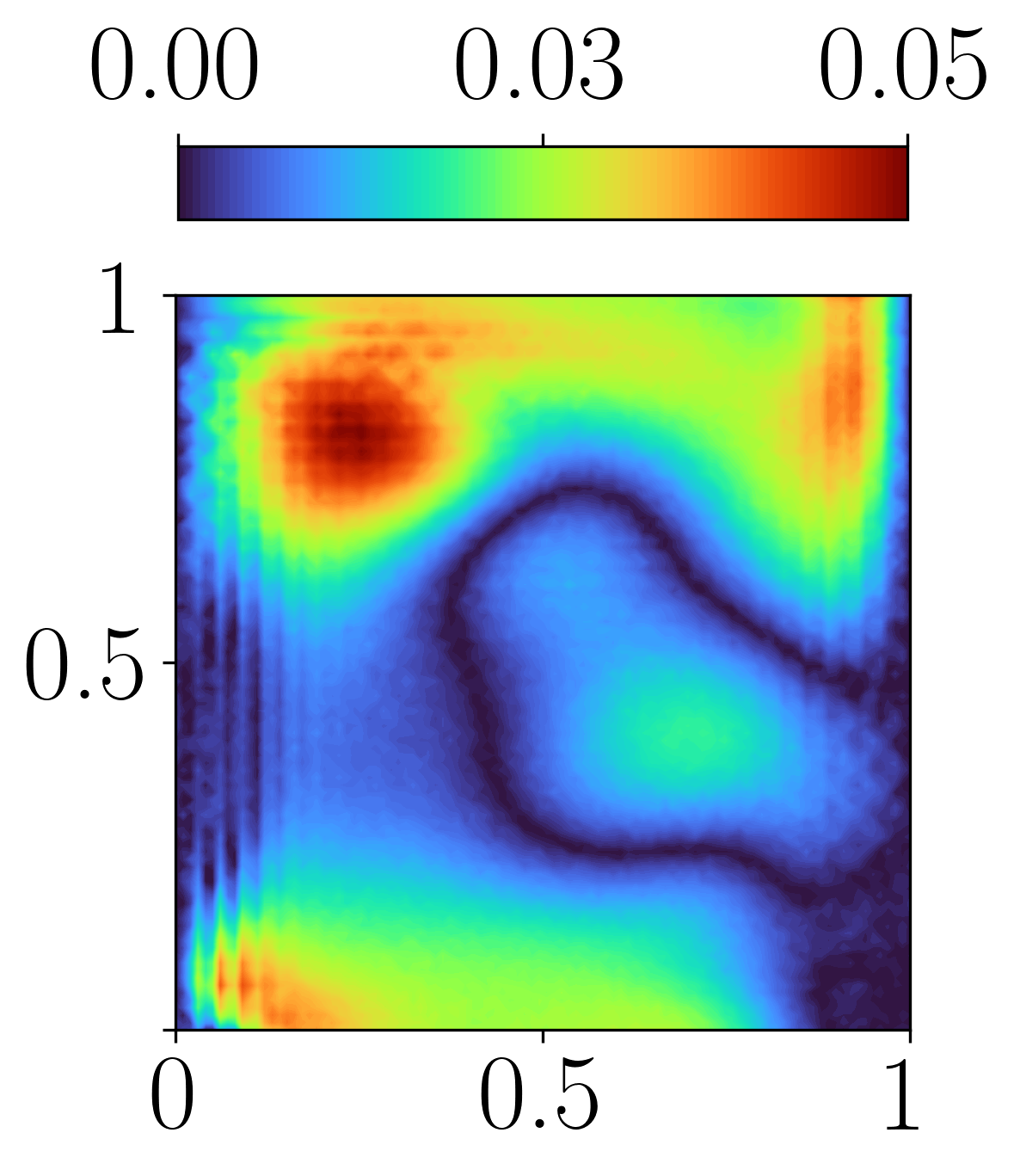}
         \caption{$t = 10$}
     \end{subfigure}
     \begin{subfigure}[t]{0.24\textwidth}
         \centering
         \includegraphics[width=\textwidth]{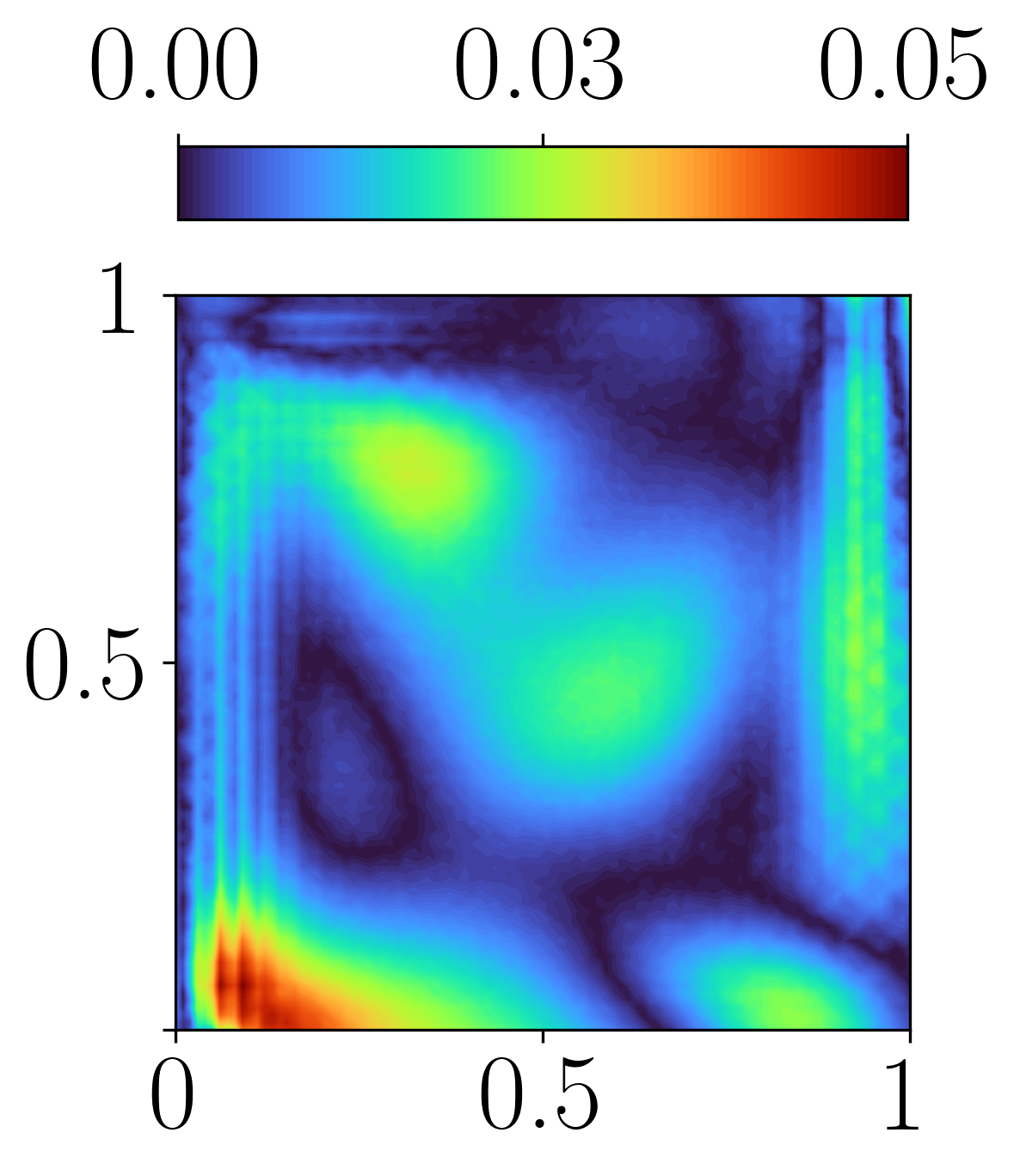}
         \caption{$t = 20$}
     \end{subfigure}
     \begin{subfigure}[t]{0.24\textwidth}
         \centering
         \includegraphics[width=\textwidth]{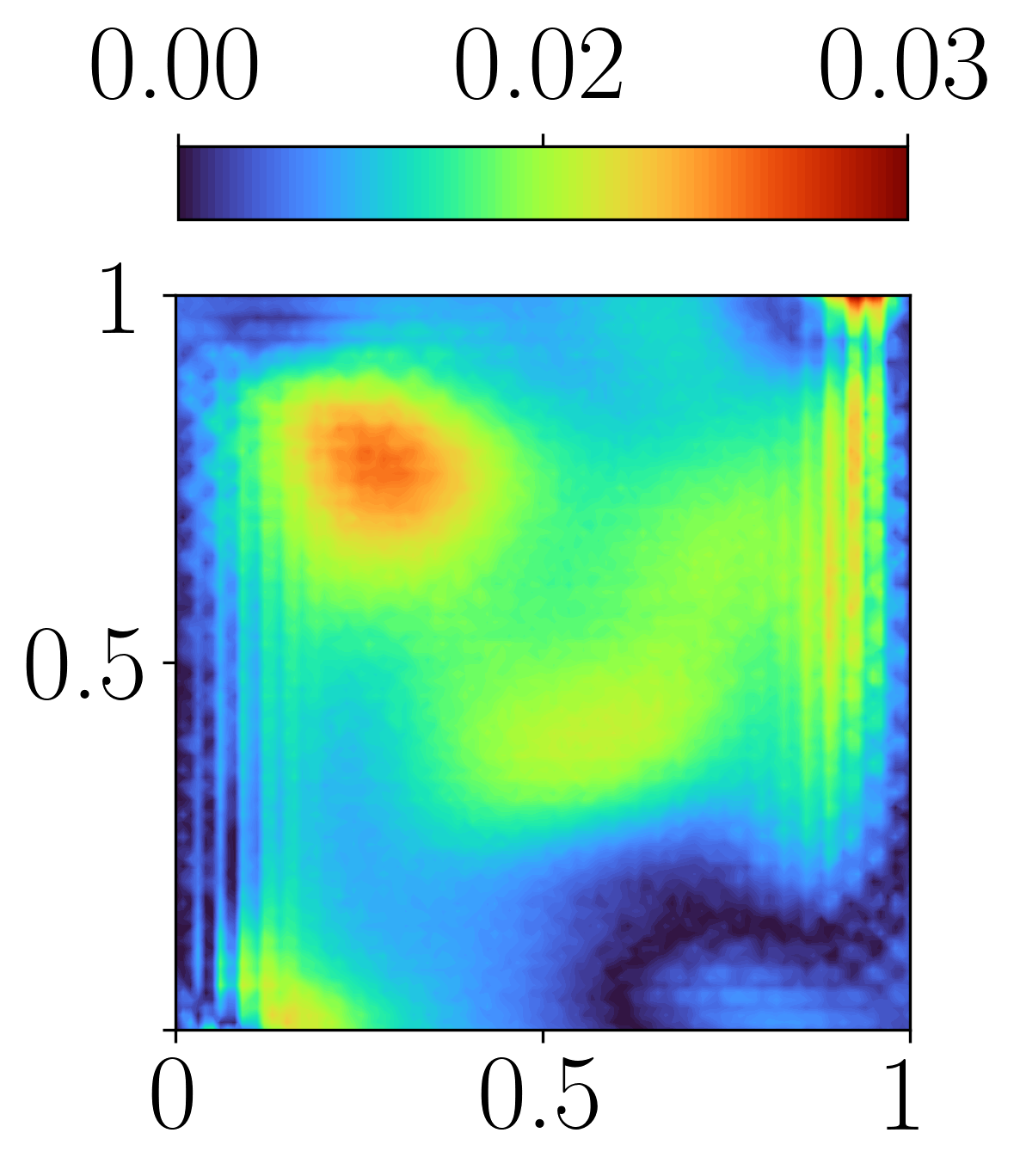}
         \caption{$t = 40$}
     \end{subfigure}
     \begin{subfigure}[t]{0.24\textwidth}
         \centering
         \includegraphics[width=\textwidth]{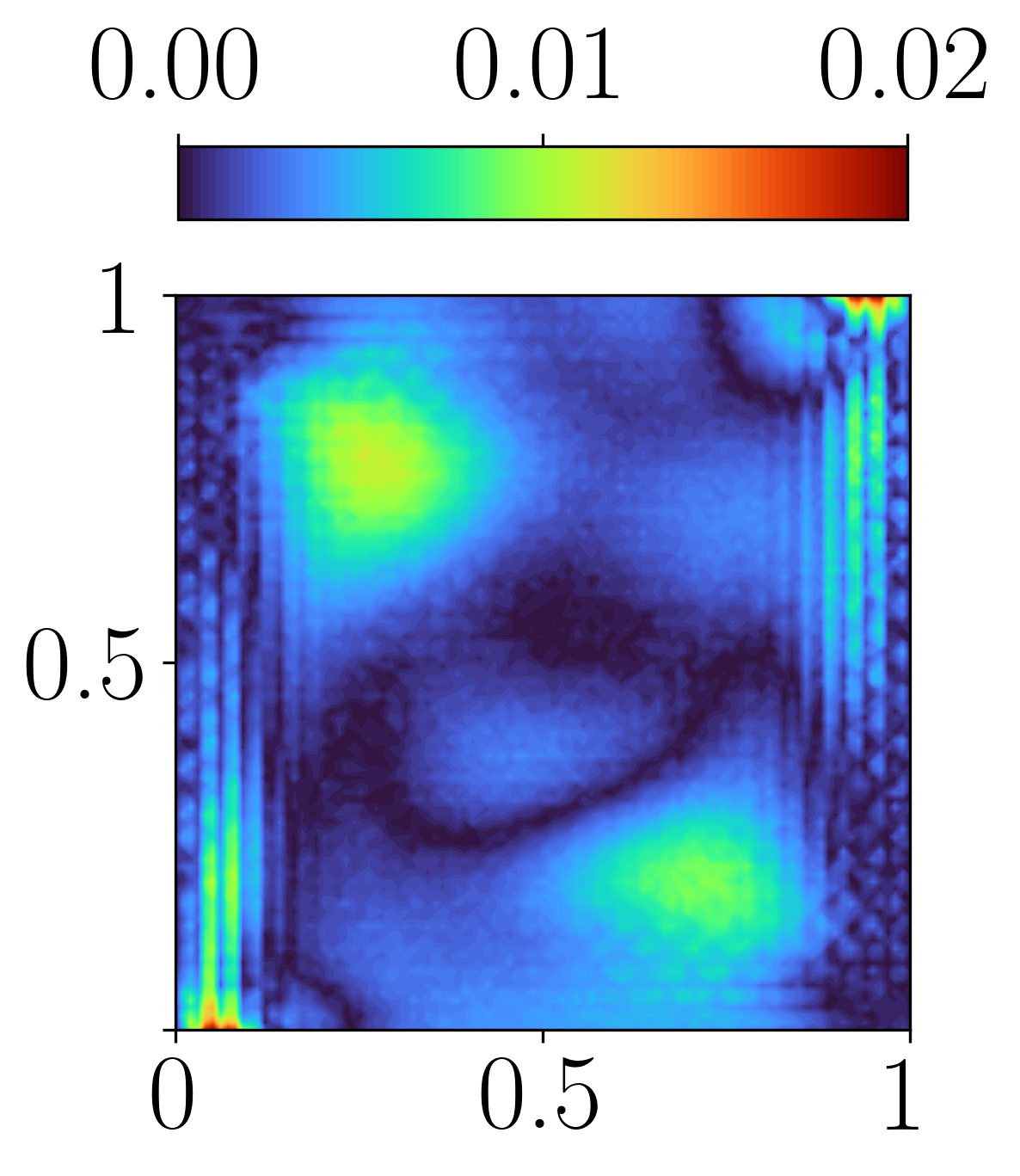}
         \caption{$t = 60$}
     \end{subfigure}
     \caption{Natural convection in a square cavity: Temporal evolution of temperature profiles (top row) and the corresponding point-wise absolute errors of the temperature predictions (bottom row) with $f(t) = e^{-\lambda t}(a + \cos(\omega t))$.}
     \label{fig:Convection_T_contour}
\end{figure}

Furthermore, we increase the number of hidden layers for FNN$_4$. 
The hyperbolic tangent $(\tanh)$ activation function is applied 
universally across all layers, with the sole exception of the 
FNN$_3$ output layer, which has linear activation. The specific network architectures are detailed below:
\begin{itemize}
    \item FNN$_1$: $[2, 50\times 2, 25]$,
    \item FNN$_2$: $[1, 50 \times 2, 25]$,
    \item FNN$_3$: $[25, 50 \times 3, 4]$,
    \item FNN$_4$: $[50, 50 \times 5, 4]$.
\end{itemize}

An important distinction from the previous examples is that 
$\text{Re} = 500$, which makes advection dominant in both the 
momentum and energy equations. The dominant advection causes the 
velocity components to fluctuate and peak before settling to the 
steady-state solution. Furthermore, the system takes a significant 
time to reach the steady state due to small diffusion. To capture 
the initial velocity fluctuations due to advection, we train 
the network for a longer time interval, i.e., $t \in [0, 10]$. The 
model's output is extrapolated beyond $t = 10$. 

We train FNN$_1$ and FNN$_3$ using 10,000 random collocation points 
to obtain the steady-state solution for the velocity, pressure, and temperature.
Once these networks are trained, we freeze the weights of FNN$_1$ and FNN$_2$ and train 
FNN$_2$ and FNN$_4$ with 100,000 random collocation points. During 
each optimization step, we employ a mini-batch strategy by 
subsampling 3,000 collocation points and using 3,000 boundary 
points. We enforce the initial 
condition using 1,000 randomly sampled points. 

Figure~\ref{fig:convection_l_infty_n_ft} presents the $L_\infty$ 
error of the temperature distribution for various choices of $f(t)$, 
alongside the learned profiles. Comparatively larger errors are 
observed within the training interval, where the velocity rapidly 
increases and peaks, forcing the temperature equation to resolve 
strong advection. During extrapolation, the damped oscillatory 
profile generally exhibits the lowest error. The mechanism behind 
this advantage is visible in Figure~\ref{fig:convection_ft}: the 
damped oscillatory profile starts at $f(0) = a + 1 \approx 2.0$ 
and decays sharply, whereas the exponential profile is anchored at 
$f(0) = 1$. The additional amplitude parameter $a$ effectively 
decouples the initial magnitude of the transient correction from 
the decay rate, allowing the optimizer to learn a fast decay 
without prematurely suppressing the contribution of 
$g_\theta(\bm{x}, t)$ during the training interval. In contrast, 
the purely exponential model $f(t) = e^{-\lambda t}$ faces a 
trade-off: because $f(0) = 1$ is fixed, a sharper decay rate would 
reduce $f(t)\,g_\theta$ too rapidly to match the peaking dynamics 
within $t \in [0, 10]$. The optimizer is therefore forced to learn 
a slower decay rate, which improves the fit during training but 
results in a slower approach to the steady state during 
extrapolation. Consistent with the previous examples, the algebraic 
and constant profiles perform worse than both the damped oscillatory 
and exponential models.

Figure~\ref{fig:Convection_T_contour} illustrates the system's 
transient behavior and evaluates the model's predictive accuracy 
from $t = 10$ to $t = 60$. The top row (a--d) captures the temporal 
evolution of the temperature field, showing the thermal front 
progressively spreading from the left boundary into the interior 
domain and developing the characteristic S-shaped isotherms of the 
differentially heated cavity. This evolution is primarily driven by 
temperature advection from the clockwise-circulating fluid. The 
bottom row (e--h) presents the corresponding pointwise absolute 
errors for the temperature predictions. The errors decrease 
monotonically during extrapolation---from $\mathcal{O}(0.05)$ at 
$t = 10$ to $\mathcal{O}(0.02)$ at $t = 60$---confirming that the 
temporal profile $f(t)$ progressively suppresses the transient 
correction as intended. 


\subsection{Example 4: Three-Dimensional Conjugate Heat Transfer}
\label{sec:Eg4}
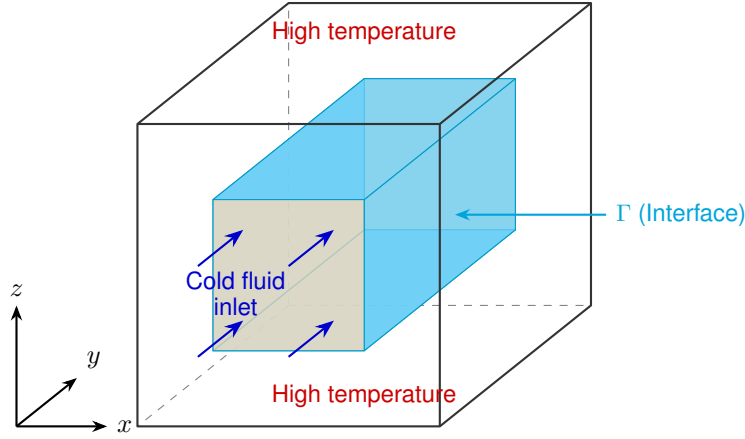
\begin{figure}
    \centering
    \begin{tikzpicture}[
        scale=4,
        x={(1cm,0cm)},      
        y={(0.5cm,0.4cm)},  
        z={(0cm,1cm)}       
    ]
    
        \draw[dashed, black!50] (0,1,0) -- (1,1,0);
        \draw[dashed, black!50] (0,1,0) -- (0,1,1);
        \draw[dashed, black!50] (0,1,0) -- (0,0,0);
    
        \tikzstyle{channel}=[fill=orange!30, fill opacity=0.7, draw=cyan!80!black, thin, join=round]
        
        \tikzstyle{interface}=[fill=cyan!50, fill opacity=0.8, draw=cyan!80!black, thin, join=round]
    
        
        \filldraw[channel] (0.25, 1, 0.25) -- (0.75, 1, 0.25) -- (0.75, 1, 0.75) -- (0.25, 1, 0.75) -- cycle;
        
        \filldraw[interface] (0.25, 0, 0.25) -- (0.75, 0, 0.25) -- (0.75, 1, 0.25) -- (0.25, 1, 0.25) -- cycle;
        
        \filldraw[interface] (0.25, 0, 0.25) -- (0.25, 1, 0.25) -- (0.25, 1, 0.75) -- (0.25, 0, 0.75) -- cycle;
        
        \filldraw[interface] (0.75, 0, 0.25) -- (0.75, 1, 0.25) -- (0.75, 1, 0.75) -- (0.75, 0, 0.75) -- cycle;
        
        \filldraw[interface] (0.25, 0, 0.75) -- (0.75, 0, 0.75) -- (0.75, 1, 0.75) -- (0.25, 1, 0.75) -- cycle;
        
        \filldraw[channel] (0.25, 0, 0.25) -- (0.75, 0, 0.25) -- (0.75, 0, 0.75) -- (0.25, 0, 0.75) -- cycle;
    
        \tikzstyle{edge}=[thick, black!80, join=round]
        
        \draw[edge] (0,0,0) -- (1,0,0) -- (1,1,0) -- (1,1,1) -- (0,1,1) -- (0,0,1) -- cycle;
        
        \draw[edge] (1,0,0) -- (1,0,1);
        \draw[edge] (0,0,1) -- (1,0,1);
        \draw[edge] (1,0,1) -- (1,1,1);
    
        \foreach \x/\z in {0.35/0.35, 0.65/0.35, 0.35/0.65, 0.65/0.65} {
            \draw[-{Stealth[length=2.5mm]}, thick, blue!80!black] (\x, -0.3, \z) -- (\x, 0, \z);
        }
        
        \node[blue!80!black, align=center, font=\sffamily\small] at (0.5, -0.35, 0.58) {Cold fluid \\ inlet};
        
        \node[red!80!black, align=center, font=\sffamily\small] at (0.5, 0.5, 1.1) {High temperature};
        
        \node[red!80!black, align=center, font=\sffamily\small] at (0.5, 0.5, -0.1) {High temperature};

        \draw[-{Stealth[length=2mm]}, thick, cyan!90!black] (1.3, 0.5, 0.5) node[right, font=\sffamily\small] {$\Gamma$ (Interface)} -- (0.8, 0.5, 0.5);

        \draw[-{Stealth[length=2mm]}, thick, black] (-0.4, 0, 0) -- (-0.1, 0, 0) node[right] {$x$};
        \draw[-{Stealth[length=2mm]}, thick, black] (-0.4, 0, 0) -- (-0.4, 0.4, 0) node[above right] {$y$};
        \draw[-{Stealth[length=2mm]}, thick, black] (-0.4, 0, 0) -- (-0.4, 0, 0.4) node[above] {$z$};
        
    \end{tikzpicture}
    \caption{Three-dimensional domain for conjugate heat transfer}
    \label{fig:3D_domain}
\end{figure}

This example presents a test case for conjugate heat transfer involving a solid domain intersected by a fluid channel. Geometrically, the system consists of a unit cube, $\Omega = (0, 1)^3$, containing a square fluid channel defined by $\Omega_f = (0.25, 0.75) \times (0, 1) \times (0.25, 0.75)$, as shown in Figure~\ref{fig:3D_domain}.
As the fluid flows through the channel, it convects heat away from the surrounding solid to the outlet. The governing equations for the fluid channel are given in~\eqref{eq:coupled}, while the standard heat equation is solved for the solid. The only difference is that this example is three-dimensional, whereas the governing equations in~\eqref{eq:coupled} are written in two-dimensional form. Let $\Gamma = \partial \Omega_f \setminus \{y=0, 1\}$ denote the fluid-solid interface and $\theta$ denote the global continuous temperature field, defined as $\theta_s$ within the solid domain ($\Omega \setminus \Omega_f$) and $\theta_f$ within the fluid domain ($\Omega_f$). For the fluid, the thermal diffusivity is $\kappa_f := \text{Re}^{-1} \text{Pr}^{-1}$, and for the solid, it is $\kappa_s$. We solve the system subject to the following boundary and initial conditions:
\begin{subequations}
    \begin{alignat}{2}
        \bu &= \left[\begin{matrix}0 \\ (1 - 16(x - 0.5)^2)(1 - 16(z-0.5)^2)\\ 0\end{matrix}\right], \quad &&\text{on } \partial\Omega_f \cap \{y=0 \}, \\
        \theta &= 1, \quad &&\text{on } \partial \Omega \cap \{z = 0, 1\}, \\            
        \theta &= 0, \quad &&\text{on } \partial \Omega_f \cap \{y = 0\}, \\
        (2\text{Re}^{-1}\bm{\varepsilon}(\bu) - p\bm{I})\bm{n}\  &= \mathbf{0}, \quad &&\text{on } \partial \Omega_f \cap \{y = 1\}, \\
        \frac{\partial \theta}{\partial \bm{n}} &= 0, \quad &&\text{on } \partial \Omega \setminus \big( (\partial \Omega_f \cap \{y = 0\}) \cup \{z = 0, 1\} \big), \\
        \bu &= \mathbf{0}, \quad &&\text{on } \Gamma, \\
        \theta_s &= \theta_f, \quad &&\text{on } \Gamma, \label{eq:temp_continuity}\\
        \kappa_s \frac{\partial \theta_s}{\partial \bm{n}} &= \kappa_f \frac{\partial \theta_f}{\partial \bm{n}}, \quad &&\text{on } \Gamma, \label{eq:flux_continuity}\\
        \bm{u}(\cdot, 0) &= \left[\begin{matrix} 0\\ (1 - 16(x - 0.5)^2)(1 - 16(z-0.5)^2)\\ 0 \end{matrix} \right], \quad &&\text{in } \Omega_f, \\
        \theta(\cdot, 0) &= 0, \quad &&\text{in } \Omega
    \end{alignat}
\end{subequations}
where~\eqref{eq:temp_continuity} and~\eqref{eq:flux_continuity} impose temperature and flux continuity, respectively. For the parameter values, we fix Re = 100, Pr = 1, Ri = 5, $\kappa_f = \text{Re}^{-1} \text{Pr}^{-1} = 0.01$, and $\kappa_s = 0.02$.

This experimental setup comprises distinct solid and fluid domains separated by an interface with a discontinuous thermal diffusion coefficient. Therefore, we employ two separate sets of neural networks—FNN$_i^r$, where $i = \{1, 2, 3, 4\}$ and $r = \{f, s\}$ represents the fluid and solid domains, respectively. Both sets are trained simultaneously, with the fluid-solid interface boundary conditions from \eqref{eq:temp_continuity} and \eqref{eq:flux_continuity} imposed as soft constraints. To evaluate the performance of the proposed method under varying training conditions, we investigate two scenarios: (1) training all FNN$_i^r$ networks with physics-informed losses; and (2) a hybrid approach training FNN$_1^r$ and FNN$_3^r$ using FEM solutions, while training FNN$_2^r$ and FNN$_4^r$ with physics-informed losses. For all scenarios, FNN$_2^r$ and FNN$_4^r$ are trained over the time interval $t \in [0, 1.5]$.


The network architectures are defined below:
\begin{itemize}
    \item FNN$_1^r$: $[3, 50 \times 2, 25]$
    \item FNN$_2^r$: $[1, 50 \times 2, 25]$
    \item FNN$_3^r$: $[25, 50 \times 3, 5]$
    \item FNN$_4^f$: $[50, 50 \times 5, 5], \quad \text{FNN}_4^s: [50, 50 \times 5, 1]$
\end{itemize}

\subsubsection{Training FNN$_i^r$ using physics-informed losses}
As in previous examples, all network blocks are trained using the physics-informed loss formulations. To train FNN$_1^r$ and FNN$_3^r$, the dataset consists of 10,000 random points on each boundary face (including the exterior and fluid-solid interfaces) and 250,000 random collocation points within each fluid and solid domain. During training, we use mini-batches comprising 1,000 boundary points per face and 10,000 collocation points. For training FNN$_2^r$ and FNN$_4^r$, the dataset is expanded over the time interval $t \in [0, 1.5]$ and includes 50,000 boundary points per face, 50,000 random initial points, and 250,000 random collocation points per domain. The mini-batch sizes for this configuration are set to 1,000 boundary points per face, 1,000 initial points, and 10,000 collocation points.

\begin{figure}[h!]
     \centering
     \begin{subfigure}[t]{0.49\textwidth}
         \centering
         \includegraphics[width=\textwidth]{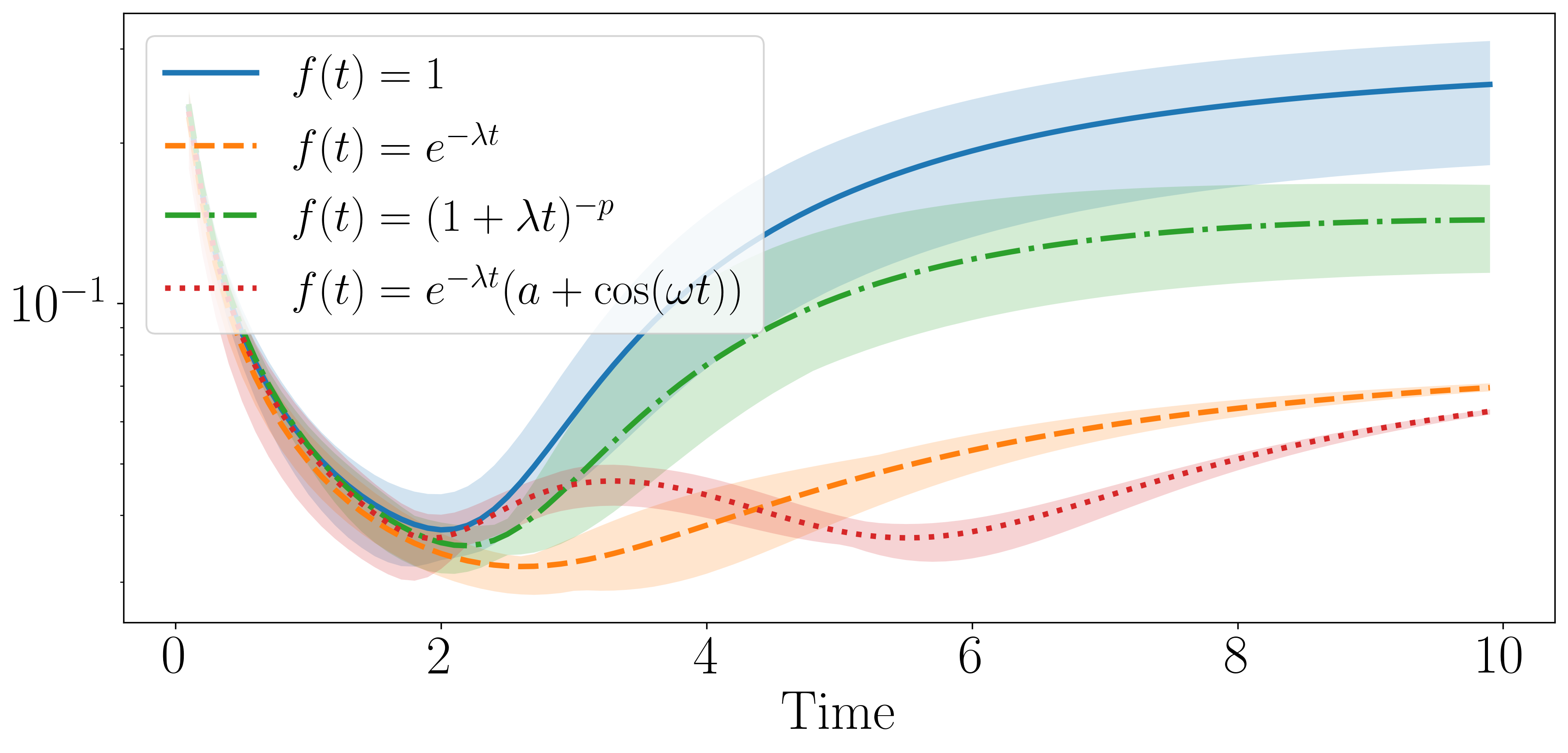}
         \caption{$\frac{\|\theta - \theta^h\|_{L_2}}{\|\theta\|_{L_2}}$}
        \label{fig:3D_PINN_SS_L_2}
     \end{subfigure}
     \begin{subfigure}[t]{0.49\textwidth}
         \centering
         \includegraphics[width=\textwidth]{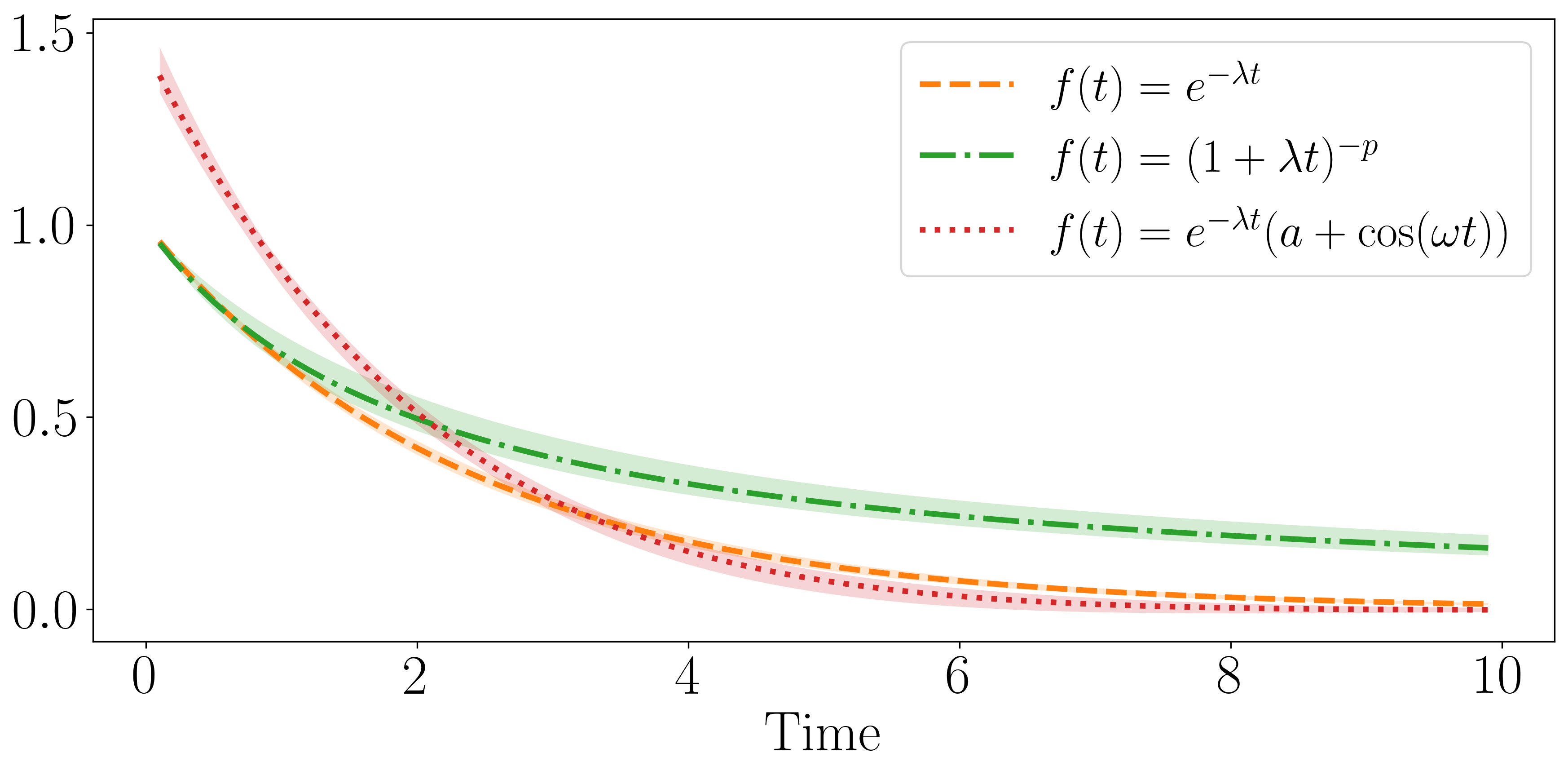}
         \caption{$f(t)$}
         \label{fig:3D_PINN_SS_ft}
     \end{subfigure}
     \caption{Conjugate heat transfer in three dimensions: (a) Relative $L_2$ error over time for the temperature field, evaluated across various temporal functions $f(t)$. The network is trained on $t \in [0, 1.5]$ and extrapolated to $t = 10$. All the networks FNN$_i^r$, where $i = \{1, 2, 3, 4\}$ and $r = \{f, s\}$, are trained with physics-informed losses. (b) The learned temporal functions $f(t)$.}
     \label{fig:3D_PINN_SS_L_2_n_ft}
\end{figure}

Figure~\ref{fig:3D_PINN_SS_L_2_n_ft} shows the error of the temperature for various choices of $f(t)$ when the model is trained on $t\in[0, 1.5]$. The y-axis in the figure shows the relative $L_2$ error, which is computed as $\|\theta - \theta^h\|_{L_2}/\|\theta\|_{L_2}$. The damped oscillatory function exhibits superior long-term performance, overtaking the purely exponential decay after $t=4$. As illustrated in Figure~\ref{fig:3D_PINN_SS_ft}, although the actual oscillation of this function is negligible, the profile benefits significantly from its flexible amplitude control. This freedom enables the damped oscillatory model to learn a larger initial amplitude and subsequently achieve a steeper decay rate. Conversely, the purely exponential decay model ($f(t) = e^{-\lambda t}$) is strictly constrained to an initial amplitude of one. To adequately fit the training data under this constraint, the network is forced to learn a much slower decay rate, which inherently degrades its ability to extrapolate effectively beyond the training domain. Consistent with previous examples, the constant and algebraic temporal profiles perform substantially worse than both the exponential and damped oscillatory models.

Finally, compared to earlier examples, we observe an increased error in the steady-state regime. This discrepancy arises from inaccuracies inherent in the steady-state condition learned via the physics-informed loss formulation. Because the ansatz inherits whatever accuracy the prescribed steady state provides, and here that state is itself learned imperfectly, the network ultimately converges to an erroneous steady-state solution rather than the true physical solution. We emphasize that this limitation reflects the quality of the supplied steady state rather than the ansatz itself, which is agnostic to the source of the steady-state component.

\begin{figure}[h!]
     \centering
     \begin{subfigure}[t]{0.49\textwidth}
         \centering
         \includegraphics[width=\textwidth]{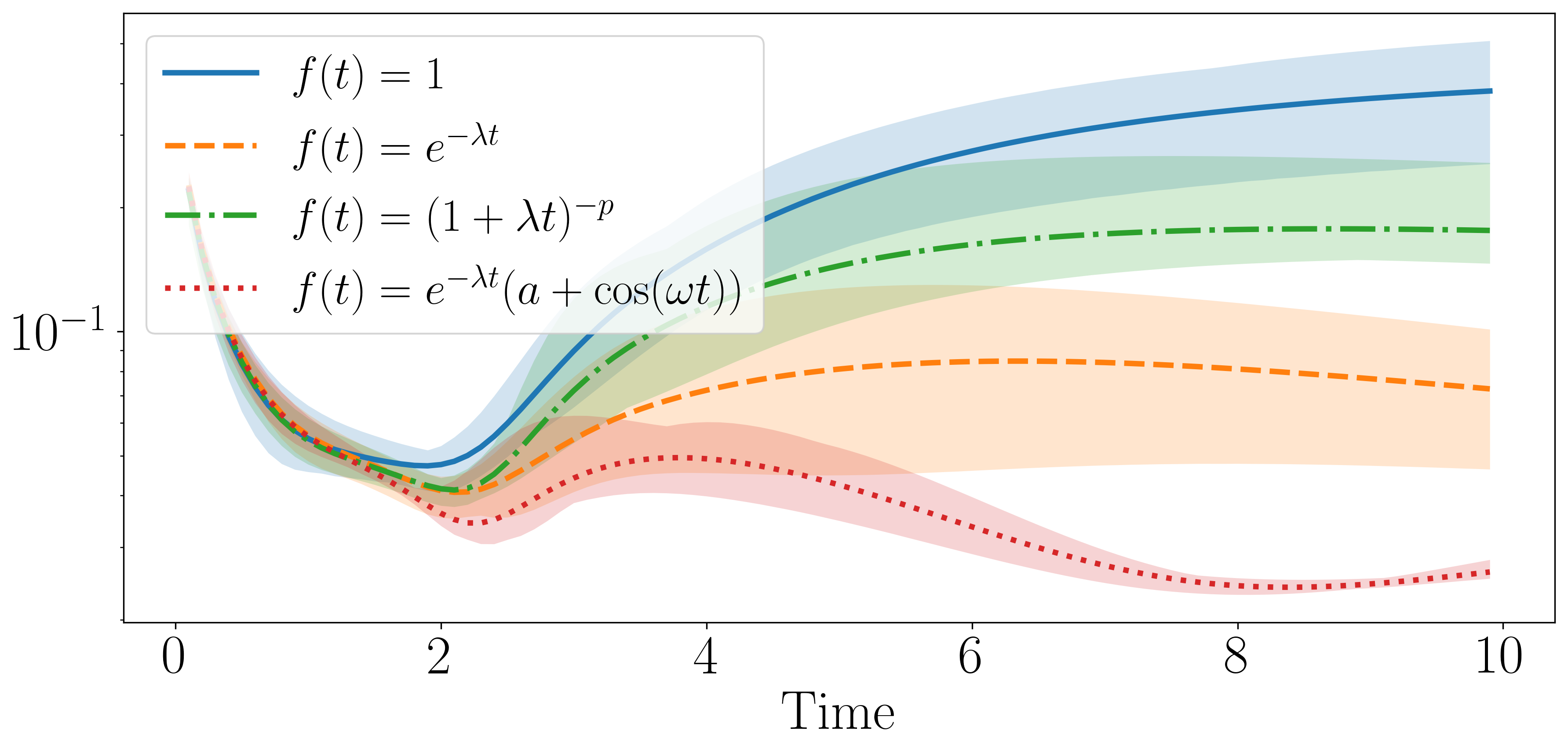}
         \caption{$\frac{\|\theta - \theta^h\|_{L_2}}{\|\theta\|_{L_2}}$}
        \label{fig:3D_Data_SS_L_2}
     \end{subfigure}
     \begin{subfigure}[t]{0.49\textwidth}
         \centering
         \includegraphics[width=\textwidth]{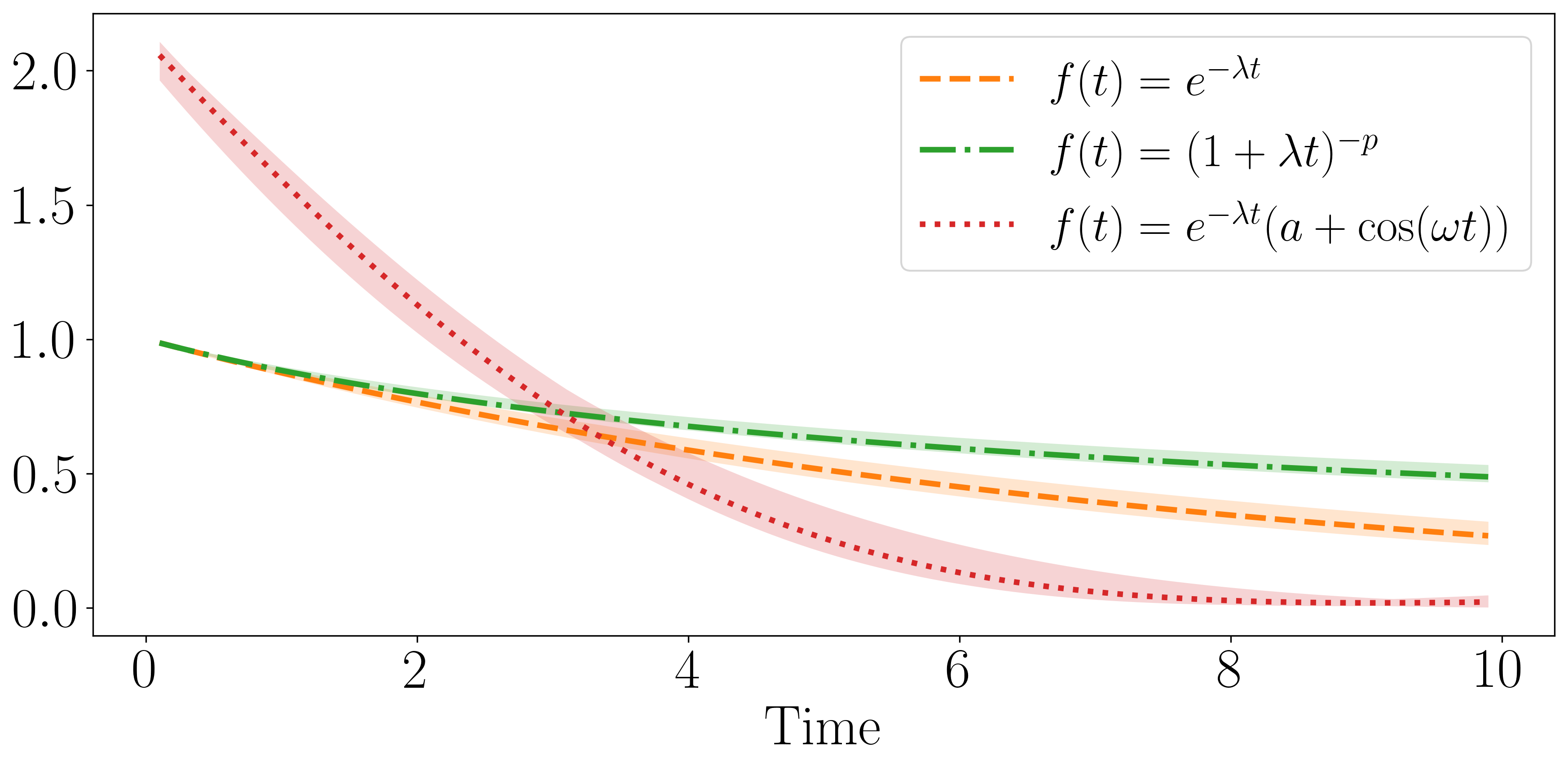}
         \caption{$f(t)$}
         \label{fig:3D_Data_SS_ft}
     \end{subfigure}
     \caption{Conjugate heat transfer in three dimensions: (a) Relative $L_2$ error over time for the temperature field, evaluated across various temporal functions $f(t)$. The network is trained on $t \in [0, 1.5]$ and extrapolated to $t = 10$. FNN$_{1, 3}^r$ are trained with FEM solutions and FNN$_{2, 4}^r$ are trained with physics-informed losses, where $r = \{f, s\}$. (b) The learned temporal functions $f(t)$.}
     \label{fig:3D_Data_SS_L_2_n_ft}
\end{figure}
\subsubsection{Training FNN$_{1,3}^r$ using FEM solutions and FNN$_{2, 4}^r$ with physics-informed losses}
To train FNN$_1^r$ and FNN$_3^r$, we utilize training data generated from the enriched Galerkin FEM model~\cite{poudel2025pressure}. For training FNN$_2^r$ and FNN$_4^r$, we use the physics-informed losses with the same number of random points as before: in the time interval $t \in [0, 1.5]$, we use 50,000 boundary points per face, 50,000 random initial points, and 250,000 random collocation points per domain, with mini-batch sizes set to 1,000 boundary points per face, 1,000 initial points, and 10,000 collocation points.

Figure~\ref{fig:3D_Data_SS_L_2_n_ft} shows the error of the temperature for various choices of $f(t)$ when the model is trained on $t\in[0, 1.5]$. The damped oscillatory function exhibits significantly superior performance. Similar to the previous examples, the damped oscillatory model's flexible amplitude allows it to learn a steeper decay rate, as shown in Figure~\ref{fig:3D_Data_SS_ft}. In contrast, the purely exponential model ($f(t) = e^{-\lambda t}$) is fixed to an initial amplitude of one, forcing a slower decay rate that limits its extrapolation ability.

In the previous setup (Figure~\ref{fig:3D_PINN_SS_L_2_n_ft}), relying solely on physics-informed losses resulted in the model converging to an erroneous steady-state solution during extrapolation. However, in this hybrid approach, we observe a drastically reduced error in the extrapolation regime. By anchoring the training of specific network blocks (FNN$_1^r$ and FNN$_3^r$) with high-fidelity data generated from the enriched Galerkin FEM model, the overall architecture is no longer solely reliant on the imperfect steady-state conditions learned purely via physics-informed loss formulations. The inclusion of the FEM solution provides a robust ground truth that stabilizes the model, preventing the extrapolation errors and ensuring the network tracks much closer to the true physical solution over extended time.


\begin{figure}[ht]
     \centering
     \begin{subfigure}[t]{0.24\textwidth}
         \centering
         \includegraphics[width=\textwidth]{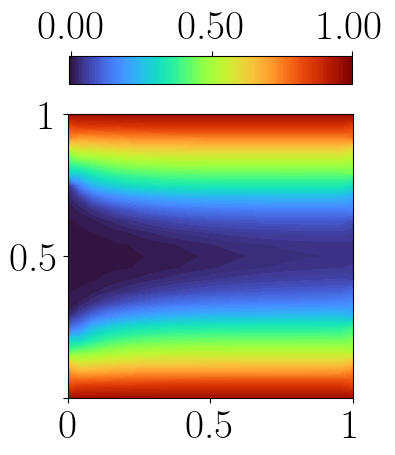}
         \caption{$t = 1.5$}
     \end{subfigure}
     \begin{subfigure}[t]{0.24\textwidth}
         \centering
         \includegraphics[width=\textwidth]{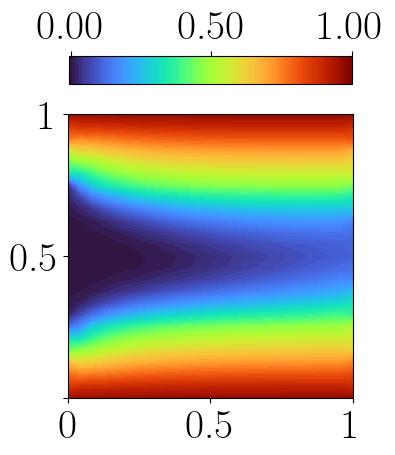}
         \caption{$t = 2.5$}
     \end{subfigure}
     \begin{subfigure}[t]{0.24\textwidth}
         \centering
         \includegraphics[width=\textwidth]{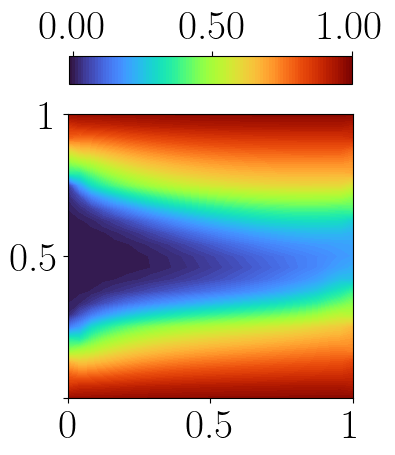}
         \caption{$t = 5$}
     \end{subfigure}

     \begin{subfigure}[t]{0.24\textwidth}
         \centering
         \includegraphics[width=\textwidth]{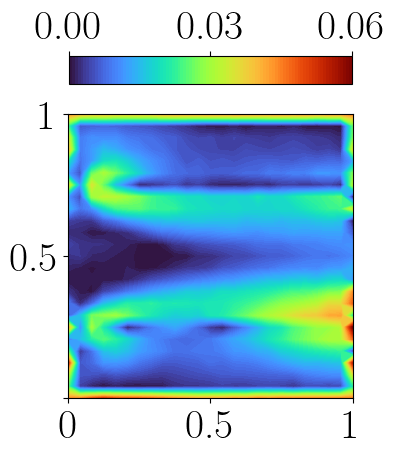}
         \caption{$t = 1.5$}
     \end{subfigure}
     \begin{subfigure}[t]{0.24\textwidth}
         \centering
         \includegraphics[width=\textwidth]{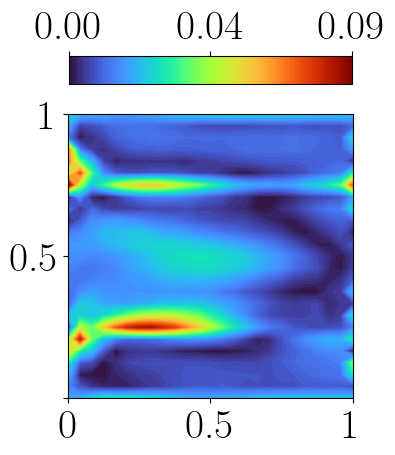}
         \caption{$t = 2.5$}
     \end{subfigure}
     \begin{subfigure}[t]{0.24\textwidth}
         \centering
         \includegraphics[width=\textwidth]{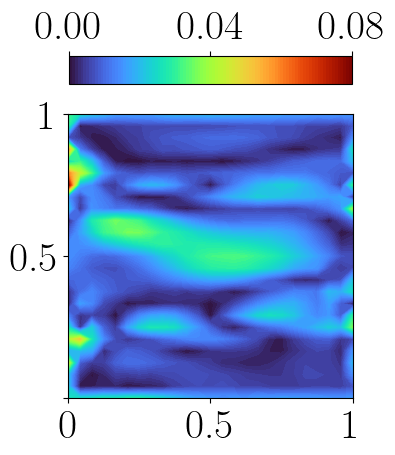}
         \caption{$t = 5$}
     \end{subfigure}
     \caption{Conjugate heat transfer in three dimensions: Temporal evolution of temperature profiles (top row) on a YZ plane at x = 0.5 and the corresponding point-wise absolute errors of the temperature predictions (bottom row) with $f(t) = e^{-\lambda t}(a + \cos(\omega t))$.}
     \label{fig:3D_Temp_Error_in_2D}
\end{figure}

\begin{figure}[h!]
     \centering
     \begin{subfigure}[t]{0.32\textwidth}
         \centering
         \includegraphics[width=\textwidth]{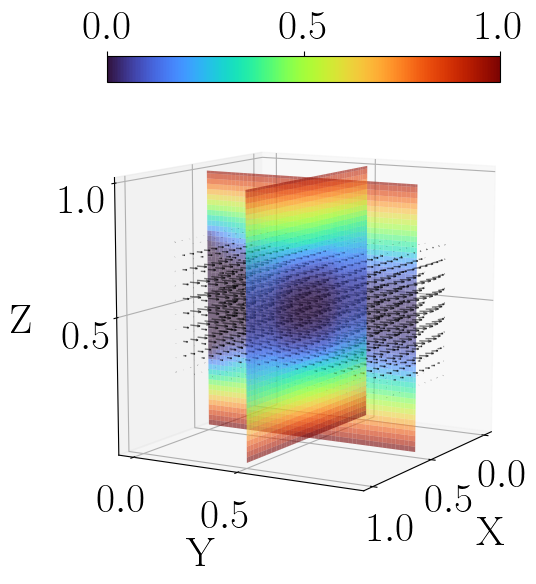}
         \caption{$t = 1.5$}
     \end{subfigure}
     \begin{subfigure}[t]{0.32\textwidth}
         \centering
         \includegraphics[width=\textwidth]{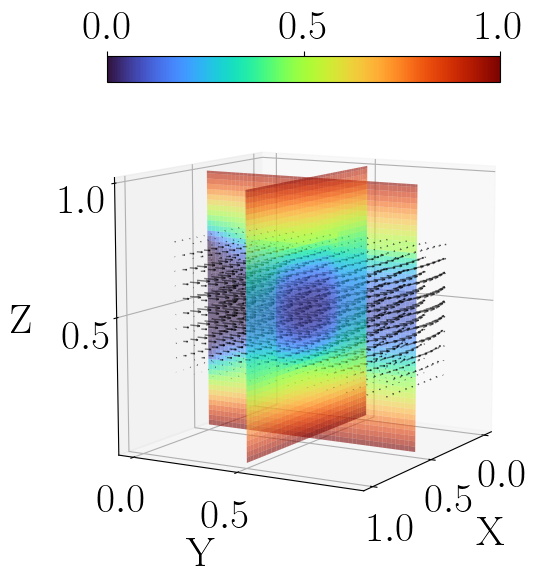}
         \caption{$t = 2.5$}
     \end{subfigure}
     \begin{subfigure}[t]{0.32\textwidth}
         \centering
         \includegraphics[width=\textwidth]{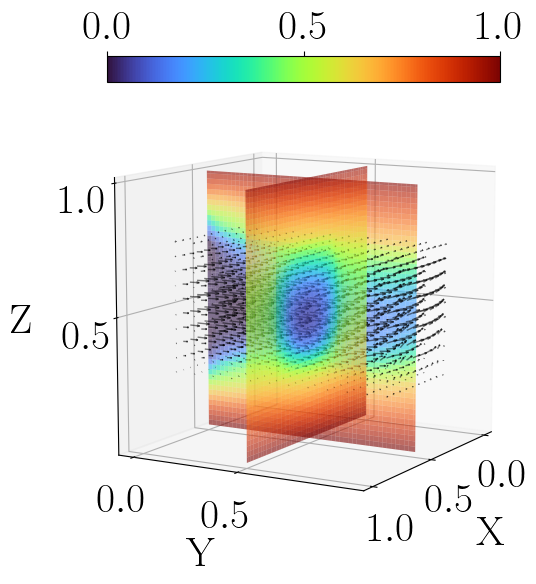}
         \caption{$t = 5$}
     \end{subfigure}

     \begin{subfigure}[t]{0.32\textwidth}
         \centering
         \includegraphics[width=\textwidth]{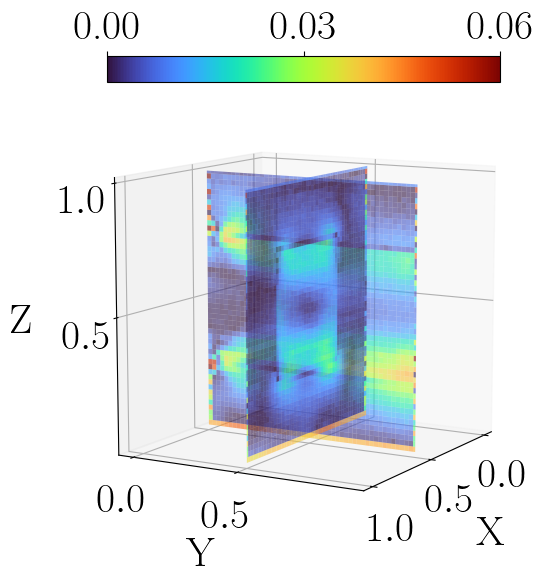}
         \caption{$t = 1.5$}
     \end{subfigure}
     \begin{subfigure}[t]{0.32\textwidth}
         \centering
         \includegraphics[width=\textwidth]{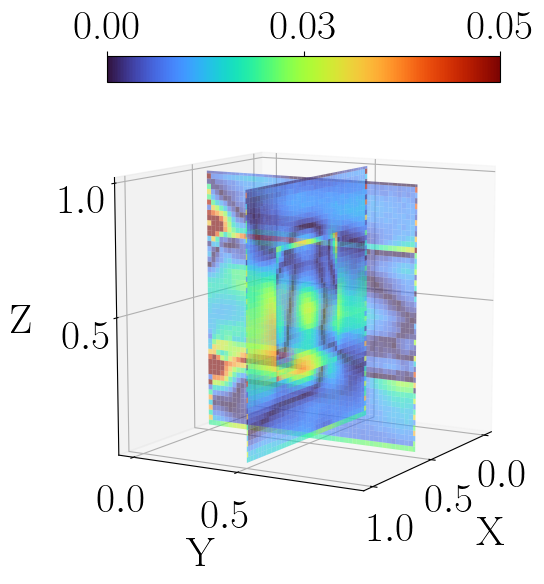}
         \caption{$t = 2.5$}
     \end{subfigure}
     \begin{subfigure}[t]{0.32\textwidth}
         \centering
         \includegraphics[width=\textwidth]{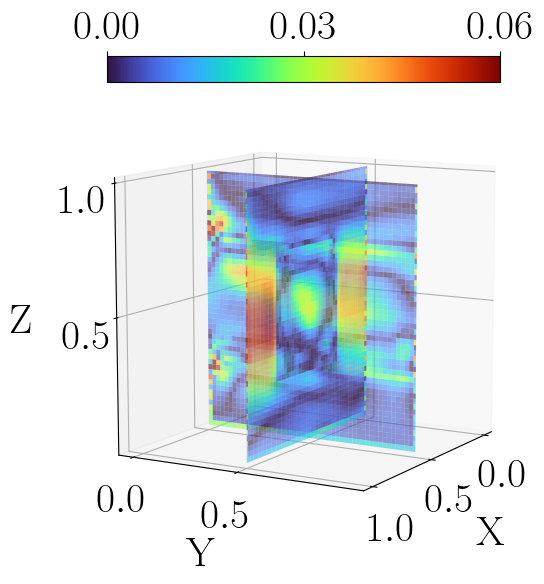}
         \caption{$t = 5$}
     \end{subfigure}
     \caption{Conjugate heat transfer in three dimensions: The top row illustrates the temporal evolution of temperature profiles and the accompanying velocity field on the YZ-plane ($x = 0.5$) and XZ-plane ($y = 0.5$). The bottom row displays the corresponding point-wise absolute errors for the temperature predictions with $f(t) = e^{-\lambda t}(a + \cos(\omega t))$.}
     \label{fig:3D_Temp_Error}
\end{figure}
Figure~\ref{fig:3D_Temp_Error_in_2D} illustrates the system's 
transient behavior and evaluates the model's predictive accuracy 
from $t = 1.5$ to $t = 5.0$. The top row (a--c) shows the temperature evolution over time on a YZ plane at $x = 0.5$. We observe that the temperature diffuses from the top and bottom boundaries $(z = \{0, 1\})$ and the cold fluid entering the left boundary $(y = 0)$ advects the hot fluid. As time progresses, the hot fluid experiences a buoyant force that drives it towards the z-direction. We observe that the temperature profile deviates upward near the outlet at $t = 5$. This phenomenon becomes clearer when we visualize the temperature profile alongside the velocity field, as in Figure~\ref{fig:3D_Temp_Error}. As time progresses, the velocity vectors point towards the z-direction as the temperature diffuses through the fluid. Furthermore, both figures display the point-wise absolute error for extrapolation across different time instances. This error is predominantly concentrated near the fluid-solid interface and boundaries. The concentration of error near the interface is because of the the discontinuity in the thermal diffusion coefficient. This mathematical discontinuities present a significant predictive challenge for the network. The overall error magnitude remains remarkably stable and bounded within a similar range over time. Ultimately, this highlights the method's robust predictive capabilities, proving it highly effective at accurately extrapolating the solution well beyond the training time.

\section{Conclusion}
In this work, we introduced a steady-state informed neural network representation for dissipative PDE systems. Standard neural networks often struggle with temporal extrapolation, but the proposed approach addresses the limitation by exploiting a structural property of such PDEs: they eventually converge to a stationary equilibrium. We decompose the solution into a stationary steady-state component and a transient correction modulated by a time-dependent decay profile, which embeds the asymptotic behavior directly into the architecture.

Using a series of numerical experiments, we demonstrate the method's performance across various PDEs, from the one-dimensional heat equation to a fully coupled three-dimensional fluid-and-heat system. Numerical experiments confirm that the proposed architecture enhances the extrapolation capability of neural networks well beyond the training horizon. The selection of the temporal profile $f(t)$ is critical and problem-dependent. We observed that, in general, exponential decay (either oscillatory or non-oscillatory) provides robust performance. Ultimately, this methodology offers an effective modeling strategy for predicting long-term thermo-fluid dynamics in practical engineering applications.

\section*{Acknowledgments}
Poudel and Lee were partially supported by the U.S. Department of Energy, Office
of Science, Energy Earthshots Initiatives under Award Number DE-SC0024703 and by the U.S. National Science Foundation under Grant DMS-2208402.

\bibliographystyle{unsrt}  
\bibliography{references}  

\end{document}